\documentclass[aps,pra,showpacs,notitlepage,onecolumn,superscriptaddress,nofootinbib]{revtex4-1}
\usepackage[utf8]{inputenc}
\usepackage[tmargin=1in, bmargin=1in, lmargin=1in, rmargin=1in]{geometry}
\usepackage{amsmath, amssymb, amsthm}
\usepackage{graphicx}
\usepackage[svgnames]{xcolor}
\usepackage{enumitem}
\usepackage{titlesec}
\usepackage{datetime}
\usepackage{hyperref}
\usepackage{import}
\usepackage{mathtools}
\usepackage{tikz}

\usepackage{makecell} 
\setcellgapes{2pt}

\usepackage{tikz-cd}

\definecolor{crimson}{RGB}{186,0,44}
\definecolor{moss}{RGB}{0, 186, 111}

\hypersetup{
    colorlinks,
    linkcolor={crimson},
    citecolor={crimson},
    urlcolor={crimson}
}

\theoremstyle{definition}
\newtheorem{definition}{Definition}[section]
\newtheorem{lemma}{Lemma}[section]
\newtheorem{theorem}{Theorem}[section]
\newtheorem{corollary}{Corollary}[theorem]
\newtheorem{remark}{Remark}[section]

\newtheorem{example}{Example}[section]

\newtheorem*{definition*}{Definition}
\newtheorem*{lemma*}{Lemma}
\newtheorem*{theorem*}{Theorem}
\newtheorem*{corollary*}{Corollary}

\newenvironment{bmatrixcolor}[1][black]
    {\colorlet{savethecolor}{.}\colorlet{bracecolor}{#1}%
    \color{bracecolor}\left[\color{savethecolor}\begin{matrix}}
    {\end{matrix}\color{bracecolor}\right]}

\makeatletter
\newcommand{\subalign}[1]{%
  \vcenter{%
    \Let@ \restore@math@cr \default@tag
    \baselineskip\fontdimen10 \scriptfont\tw@
    \advance\baselineskip\fontdimen12 \scriptfont\tw@
    \lineskip\thr@@\fontdimen8 \scriptfont\thr@@
    \lineskiplimit\lineskip
    \ialign{\hfil$\m@th\scriptstyle##$&$\m@th\scriptstyle{}##$\hfil\crcr
      #1\crcr
    }%
  }%
}
\makeatother

\begin{document}

\title{A Solovay–Kitaev theorem for quantum signal processing}
\author{Zane M.~Rossi}
\affiliation{Department of Physics, Graduate School of Science, The University of Tokyo, Hongo 7-3-1, Bunkyo-ku, Tokyo 113-0033, Japan}


\begin{abstract}
    \noindent Quantum signal processing (QSP) studies quantum circuits interleaving known unitaries (the phases) and unknown unitaries encoding a hidden scalar (the signal).
    For a wide class of functions one can quickly compute the phases applying a desired function to the signal; surprisingly, this ability can be shown to unify many quantum algorithms.
    A separate, basic subfield in quantum computing is gate approximation: among its results, the Solovay–Kitaev theorem (SKT) establishes an equivalence between the universality of a gate set and its ability to efficiently approximate other gates.

    
    In this work we prove an `SKT for QSP,' showing that the density of parameterized circuit ansätze in \emph{classes of functions} implies the existence of short circuits approximating desired functions.
    This is quite distinct from a pointwise application of the usual SKT, and yields a suite of independently interesting `lifted' variants of standard SKT proof techniques.
    Our method furnishes alternative, flexible proofs for results in QSP, extends simply to ansätze for which standard QSP proof methods fail, and establishes a formal intersection between QSP and gate approximation.
\end{abstract}

\maketitle

\section{Introduction}

\noindent Quantum signal processing (QSP) \cite{ylc_14, lyc_16_equiangular_gates, lc_17_simulation, lc_19_qubitization} and quantum singular value transformation (QSVT) \cite{gslw_19} have proven useful in unifying the presentation of most quantum algorithms, improving their performance, and simplifying their analysis \cite{mrtc_21}. These algorithms are built on a simple alternating circuit ansatz, interleaving applications of a unitary oracle encoding an unknown scalar \emph{signal} with user-specified unitaries parameterized by a finite list of real \emph{phases}. The choice of phases specifies a function applied to the signal, and in their various forms these algorithms allow precisely tunable functions from a wide class to be efficiently applied to the spectra of large linear operators,\footnote{Recovering QSVT from QSP follows from ensuring invariant two-dimensional subspaces through the cosine-sine decomposition \cite{cs_qsvt_tang_tian, pw_cs_decomp_94} (discussed in Rem.~\ref{rem:two_notions_lifting_qsp}); this allows one to apply QSP-achievable polynomials to the spectra of certain linear operators \emph{very efficiently}, capturing BQP-complete problems and thus quantum speedups \cite{gslw_19, mrtc_21}.} enabling diverse problems to be encoded in a single quantum algorithmic setting.

Speaking informally, these parameterized ansätze, which multiply known and unknown unitaries, strike a subtle balance, as these products appear to `fill' some `space' of `useful' protocols quickly (discussed below), while remaining quite structured and thus simple to analyze and optimize. To jump ahead, one is reminded of results like the Solovay–Kitaev theorem (SKT) \cite{dn_skt_overview_05, nc_textbook_11} from gate approximation, where one starts with a finite `instruction set' of implementable quantum gates known to `fill' a compact Lie group like SU(2), and leverages this to show that, in fact, one can approximate any element of this Lie group \emph{efficiently} with only products from the instruction set. In this work we investigate whether this hazy analogy has legs.

The relation between QSP phases and achievable functions is well-understood---up to choice of basis, ansatz variant, and signal parameterization, QSP unitaries have matrix elements which, taken as a function of the signal, are dense in simply-described classes of functions (e.g., those of Def.~\ref{def:qsp_skt_function_space}). Proofs of this assertion are uniformly \emph{inductive} and \emph{constructive}, yielding classical algorithms which, given a target function satisfying constraints, can output QSP phases whose unitary has a matrix element approximating the target function evaluated at the unknown signal. The common conceit of these proofs is to instantiate the following algorithms (using the conventions of \cite{amt_23}):
    \begin{align}
        P(x) \in \mathbb{R}_{n}[x] &\overset{\mathcal{A}}{\longmapsto} \Phi \in \mathbb{R}^{n + 1}, &&\text{Phase finding},\label{eq:qsp_phase}\\
        \Phi \in \mathbb{R}^{n + 1}&\overset{\mathcal{B}}{\longmapsto} U(\Phi, x) \in \text{SU(2)}, &&\text{Compilation},\label{eq:qsp_comp}\\
         U(\Phi, x) \in \text{SU(2)} &\overset{\mathcal{C}}{\longmapsto} \Im{[\langle 0|U(\Phi, x)|0\rangle]} \approx_{\varepsilon} P(x),&&\text{Projection}.\label{eq:qsp_projection}
    \end{align}
Here $\mathcal{A}$ is a classical algorithm, the \emph{phase finding algorithm}, taking a polynomial as input and outputting a list of phases, $\mathcal{B}$ is a \emph{compilation algorithm} taking phases as input and outputting a QSP unitary, and $\mathcal{C}$ is a \emph{projection algorithm} taking a unitary as input and outputting a component of a matrix element. 

The core insight of QSP is that, given various $P(x)$, a suitable $\Phi$ can be quickly found satisfying the $\varepsilon$-approximation condition ($\approx_{\varepsilon}$) of (\ref{eq:qsp_projection}), i.e., the composition $(\mathcal{C} \circ \mathcal{B} \circ \mathcal{A})$ is \emph{approximately the identity}, with respect to standard metrics, over a wide class of input polynomials.\footnote{This is a simplified view; the strengths of the theory of QSP also lie in the fact that the required length of $\Phi$ for a given $P(x)$ is surprisingly short and easy to calculate, and moreover that each of $\mathcal{A}$, $ \mathcal{B}$, and $\mathcal{C}$ have simple interpretation.} With this knowledge of $\mathcal{A}$, $\mathcal{B}$, and $\mathcal{C}$, the algorithmist can design and apply QSP/QSVT (per Rem.~\ref{rem:two_notions_lifting_qsp}) to their problem of interest.
    
However, while the understanding above can be found even in the earliest papers on QSP \cite{ylc_14}, in practice the algorithmist encounters nonidealities. Among these, instantiations of $\mathcal{A}$ implied by QSP's constructive proofs are slow and unstable.\footnote{Here meaning that the number of bits of precision required scales polynomially (not \emph{poly-logarithmically}; see the introduction of \cite{szego_nlfa_qsp_24}) with the inverse of the desired precision.} Substantial research has dramatically lowered the required precision and runtime of $\mathcal{A}$ \cite{haah_2019,chao_machine_prec_20, dong_efficient_phases_21, mw_gqsp_24}, to such an extent that a basic laptop can compute $10^7$ phases in seconds. Curiously, these state-of-the-art algorithms exhibit no obvious relation to QSP proof methods, and are based on iterative numerical optimization and fixed-point theorems \cite{dong_efficient_phases_21, dlnw_infinite_22}. A second shortcoming appears when one modifies the circuit ansatz; this has been done extensively (see Sec.~\ref{sec:prior_work}) and in each case requires updating a chain of fragile proofs. Moreover, crucially, for many extensions, e.g., the multivariable setting \cite{rc_m_qsp_22, mori_m_qsp_comment_23, nemeth_m_qsp_23}, there exist \emph{no known} sufficient modifications to simply describe the class of achievable functions, and thus no $\mathcal{A}$.

The sum of these observations is that current proof methods analyzing QSP ansätze are (1) fragile and insufficient, and (2) misaligned with how one practically and numerically treats these algorithms. In the following Sec.~\ref{sec:prob_setup}, we formalize this mismatch into a series of problem statements, toward summarizing our main result in Sec.~\ref{sec:main_results}: a new class of \emph{non-constructive}, \emph{flexible} proof methods for results in QSP, based in techniques used to prove the well-known Solovay–Kitaev theorem (SKT) for gate approximation \cite{dn_skt_overview_05}. Just as in the usual SKT, our result shows that QSP ansätze that are dense in classes of functions (often easier to show) are immediately guaranteed to be \emph{efficiently dense} (i.e., the approximating circuit is short). Together with numerical phase finding algorithms, this provides a novel path to recover $\mathcal{A}$ in Eq.~\ref{eq:qsp_phase}.

Our `SKT for QSP' serves multiple purposes: (a) as a collection of \emph{detours} around barriers in standard QSP proofs, (b) as a method compatible with the (already dominant) use of numerical phase finding algorithms, extensible to wider classes of \emph{QSP-like} ansätze, and (c) as a concrete investigation into `lifted' SKTs (see Rem.~\ref{rem:two_notions_lifting_qsp}). This connection, which goes well beyond applying the SKT in a pointwise way, opens novel, bi-directional bridges between the previously disparate subfields of QSP/QSVT and gate approximation, and demonstrates the feasibility of diverse methods to satisfy the $(\mathcal{C}\circ\mathcal{B}\circ\mathcal{A})$ condition between Eqs.~\ref{eq:qsp_phase}--\ref{eq:qsp_projection}.

We compare our construction with prior work on both QSP and the SKT in Sec.~\ref{sec:prior_work} before moving onto an overview of common techniques from the proof of the standard SKT in Sec.~\ref{sec:skt_thm} (many of which we re-purpose), followed by definitions, lemmata, and theorems composing our main results in Sec.~\ref{sec:qsp_skt}. We discuss and demonstrate the utility of our results for concrete QSP variants in Sec.~\ref{sec:skt_qsp_applications}, followed by discussion and open questions in Sec.~\ref{sec:discussion_open}.

\subsection{Problem setup} \label{sec:prob_setup}

\begin{figure}
    \centering
    \includegraphics[width=0.9\textwidth]{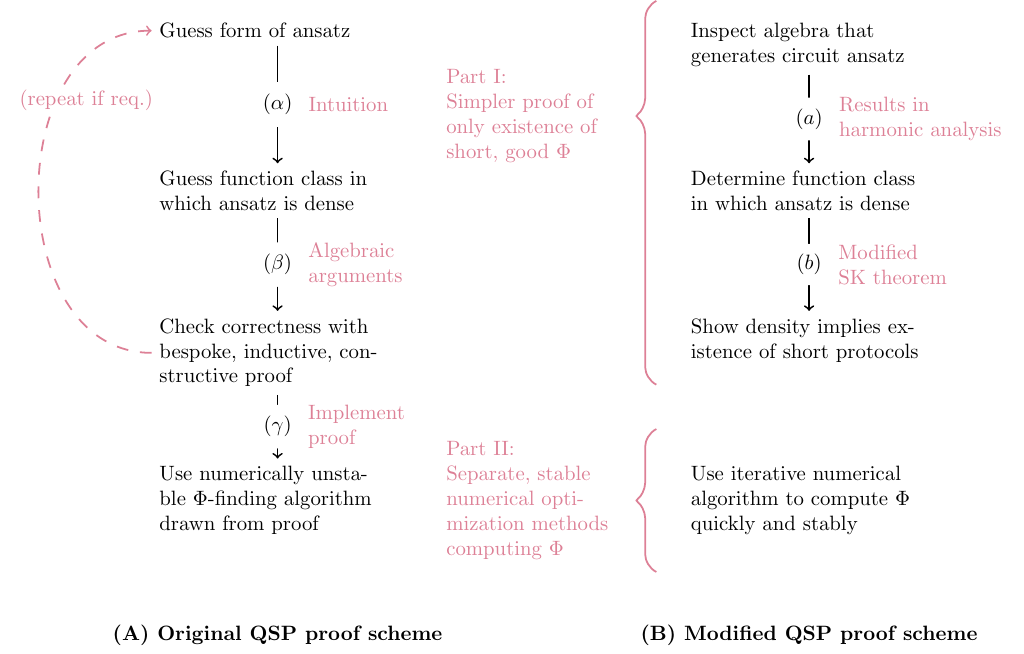}
    \caption{Diagrammatic overviews of (A) the standard proof scheme for investigating and applying QSP-like ansätze and (B) the scheme constructed in this work. Instead of guessing a circuit ansatz, guessing its functional properties, and devising constructive, algebraically-intense proofs whose corresponding algorithms have poor numerical properties, we divide the QSP proof scheme into two parts. We (1) directly connect properties of the \emph{algebra} generating an ansatz to its \emph{functional} properties to give \emph{existence} proofs of good protocols by Solovay–Kitaev (SK) methods, and then (2) apply numerical optimization methods for phase finding under the guarantee from (1).}
    \label{fig:proof_scheme_summary}
\end{figure}

\noindent Proof methods in QSP are brittle: they have to be renovated whenever the ansatz is modified, and their constructive aspect yields substandard phase-finding algorithms. The structure of such `standard' proofs in QSP is discussed in Rem.~\ref{rem:anatomy_qsp_proof} and depicted in Fig.~\ref{fig:proof_scheme_summary}; while these proofs are succinct and beautiful, refined over years, they have proven difficult to extend or warp.

In mechanical structures brittleness is resolved in multiple ways: annealing, alloys, expansion joints, etc. In each case one wants to allow for `give': decoupling internal structures so that stress can be dispersed. For proofs this might correspond to modularity---a final result might follow from a chain of statements whose input and output assumptions can each be guaranteed in multiple ways, such that even if one version of one lemma fails in one setting, its cousin can be re-proven without scrapping other parts of the argument.

For expediency we give a minimal definition and theorem below (Def.~\ref{def:qsp} and Thm.~\ref{thm:qsp_main_results}), describing main statements in QSP. Together these describe the most common path for `designing' QSP protocols, given a partial specification of the target unitary. Recall that this SU(2) analysis is the critical step to building QSVT, as the same functions can be applied to the spectra of large linear operators `block encoded' in unitary processes (Rem.~\ref{rem:two_notions_lifting_qsp}).

\begin{definition}[Quantum signal processing protocol] \label{def:qsp}
    Let $\Phi = \{\phi_0, \phi_1, \dots, \phi_k\} \in \mathbb{R}^{k + 1}$; the QSP protocol according to $\Phi$ evaluated at a signal $x \in [-1,1]$ is denoted $U(\Phi, x)$ and has the form:
    \begin{equation} \label{eq:qsp_product_form}
        U(\Phi, x) 
        \equiv
        e^{i\phi_0\sigma_z}
        \prod_{j = 1}^{k} 
        \left[
        \underbrace{
        \begin{bmatrix}
            x & i\sqrt{1 - x^2}\\
            i\sqrt{1 - x^2} & x
        \end{bmatrix}
        }_{\textstyle e^{i\arccos{(x)}\,\sigma_x}}
        \underbrace{
        \begin{bmatrix}
            e^{i\phi_j} & 0\\
            0 & e^{-i\phi_j}
        \end{bmatrix}
        }_{\textstyle e^{i\phi_j\sigma_z}}
        \right]
        =
        \begin{bmatrixcolor}[black!30!white]
            P(x) & \textcolor{black!30!white}{iQ(x)\sqrt{1- x^2}}\\
            \textcolor{black!30!white}{\ast} & \textcolor{black!30!white}{\ast}
        \end{bmatrixcolor}.
    \end{equation}
    Here the unitary $W(x) \equiv \exp{(i\arccos{(x)}\,\sigma_x)}$ is the \emph{signal unitary} (or oracle), the scalar $x$ the \emph{signal}, and the interspersing $\exp{(i\phi_j\,\sigma_z)}$ the \emph{phase unitaries} according to the programmable QSP \emph{phases} $\Phi$.
\end{definition}

\begin{theorem}[QSP main theorem (abridged) \cite{gslw_19}] \label{thm:qsp_main_results}
    Let $k \in \mathbb{N}$; there exists $\Phi \in \mathbb{R}^{k + 1}$ as in Def.~\ref{def:qsp} such that for all $x \in [-1, 1]$ the unitary $U(\Phi, x)$ encodes $P, Q \in \mathbb{C}[x]$ (polynomials in $x$ with complex coefficients) \emph{if and only if} $P, Q$ satisfy the following:
        \begin{enumerate}[label=(\arabic*)]
            \item $\text{deg}(P) \leq k$ and $\text{deg}(Q) \leq k - 1$.
            \item $P$ has parity-$(k\pmod{2})$ and $Q$ has parity-$(k-1 \pmod{2})$.
            \item $\forall x \in [-1, 1]$: $\lvert P(x)\rvert^2 + (1 - x^2)\lvert Q(x)\rvert^2 = 1$. 
        \end{enumerate}
    Moreover, let $k \in \mathbb{N}$ and $\tilde{P}, \tilde{Q} \in \mathbb{R}[x]$. Then there exist $P, Q \in \mathbb{C}[x]$ such that (a) $\tilde{P} = \Re[P]$ and $\tilde{Q} = \Re[Q]$ and (b) $P, Q$ satisfy conditions (1)-(3) above \emph{if and only if} $\tilde{P}$ and $\tilde{Q}$ satisfy conditions (1)-(2) \emph{and} $|\tilde{P}(x)|^2 + (1 - x^2) |\tilde{Q}(x)|^2 \leq 1$.
\end{theorem}

Given the above, we can see how a quantum algorithmist might `design' a QSP protocol, say by choosing $\Im[P]$; the useful statements an algorithmist wants to know about a QSP ansatz thus come in a few flavors:
    \begin{enumerate}[label=(\alph*)]
        \item \textbf{Expressiveness:} We want a \emph{simple description} of the \emph{image} of \emph{physically reasonable} parameterizations. Following Def.~\ref{def:qsp}, this means asking which $P, Q$ are possible for reasonable $\Phi$.
        
        \item \textbf{Running complexity:} We want to know the resources required (e.g., quantum gates, auxiliary qubits, etc.) to \emph{run} a quantum algorithm that (approximately) achieves this target. This might mean asking, for a given $P, Q$, the minimal required length of $\Phi$.

        \item \textbf{Programming algorithm:} We want an efficient method by which to compute the parameterization for a circuit (approximately) achieving a target. Following Def.~\ref{def:qsp}, if we choose (or partially specify) $P, Q$, how do we actually compute $\Phi$?
        
        \item \textbf{Programming complexity:} We want to know the classical resources required to compute the parameterization for a circuit achieving a target. Following Def.~\ref{def:qsp}, if we partially specify $P, Q$, how expensive is it to compute $\Phi$?
    \end{enumerate}
To tie the questions above to actual proof methods, we now gloss the anatomy of a standard proof in QSP. We also depict this structure in Fig.~\ref{fig:proof_scheme_summary}, side-by-side with a new, decoupled version that will be the main focus of this work (with constitutive theorems in Sec.~\ref{sec:main_results}).

\begin{remark}[Anatomy of a standard proof of QSP properties] \label{rem:anatomy_qsp_proof}  
    For the standard QSP ansatz, establishing \emph{necessary} properties of the image is straightforward, and we refer to this as the forward $(\implies)$ direction. Proofs of this direction induct on protocol length, establishing degree, parity, and matrix-element relations for possible QSP unitaries in terms of polynomials of the signal.

    The more difficult task, sometimes called the reverse $(\impliedby)$ direction, starts with a desired matrix element component (e.g., $\Im[P(x)]$) and determines at least one $\Phi$ whose unitary approximately agrees with this component. Proofs of such statements tend to be broken into two parts:
        \begin{align}
            \Im[P(x)] &\overset{\mathcal{D}}{\longmapsto} P(x), Q(x)\,\text{ s.t. } |P(x)|^2 + (1 - x^2)|Q(x)|^2 = 1, && \text{Completion},\\
            P_{n}(x), Q_{n}(x) &\overset{\mathcal{E}}{\longmapsto} P_{n - 1}(x), Q_{n - 1}(x), \phi_n, && \text{Layer stripping}.
        \end{align}
    The first, the classical algorithm $\mathcal{D}$, takes a desired polynomial target $\Im[P(x)]$ (or some other component of a matrix element) and finds a \emph{complementary polynomial} $Q(x)$ (as well as the missing real part of $P(x)$) such that the two together obey a constraint, `completing' the missing elements of the unitary. This `completion' depends on some basic results in the theory of the decomposition of positive polynomials, such as the Fejér-Riesz lemma \cite{polya_szego_analysis_98}. The second algorithm $\mathcal{E}$ is presented in an inductive form, taking complementary polynomials of a given degree and returning a complementary pair of lower degree \emph{plus} a single real phase $\phi_n \in \mathbb{R}$. This process is repeatedly applied (`stripping off layers') until a base case is reached, where the accrued $\{\phi_n, \phi_{n - 1}, \dots, \phi_0\}$ constitute $\Phi$. Together the algorithms $\mathcal{D}$ and $\mathcal{E}$ clearly allow one to answer the four QSP questions above.
\end{remark}

Having outlined common algorithms and methods for the proof of basic attributes of QSP, we can discuss their current status in answering the four QSP questions.
\begin{enumerate}[label=(\alph*)]
    \item \textbf{Status of expressiveness:} Sufficient conditions on $\Im[P]$ are mild: \emph{any} definite-parity (bounded on $[-1,1]$) polynomial can be realized; moreover, such polynomials are \emph{dense} in simple classes of functions.
    
    \item \textbf{Status of running complexity:} By the recursive nature inherent to $\mathcal{E}$, a polynomial target of degree $n$ means the length of $\Phi$ scales linearly with $n$.
    
    \item \textbf{Status of programming algorithm:} The programming complexity and the programming algorithm relate simply: the composition $(\mathcal{E} \circ \mathcal{D})$ is sufficient to recover the phases.

    \item \textbf{Status of programming complexity:} determining  $\Phi$ for a chosen $\Re[P]$ requires running $\mathcal{E} \circ \mathcal{D}$ and recursing. This might rely on root finding for high-degree polynomials, taking time $\text{poly}(n, 1/\varepsilon)$ for a degree-$n$ target to precision $\varepsilon$, meaning numeric instability.
\end{enumerate}
That standard QSP proof methods work so well is a little mysterious. It just so happens that the necessary and sufficient conditions for possible $P, Q$ basically coincide (meaning a reasonable inductive hypothesis can be guessed), \emph{and} achievable $P, Q$ are dense in a simply describable class of functions. Practically all elements of the proof of attributes of the QSP ansatz rely on the simplicity of $\mathcal{D}$ and $\mathcal{E}$, whose felicitous existence and efficiency depend on specific functional analytic results.

The success of QSP hides that the functional properties of its ansatz were basically \emph{guessed and checked}. We depict this in Fig.~\ref{fig:proof_scheme_summary}, and propose a parallel, decoupled version for which, rather than guessing and checking properties of a given ansatz, one can \emph{directly determine} its functional properties.

\begin{remark}[Motivation and outlook]
    Before finally summarizing our main results we further situate (1) why alternative proof methods for QSP-like ansätze are desirable, and (2) why it is reasonable that gate approximation techniques might work.
    \begin{enumerate}[label=(\arabic*)]
        \item \textbf{Gate universality is generic and gate approximation is efficient:} Observed early in \cite{lloyd_gate_set_universal_95}, \emph{most} quantum gate sets are universal, and compiling one gate set to another is almost always efficient \cite{dn_skt_overview_05}. It is an interesting question then whether QSP, viewed as a lifted form of gate approximation (Fig.~\ref{fig:skt_lifting_version}), permits the same type of `generically universal and efficient' statements for \emph{functions}.
    
        \item \textbf{The best methods for computing QSP phases are numerical:} The algebraic algorithms for phase-finding in QSP have been superseded by Newton's-method approaches \cite{dong_efficient_phases_21, sym_qsp_21}. These iterative algorithms compute basic gradients and are simple to generalize. This poses a new, useful couplet: (1) a proof of \emph{existence} of a short $\Phi$, and (2) a fast numerical $\Phi$-finding algorithm.
    
        \item \textbf{It is easier to work with Lie algebras than Lie groups:} QSP exhibits favorable space and time complexity compared many block encoding algorithms. Formalizing this idea has been the focus of recent work casting QSP as a type of non-linear Fourier analysis (NLFA) \cite{amt_23, szego_nlfa_qsp_24}. Many basic statements in linear Fourier analysis are severely weakened in the nonlinear setting. It would be great, then, if properties of QSP could be derived from `linearized' limits where the Lie algebra is manipulated directly. This simple idea underlies the usual SKT \cite{dn_skt_overview_05, nc_textbook_11}.
    
        \item \textbf{Algorithms for manipulating block encodings should be modular:} Standard QSP proofs tightly bound resource complexity. This work explores whether, by sacrificing some of this refinement, our methods can work in extended settings. A secondary benefit is that we can allow for the application of the many algorithms that manipulate block encodings, like LCU \cite{bck_ham_lcu_15}, linear combination of Hamiltonian simulation (LCHS) \cite{dll_lchs_23}, QEVT \cite{low_su_qevt_24}, and others \cite{mw_gqsp_24, mrclc_parallel_qsp_24, martyn_rall_halving_24}. Each of these can perform better in specific settings, and thus ought to be interoperable.
    \end{enumerate}
\end{remark}

\subsection{Main results} \label{sec:main_results}

\noindent Here we minimally present our main theorems and methods---as suggested by the title, this involves instantiating common objects and ideas from both the theory of quantum signal processing and Solovay–Kitaev-like theorems. Information on prior work is given in Sec.~\ref{sec:prior_work}, and some full definitions, lemmata, and theorems from the main Secs.~\ref{sec:qsp_skt} and \ref{sec:skt_qsp_applications} are suppressed. 

We have to do some work to make SKTs and QSP compatible---consequently there are many sub-results to which which recent SKT methods might be employed to improve constant factors (see Sec.~\ref{sec:discussion_open}). We try to apply clear constructions rather than optimal ones so that someone new to any subset of the subfields referenced might appreciate which aspects of our methods were known versus which were non-trivial.
\begin{remark}[Overview of methods]
    This work is broken up into a few parts: the first two establish a pair of related mathematical results allowing SKT-based methods to apply to the analysis of QSP-like circuit ansätze, while the final part investigates practical algorithmic properties of these results.
    \begin{enumerate}[label=(\arabic*)]
        \item We want to establish the density (Def.~\ref{def:pi_density_qsp}) of QSP circuit ansätze in classes of functions. This requires firm mathematical definitions of what we mean by \emph{density}, \emph{ansatz}, and \emph{class of functions}. While showing density may sound as difficult as recovering the whole theory of QSP, we will not need such approximations to be \emph{efficient}, and so can uniquely consider compositions of simple subroutines strictly easier to construct than QSP (e.g., as in Thm.~\ref{thm:abs_sum_sym_qsp_density}).

        \item Separately, as per the standard Solovay–Kitaev theorem, we then show that the density property proven above implies something algorithmically useful: efficient approximations (i.e., that the existence of good approximating circuits implies the existence of \emph{good and short} circuits). 
        
        A difficulty for this argument is that the group commutators usually employed in SKT proofs fail to preserve critical structure required of our QSP ansätze. Building a modified group commutator that preserves this structure, as well as preserves certain continuity properties, constitutes the bulk of our work. The main takeaway is this: while applying the SKT in a pointwise way is \emph{not} sufficient for our approximation of \emph{functions}, many SKT tools \emph{can} be repurposed.
        
        \item Finally, after establishing an SKT for QSP (Thm.~\ref{thm:qsp_skt}, informally stated in Thm.~\ref{thm:informal_qsp_skt}), we analyze specific ansätze. Namely, our result establishes many couplets of theorems showing (a) the density of a particular circuit ansatz in a particular function class, and (b) that this density implies the existence of \emph{short} approximations. We show variants of this (a)--(b) couplet for differing ansätze and function classes in Sec.~\ref{sec:skt_qsp_applications}, and discuss lower bounds and connections to other block encoding methods.
    \end{enumerate}
\end{remark}

\begin{figure}
    \centering
    \includegraphics[width=0.8\textwidth]{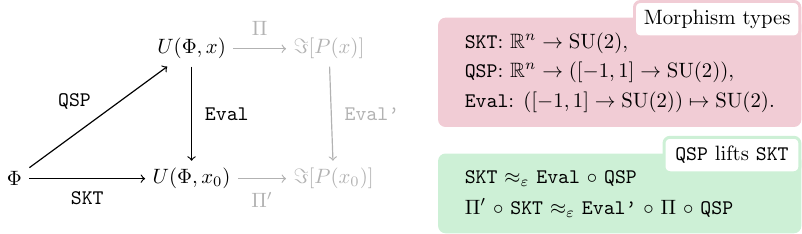}
    \caption{Saying QSP lifts the Solovay–Kitaev theorem (SKT) is meant as in Def.~\ref{def:lift}; we define three morphisms, (1) \texttt{SKT} taking $\Phi \in \mathbb{R}^n$ to a gate sequence $U(\Phi, x_0)$, (2) \texttt{QSP} taking $\Phi$ to a \emph{function} $U(\Phi, x)$ from $[-1,1]$ to SU(2), and (3) \texttt{Eval} evaluating a QSP protocol at $x = x_0$. Lifting means \texttt{SKT} can be rewritten\footnote{We use the (for now vague) notation $\approx_\varepsilon$ to indicate we will accept approximate lifting to arbitrary precision $\varepsilon$.} as the composition \texttt{Eval} $\circ$ \texttt{QSP}, i.e., that \texttt{SKT} \emph{factors through} \texttt{QSP}. In actuality (indicated in light-gray) we ask for something a little weaker, namely that this factorization holds for a \emph{component of a matrix element} of SU(2) unitaries in the computational basis, selected by a pair of projections: $\Pi: C(X, \text{SU(2)}) \rightarrow C(X, \mathbb{R})$ and $\Pi': \text{SU(2)} \rightarrow \mathbb{R}$. In other words, we ask to satisfy SKT conditions for a continuum of $x_0 \in [-1,1]$ \emph{simultaneously} for the same $\Phi$.}
    \label{fig:skt_lifting_version}
\end{figure}

In many places we have said QSP can be seen as a `lifted' version of the SKT. This \emph{lifting} (depicted in Fig.~\ref{fig:skt_lifting_version}) is different from the \emph{lifting} taking QSP to QSVT (see Rem.~\ref{rem:two_notions_lifting_qsp}). Both, however, are instances of the same mathematical idea: a specific relation among morphisms.

\begin{definition}[Lift] \label{def:lift}
    Let $f: X \rightarrow Y$ and $g: Z \rightarrow Y$ morphisms.\footnote{Morphisms have a strict definition, but can be thought of as arrows acting on objects which allow composition (arrows can be strung together tip-to-tail). Composition is associative, and has an identity (an arrow from an object to itself).} Then a lifting of $f$ to $Z$ is a morphism $h: X \rightarrow Z$ such that $f = g \circ h$, or equivalently that $f$ \emph{factors through $h$}. The category-theoretic diagram makes the statement/name more obvious:
        \begin{equation}
            \begin{tikzcd}
            & Z \arrow[d, "g"]\\
            X \arrow[r, "f"] \arrow[ur, "h"] & Y
            \end{tikzcd}
        \end{equation}
    I.e., the action of $f$ is `lifted' such that it goes through a specific path (summarized by $h$), where composing $h$ with $g$ exactly reproduces the morphism $f$. Oftentimes we might think of $g$ as some projection, or loss of information, where $h$ `acts like' $f$ in some larger, more complicated space. In our setting, depicted in Fig.~\ref{fig:skt_lifting_version}, $h$ will concern the behavior of SU(2)-valued functions while $f$ concerns products of unitaries.
\end{definition}

Before giving higher-level versions of the two main results of this work (Thms.~\ref{thm:informal_qsp_skt} and \ref{thm:informal_qsp_density}), we define modified versions of common mathematical objects used in the study of gate approximation, including some auxiliary objects to shift from SU(2) to SU(2)-valued functions. The standard SKT versions of some of these objects is given in the pedagogical Sec.~\ref{sec:skt_thm}.

\begin{definition}[QSP instruction set] \label{def:qsp_instruction_set}
    Let $X$ a compact metric space, $\Pi$ a projector mapping $C(X, \text{SU(2)})$, the space of continuous SU(2)-valued functions on $X$, to $C(X, \mathbb{R})$, and $F(X) \subseteq C(X, \mathbb{R})$.\footnote{Sometimes for brevity we will just write $F$ for $F(X)$.} Finally let \emph{evaluation} take functions defined on $X$ and return their value at a chosen $x \in X$.
    
    A QSP instruction set $\Sigma = \{W, P_0, P_1, \dots, P_n\}$ for $(\Pi, F, X)$ is a finite set of unitaries (quantum gates) and unitary-valued functions satisfying the following:
        \begin{enumerate}[label=(\alph*)]
            \item All elements of $\Sigma$, when evaluated, return results in SU(2).
            
            \item There is a privileged element $W \in \Sigma$, called the \emph{oracle}, which has type $X \rightarrow \text{SU(2)}$. Evaluating $W$ at $x \in X$ returns $W(x)$.
            
            \item The remaining elements, denoted $\{P_0, P_1, \dots, P_n\}$, the \emph{phases}, are elements of SU(2) and evaluate to themselves. This subset is closed under inverses, and finite products of its elements are dense in SU(2).
            
            \item The set $\Sigma$, under evaluation at any $x \in X$, is implicitly expanded to be closed under inverses.
            
            \item The set $\Sigma^\ast$ of finite products of elements of $\Sigma$ is $\Pi$-dense (Def.~\ref{def:pi_density_qsp}) in $F(X)$. I.e., for any $\varepsilon > 0$ and $f \in F(X)$, there exists an element of $\Sigma^\ast$ that is $\varepsilon$-close, under $\Pi$, to $f$ in the supremum norm.
            
            Note that by \emph{products} we mean those induced by the ambient group applied pointwise over $X$. In other words we group-multiply the group-valued outputs of two functions of the same variable and take this product as a new group-valued function of that single, shared variable: this has type
                \begin{align}
                    (X \rightarrow \text{SU(2)}) \times (X \rightarrow \text{SU(2)}) \rightarrow (X \rightarrow \text{SU(2)}),
                \end{align}
            where phases $P_j$ in the instruction set can be viewed as constant $P_j$-valued functions over $X$. This product inherits associativity, identities, and inverses from SU(2); such objects are commonly studied in the context of groups of continuous, group-valued functions $C(X, G)$ endowed with the product topology, but this is beyond our scope \cite{top_groups_90}.
        \end{enumerate}
\end{definition}

The definition of our QSP instruction set is almost identical to that of the standard SKT \cite{dn_skt_overview_05} save for the addition of a privileged parameterized element (an SU(2)-valued function) called the oracle. These products, evaluated \emph{pointwise}, would reduce to a weaker variant of the standard SKT if we only cared about a single $x \in X$. Consequently our contributions focus on ensuring various properties \emph{across signals} in $X$. We now define a few more helpful objects unique to QSP instruction sets.

\begin{definition}[Phase-length and oracle-length] \label{def:phase_oracle_length}
    Let $\Sigma = \{W, P_0, P_1, \dots, P_n\}$ a QSP instruction set for some $(\Pi, F, X)$. Then for every $\sigma \in \Sigma^\ast$ there is an associated tuple of integers $(n, m)$ comprising the \emph{phase length} $n$ and \emph{oracle length} $m$ which count the number of appearances of elements from the subsets $\{P_0, P_1, \dots, P_n\} \subsetneq \Sigma$ and $\{W\} \subsetneq \Sigma$ in $\sigma$ respectively.
\end{definition}

\begin{definition}[\texorpdfstring{$\Pi$}{Pi}-density of QSP protocol] \label{def:pi_density_qsp}
    Let $\Pi: \text{SU(2)} \rightarrow \mathbb{R}$. A QSP instruction set (Def.~\ref{def:qsp_instruction_set}) is said to be $\Pi$-dense in a class of functions $F(X)$ if for all $\varepsilon > 0$ and all $f \in F(X)$, there exists an element of $\Sigma^\ast$ that $\varepsilon$-uniformly approximates $f$ under $\Pi$ (i.e., over all $x \in X$). Normally $\Pi$ has physical significance, e.g., the projection onto a component of a matrix element after evaluation\footnote{Note that while evaluation and our group product commute, $\Pi$ and our group product do not, and thus $\Pi$ is always applied last. Also note we sometimes think of $\Pi$ as performing $G \rightarrow \mathbb{R}$ and other times as performing $C(X, G) \rightarrow C(X, \mathbb{R})$---this causes no issue as evaluation and $\Pi$ \emph{do} commute.} at $x \in X$:
        \begin{equation}
            \Pi : U(\Phi, x) \mapsto \Im[\langle 0 | U(\Phi, x) | 0\rangle].
        \end{equation}
    Usually also $F(X)$ is a well-behaved class of functions (e.g., the common QSP function space in Def.~\ref{def:qsp_skt_function_space}: continuous, bounded, definite parity, real functions with bounded Lipschitz constant over $[-1,1]$).
\end{definition}

We can now state our (condensed) main theorems. QSP ansätze come in many flavors \cite{mrtc_21}; we suppress some of these specifics to emphasize similarity to the standard SKT. In actuality, as is the focus of Sec.~\ref{sec:skt_qsp_applications} and a benefit of our approach, we can provide a variety of related statements for differing ansätze.

\begin{theorem}[A Solovay–Kitaev theorem for QSP; condensed Thm.~\ref{thm:qsp_skt}] \label{thm:informal_qsp_skt}
    Let $\Sigma = \{W, P_0, P_1, \dots, P_n\}$ a \emph{QSP instruction set} for $(\Pi, F, X)$ (Def.~\ref{def:qsp_instruction_set}) where $F$ is the common QSP function space (Def.~\ref{def:qsp_skt_function_space}). 
    Moreover, let $\mathfrak{G}$ a \emph{compatible commutator} for $(\Pi, F, X)$ (Def.~\ref{def:compatible_comm}).
    Then for any $\varepsilon > 0$ and $f \in F(X)$ there exists an element in $\Sigma^*$ (finite pointwise products of elements from $\Sigma$; see Def.~\ref{def:qsp_instruction_set}) of oracle length and phase length (Def.~\ref{def:phase_oracle_length}) $\mathcal{O}(\log^{c}{(\varepsilon^{-1})})$ ($c$ a constant) which $\varepsilon$-uniformly approximates $f$ over $X$, under $\Pi$.
\end{theorem}

\begin{theorem}[Properties of symmetric QSP] \label{thm:informal_qsp_density}
    Let $\Sigma = \{W, P_0, P_1, P_1^\dagger\}$ where $W(x) = \exp\{i\arccos{x}\,\sigma_x\}$ for $x \in X \equiv [-1,1]$ and $P_0, P_1$ are $z$-rotations $R_z(\theta) = \exp\{i\theta\sigma_z\}$ by angles $\{\pi/2, \phi\}$ respectively, where $\phi$ is an irrational multiple of $\pi$. Let $\Pi$ the projection onto the imaginary part of the top-left matrix element in the computational basis. 
    Then $\Sigma$ is $\Pi$-dense in $F(X) \subsetneq C(X, \mathbb{R})$, where $F(X)$ are continuous functions over $X = [-1,1]$ with norm bounded by one of definite parity. Equivalently, $\Sigma$ is an \emph{instruction set} for $(\Pi, F, X)$. Moreover, for any $f \in F(X)$, there exists a \emph{symmetric} element of $\Sigma^\ast$ which $\varepsilon$-approximates $f$, under $\Pi$, over $X$, where symmetric means $\sigma_y U(\Phi, x) \sigma_y = U(\Phi, x)^\dagger$ for any $U(\Phi, x) \in \Sigma^\ast$ at any $x \in X$. Here $\{\sigma_x, \sigma_y, \sigma_z\}$ are the usual Pauli matrices.
    In this case, a \emph{compatible commutator} (Def.~\ref{def:compatible_comm}) $\mathfrak{G}$ for $(\Pi, F, X)$ is given by the \emph{nested commutator} (Lem.~\ref{lem:group_comm_planar_qsp}).
\end{theorem}

While we provide only two condensed theorem statements here, we spend significant time later in this work on corollaries and extensions. This includes numerous lemmata on the properties of symmetric QSP protocols under structured group commutators, different methods to show density in different classes of functions for differing ansätze, and investigations into lower bounds for the $c$ given in Thm.~\ref{thm:informal_qsp_skt}, as well as finer estimates depending on the chosen $F(X)$; this is contained to Sec.~\ref{sec:skt_qsp_applications}.

Before going on to proofs of these results in Sec.~\ref{sec:qsp_skt}, we review how this work relates to prior efforts on QSP and Solovay–Kitaev-like theorems. We also provide a section glossing key techniques from the proof of the standard SKT (Sec.~\ref{sec:skt_thm}), whose methods are repurposed variously and might scaffold intuition for the surprising results of gate approximation for the curious reader.

\subsection{Prior work} \label{sec:prior_work}

\noindent As this work incorporates disparate quantum computing subfields, we've broken this section in two. Work on QSP and gate approximation (let alone the Solovay–Kitaev theorem itself) is extensive; we aim to sketch the most relevant, recent, or pedagogically helpful work.

\subsubsection{Quantum signal processing: classical and quantum subroutines}

\noindent A significant strain of QSP research focuses on finding efficient, numerically stable classical algorithms for computing phases. That is, algorithms which, given some classical data specifying a target function, compute a list of real numbers (the QSP phases) such that the unitary generated according to these phases contains (e.g., as a component of a matrix element) an approximation to this function up to a desired precision.

The first algorithms devised for computing QSP phases were numerically unstable \cite{ylc_14, lyc_16_equiangular_gates, lc_17_simulation}, relying on root-finding subroutines limiting protocols to dozens of phases. Multiple works have appeared since extending computability past tens-of-thousands of phases, utilizing novel divide-and-conquer approaches manipulating Laurent polynomials \cite{haah_2019}, modified Fourier transforms \cite{chao_machine_prec_20, dong_efficient_phases_21}, as well as a variety of iterative optimization techniques \cite{mw_gqsp_24, dong_efficient_phases_21, dlnw_robust_iter_23}.

While many of these improvements exceed the needs of experimentalists, the state-of-the-art is a family of iterative phase-finding methods (introduced in \cite{dlnw_infinite_22}) and linear-systems based approaches \cite{amt_23, szego_nlfa_qsp_24}) with competitive performance extending to millions of phases and (for the latter type) \emph{provable} stability/convergence guarantees for target functions only mildly more restricted than those achievable by standard QSP.\footnote{Original variants considered targets whose Fourier coefficients had constant-bounded $\ell_1$ norm \cite{amt_23}, while recent methods allow slightly sub-normalized supremum-norm or Szegő functions \cite{szego_nlfa_qsp_24}.} These guarantees are unique among QSP phase-finding algorithms, and in sum indicate that the classical companion algorithms required by QSP are remarkably well-understood.

The maturity of QSP phase-finding algorithms is restricted, however, to a narrow class of circuit ansätze. Simultaneous with the work discussed above, multiple attempts have been made to extend QSP to new settings by modifying its circuit form or combining its insights with other block encoding manipulation methods; this includes G-QSP (allowing arbitrary SU(2) gates instead of $z$-rotations) \cite{mw_gqsp_24}, QEVT (relying on quantum linear systems solvers to efficiently generate special `history states') \cite{low_su_qevt_24}, QSP with antisymmetric phase lists \cite{mf_recursive_24}, multivariable QSP (allowing multiple oracles) \cite{rc_m_qsp_22, glw_m_qsp_24, laneve_m_qsp_25}, LCHS (for simulation of open, time-dependent quantum systems) \cite{dll_lchs_23}, infinite QSP \cite{dlnw_infinite_22}, randomized QSP \cite{martyn_rall_halving_24}, parallelized QSP \cite{mrclc_parallel_qsp_24}, and modularly composable QSP-based `gadgets' \cite{rcc_modular_qsp_23}. In many of these extensions, especially the multiple-oracle variants, the `completion and layer-stripping' paradigm discussed in Rem.~\ref{rem:anatomy_qsp_proof} breaks down completely, necessitating new techniques to show the existence of good protocols. QSP and QSVT, by merit of their success in unifying the description and analysis of most quantum algorithms \cite{mrtc_21}, appear to be a good foundation for studying quantum advantage generally---consequently, supplementing the toolkit to recover proofs of useful properties under modification is a first step toward diversifying quantum algorithms.

\subsubsection{The Solovay–Kitaev theorem and gate approximation}

\noindent A basic concern in gate approximation or compilation is, given a set of implementable gates, to understand properties of the closure of this set under differing notions of composition (e.g., unitary products). While oftentimes overlooked by the algorithmist, such results are central to a coherent notion of complexity classes for quantum computation (as well as the persistence of polynomial time-and-space separations between classical and quantum algorithms). Understanding which gate sets are (efficiently) inter-convertible allows implementation agnosticism, separating questions in algorithms from those of architecture.

Among the earliest and most basic results in the theory of gate approximation, the Solovay–Kitaev theorem (Thm.~\ref{thm:skt}) \cite{kitaev_q_algs_ecc_97, ksv_textbook_02, dn_skt_overview_05, nc_textbook_11} establishes an equivalence between an instruction set's universality (e.g., density in a compact Lie group), and the efficiency of such approximations. The original proof of this result is very-nearly folkloric, and numerous alternative proofs have since appeared. Even now research related to the SKT and gate approximation is ongoing, and can be split into a few categories:
    \begin{enumerate}[label=(\arabic*)]
        \item \textbf{Improving constants:} The original statement of the SKT gives an asymptotic gate complexity for an approximation to precision $\varepsilon$ in the operator norm for SU(2) or SU(d), as well as an asymptotic time complexity for the classical algorithm computing these gate sequences. The methods in \cite{dn_skt_overview_05} show that $\mathcal{O}(\log^{(3 + \delta)}{\varepsilon^{-1}})$ gates are sufficient for any $\delta > 0$, and various methods like those in \cite{hrc_efficient_discrete_02} can show that $\mathcal{O}(\log{\varepsilon^{-1}})$ are necessary (and can be arbitrarily approached for certain instruction sets non-constructively). Sophisticated techniques have been employed to improve the exponent in the constructive setting, most recently to $\log_{\phi}(2) \approx 1.44$ using higher-order commutators \cite{kuperberg_skt_23, et_shortest_comm_13}. In practice, there has also been significant theoretical and heuristic work to improve constants multiplying these logarithmic terms on near-term devices. 
        
        \item \textbf{Proving SKTs for restricted instruction sets or approximation tasks:} In practical settings the instruction set often has specific structure beyond its universality (e.g., Clifford+T, or certain `magic gates' \cite{rs_ancilla_free_approx_16, sarnak_letter_15}). Alternatively, the set of gates one wishes to approximate may not be all of SU(d) or SU(2), but instead some useful subset like $Z$-rotations \cite{rs_ancilla_free_approx_16}, or with special resource restrictions (e.g., allowing or disallowing auxiliary qubits, or intermediate measurement). Each of these choices can impact the necessary and sufficient complexity of approximation.

        \item \textbf{Proving SKT-like results for relaxed instruction sets:} In the obverse of the above, assumptions on the gate set or approximation space may be made \emph{less specific}, for instance considering instruction sets not closed under inverse \cite{bt_inverse_free_21}, or the approximation of non-compact Lie groups \cite{kuperberg_skt_23}. Proving SKTs for these gate sets is deceptively difficult, producing novel sub-lemmata on `self-correcting sequences' and `inverse factories.'
    \end{enumerate}
We discuss the above ongoing research because the work we present here is similar, considering a modified instruction set over certain compact subsets of \emph{group-valued functions}. To our knowledge this is the first instance of an SKT over such objects. And while our approach establishes an initial connection between SKT methods and the quantum algorithms QSP/QSVT, optimizing resource complexities, as well as investigating restricted or relaxed instruction sets, are very much open questions (see Sec.~\ref{sec:discussion_open}).

\section{On the SU(2) Solovay–Kitaev theorem} \label{sec:skt_thm}

\noindent The Solovay–Kitaev theorem (SKT) is a basic result in the theory of quantum computation ensuring the existence of \emph{good} discrete sets of single-qubit (or qudit) gates with which one can \emph{efficiently} approximate \emph{any} desired single-qubit (or qudit) gate. This simplifies the analysis of quantum algorithmic subroutines and is especially helpful in the context of error-correction where gates are necessarily discretized. It is also this author's opinion that the SKT is surprising and beautiful in disproportion to the length of its proof, and scaffolds intuition about the structure of SU(d).

We mainly refer to the excellent review of the SKT by Dawson and Nielsen \cite{dn_skt_overview_05}, though the history of the origin and formalization of the SKT is various. Specifically, \cite{dn_skt_overview_05} covers gaps originally left as exercise in the standard textbook \cite{nc_textbook_11}, though the interested reader should also check out the earlier presentation (which features two orthogonal proofs!) by Kitaev, Shen, and Vyalyi in the more mathematically-oriented \cite{ksv_textbook_02}, or even the earliest published treatment\footnote{There is a claim made in \cite{dn_skt_overview_05} that Solovay announced the result on an email listhost in 1995.} in \cite{kitaev_q_algs_ecc_97}. The study of SK-type theorems is healthy and ongoing ( Sec.~\ref{sec:prior_work}), with recent results greatly relaxing assumptions on the initial discrete gate set \cite{bt_inverse_free_21}, and continuous improvements appearing for algorithms for constructing gate-approximations in realistic settings \cite{rs_ancilla_free_approx_16}.

We give a short exposition of the constitutive definitions and lemmata toward the proof of the simplest SU(2)-variant of the SK theorem; many of these will be modified and reapplied in the QSP setting in Sec.~\ref{sec:qsp_skt}, for which good geometrical intuition for the original SKT proof helps. Our intent is not to cover bleeding-edge methods but instead the basic tools required and conditions sufficient to show the bones of SU(2) SKTs: \emph{if a set of gates is dense in SU(2), then it is guaranteed to generate SU(2) quickly}.

\begin{definition}[Instruction set, from \cite{dn_skt_overview_05}] \label{def:ins_set}
    An \emph{instruction set} $\mathcal{G}$ for a qubit is a set of quantum gates satisfying: (1) all gates $g \in \mathcal{G}$ are in SU(2), (2) the set $\mathcal{G}$ is closed under inverses, and (3) $\mathcal{G}$ is \emph{universal} for SU(2), that is, that the group generated by $\mathcal{G}$ is dense in SU(2), equivalently for any $\varepsilon > 0$ and $U \in \text{SU(2)}$ there exists a finite product $V = g_1 g_2 \cdots g_n \in \mathcal{G}^\ast$ such that $V$ $\varepsilon$-approximates $U$ in the operator norm.
\end{definition}

\begin{theorem}[Solovay–Kitaev theorem, from \cite{dn_skt_overview_05}] \label{thm:skt}
    Let $\mathcal{G}$ an instruction set (Def.~\ref{def:ins_set}) for SU(2), and let $\varepsilon > 0$. Then there is a constant $c$ such that for any $U \in \text{SU(2)}$ there exists a product $V$ of gates from $\mathcal{G}$ of length $\mathcal{O}(\log^{c}(\varepsilon^{-1}))$ where $d(V, U) < \varepsilon$.
\end{theorem}

One of the core components of the proof of the usual SKT is the \emph{balanced group commutator}. Here we discuss the definition and properties of this object, why it is useful for proofs of the SKT, and how such objects naturally suggest a relation to QSP.

\begin{figure}
    \centering
    \includegraphics[width=0.8\textwidth]{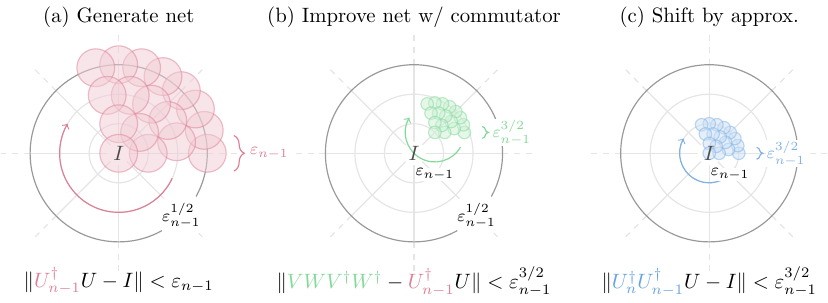}
    \caption{Visualizing the major steps used in the standard proof(s) of the Solovay–Kitaev theorem \cite{nc_textbook_11, dn_skt_overview_05}. The base assumption of the recursive argument is (a) an $\varepsilon_{n-1}$-net\footnote{Only a sector is shown of each net, where the rest would be filled in following the arrows, to improve visual clarity. We also exaggerate the size of the open sets, which shrink exponentially.} for a ball (in the operator norm) about the identity of SU(2) of radius $\varepsilon_{n-1}^{1/2}$. It turns out that (b) balanced group commutators (see discussion in Ex.~\ref{ex:balanced_comm_construction_qsp} and Rem.~\ref{rem:skt_proof_overview}) of pairs of elements $(V, W)$ in this net can form unexpectedly-good $\varepsilon_{n-2}^{3/2}$-approximations of its elements (see Lem.~\ref{lem:approx_group_com}). Finally, shifting (c) by the inverse of the improved approximation $VWV^\dagger W^\dagger = U_{n} \approx U_{n-1}^\dagger U$ leaves us with a better, $\varepsilon_{n} = \varepsilon_{n-1}^{3/2}$-net for a ball of diameter $\varepsilon_{n-1}$ about the identity. This process can then be repeated, approaching $U$.}
    \label{fig:skt_proof_overview}
\end{figure}

\begin{example}[Exposition on a special case of a balanced commutator] \label{ex:balanced_comm_construction_qsp}
    A basic tool used in most proofs of the SKT is the \emph{balanced group commutator}. This allows us to approximate group elements near the identity in SU(2) by products of other group elements that are (1) \emph{further from the identity}, and (2) \emph{themselves approximations}. In other words, given $U$ such that $d(I, U) < \varepsilon$, we can find $V, W$ such that
        \begin{equation}
            VWW^\dagger V^\dagger = U \text{ and } d(I, V), d(I, W) = \mathcal{O}(\sqrt{\varepsilon}),
        \end{equation}
    It might seem that we have gotten nowhere by decomposing $U$ in this way; however, a critical insight for the balanced commutator is that we don't just use the ideal $V, W$ but rather \emph{approximations} $\tilde{V}, \tilde{W}$ to these unitaries (i.e., those guaranteed by `nets' (open covers) given by our initial density assumptions). Surprisingly, one can show that the group commutator of such approximate elements is a \emph{more accurate approximation} to the ideal $U$ than expected. This property depends on the fact that we are working near the identity, and is discussed in Lem.~\ref{lem:approx_group_com}. This allows us to `bootstrap' the net we assume exists by the original density statement to a better net very quickly by recursing. We will use a modified version of this fact (Lem.~\ref{lem:approx_nested_comm}) in the QSP setting, and depict this `net refinement' visually in Fig.~\ref{fig:skt_proof_overview}. Studying properties of such nets, as well as modified group commutators, underlies most of the improvements since \cite{dn_skt_overview_05}.

    Considering only the problem of ensuring $V, W$ for a moment, the overview section in \cite{dn_skt_overview_05} gives a cursory argument, working without loss of generality\footnote{Namely, that eventually we will be interested in the \emph{inverse} problem, finding suitably $V, W$ given $U$, for which the $x, y$-axes will be taken by similarity transformations to some other axes, dictated by the desired $U$.} in the simpler setting where $V, W$ are rotations by an angle $\phi$ generated by the Pauli $X$ and $Y$ operators respectively. In this case the group commutator is
        \begin{align}
            [e^{i\phi X}, e^{i\phi Y}] &= e^{i\phi X} e^{i\phi Y} e^{-i\phi X} e^{-i\phi Y}, \\
            &= e^{i\phi X} e^{-i(\pi/4) Z} e^{i\phi X} e^{i(\pi/4) Z} e^{-i\phi X} e^{-i(\pi/4) Z} e^{-i\phi X} e^{i(\pi/4) Z},\\
            &= e^{i\phi X} e^{-i(\pi/4) Z} e^{i\phi X} e^{i(3\pi/4) Z} e^{i\phi X} e^{-i(\pi/4) Z} e^{i\phi X} e^{-i(\pi/4) Z}.
        \end{align}
    Namely, the group commutator in this simplified setting looks just like a QSP protocol with phases
        \begin{equation}
            \Phi = \{0, -\pi/4, 3\pi/4, -\pi/4, -\pi/4\}.
        \end{equation}
    While we are mainly interested in the small-angle limit of this protocol (i.e., as $\phi \rightarrow 0$ or $\cos{\phi} \equiv x \rightarrow 1$), it is worthwhile to show the associated QSP unitary:
        \begin{equation}
            U(\Phi, x) =
            \begin{bmatrix}
                P(x) & \ast\hphantom{\ast} \\
                \ast & \ast\hphantom{\ast}
            \end{bmatrix},
        \end{equation}
    where
        \begin{equation}
            \Re[P(x)] = -1 + 4 x^2 - 2 x^4
            \;\text{ and }\;
            \Im[P(x)] = -2 x^2 (1 - x^2).
        \end{equation}
    From this we see that for $x \rightarrow 1$ ($\phi \rightarrow 0$) the unitary approaches the identity, with limit
        \begin{equation}
            \lim_{\phi \rightarrow 0} U(\Phi, \phi) =
            \begin{bmatrix}
                1 - 2i\phi^2 & 0\\
                0 & 1 + 2i \phi^2
            \end{bmatrix}
            + \mathcal{O}(\phi^3).
        \end{equation}
    Namely, the group commutator approximates a rotation about an axis in the $XZ$-plane by an angle $\theta = 2\phi^2$. While this is unsurprising given Lems.~\ref{lem:group_comm} and Lem.~\ref{lem:su2_group_comm}, stepping back, this reduction to QSP provides a new, \emph{functional interpretation} of the geometric action of the group commutator. Rotations by small angles ($\phi$) can be used to generate composite pulses whose angle ($\theta$) is \emph{quadratically smaller} than the original angle, and whose axis of rotation is \emph{known and fixed}.

    This `angle shrinking' property will also be crucial to proving the other desired aspect of our group commutator, namely that inserting $\varepsilon$-\emph{approximations} to the phases and signals should only perturb the output by terms proportional to $\varepsilon^{1 + b}$ for some $b > 0$. We leave the proof of this to the more streamlined Lem.~\ref{lem:approx_group_com}. An interesting open problem (see Sec.~\ref{sec:discussion_open}) exists in characterizing, in the language of QSP, the space of `good group commutators' in the style of this example.
\end{example}

\begin{lemma}[On approximating group commutators, from \cite{dn_skt_overview_05}] \label{lem:approx_group_com}
    Suppose $V, W, \tilde{V}, \tilde{W}$ are unitaries such that $d(V, \tilde{V}), d(W, \tilde{W}) < \Delta$ and also $d(I, V), d(I, W) < \delta$ where this distance is the operator norm of the difference. Then
        \begin{equation} \label{eq:dist_ideal_comm}
            d(VWV^\dagger W^\dagger, \tilde{V}\tilde{W}\tilde{V}^\dagger \tilde{W}^\dagger ) 
            <
            8\Delta\delta +
            4\Delta \delta^2 +
            8\Delta^2 + 
            4\Delta^3 + 
            \Delta^4.
        \end{equation}
    This gives rise to the following inequality, assuming that $\Delta = \varepsilon_{n-1}$ and $\delta = c\sqrt{\varepsilon_{n-1}}$, as they would be at a single step of the net refinement procedure depicted in Fig.~\ref{fig:skt_proof_overview}. Then
        \begin{equation}
            d(VWV^\dagger W^\dagger, \tilde{V}\tilde{W}\tilde{V}^\dagger \tilde{W}^\dagger ) 
            \lessapprox
            8c \varepsilon^{3/2},
        \end{equation}
    where in general this constant can be rigorously bounded and improved. We see, then, that instead of varying linearly in the approximation error of the inputs $\tilde{V}, \tilde{W}$, we have ended up with an improved approximation to the ideal product at order $\Delta\delta$. The purpose of this remark is to prime us for the similar Lem.~\ref{lem:approx_nested_comm}, whose proof we provide and which looks at a slightly more complicated product.
\end{lemma}

We've given almost all of the main insights and lemmas required to prove the SU(2) variant of the SKT; before covering this summary in Rem.~\ref{rem:skt_proof_overview} we give two helpful identities we will use often.

\begin{lemma}[Group commutator identity] \label{lem:group_comm}
    Consider an algebra whose elements $A$ are such that $e^A$ can be meaningfully defined in terms of its power series $1 + A + A^2/2! + \dots$; then the group commutator of two such exponentials can be expressed in terms of nested commutators:
        \begin{equation}
            e^{A}e^{B}e^{-A}e^{-B} = 
            \exp{\left(
                [A, B] + \frac{1}{2!}[A + B, [A, B]] + \cdots
            \right)}.
        \end{equation}
\end{lemma}

\begin{lemma}[Group commutator in SU(2)] \label{lem:su2_group_comm}
    Let $\hat{n}, \hat{m} \in \mathbb{R}^3$ with unit norm and $A = i \theta\, (\hat{n}\cdot\hat{\sigma})$ and $B = i\phi\, (\hat{m}\cdot\hat{\sigma})$ where $\theta, \phi > 0$ and $\hat{\sigma}$ is the tuple of Pauli matrices $\hat{\sigma} = \{\sigma_x, \sigma_y, \sigma_z\}$. Then the group commutator according to the complex exponentials of $A, B$ as in Lem.~\ref{lem:group_comm} has the simplified form
        \begin{equation}
            e^{A}e^{B}e^{-A}e^{-B} =
            I + 2i \theta\phi (\hat{n}\times\hat{m})\cdot \hat{\sigma} + \mathcal{O}(\theta^2 \phi^2).
        \end{equation}
    Proof follows from relations among Pauli operators, $[\sigma_i, \sigma_j] = 2i\epsilon_{ijk}\sigma_k$, for $\epsilon_{ijk}$ the Levi-Civita symbol.
\end{lemma}

\begin{remark}[Overview of SKT proof and complexity derivation] \label{rem:skt_proof_overview}
    The structure of the SKT proof is best understood visually; we give a diagram (off-scale for clarity) in Fig.~\ref{fig:skt_proof_overview} (whose sub-diagrams' labels we match here), inspired by the SKT appendix of \cite{nc_textbook_11} but following the exposition of \cite{dn_skt_overview_05}. A similar diagram can be drawn in the QSP case (Fig.~\ref{fig:qsp_net_creation}), with an updated proof scheme in Thm.~\ref{thm:qsp_skt}.
    \begin{enumerate}[label=(\alph*)]
        \item \textbf{Generate net:} The assumption for the SKT (Thm.~\ref{thm:skt}) is an instruction set. This is used to generate an $\varepsilon_0$-net (constant precision) for SU(2). By multiplying the inverse of the $\varepsilon_0$-approximation with our target unitary we can `shift' to work in a ball of radius $\sqrt{\varepsilon_0}$ around the SU(2) identity. By the compactness of SU(2) and the universality of our instruction set we can guarantee that our initial net has elements with only finite length.
        
        \item \textbf{Improve net with commutator:} This initial net is only $\varepsilon_0$ good, and naïvely taking products of elements in this net might not seem to do any better than this. However, using our group commutator (Lem.~\ref{lem:approx_group_com}) we can create anomalously good $\varepsilon_{0}^{3/2}$-approximations to elements in our ball around the origin at the cost of increasing product lengths from $\ell_0$ to $5\ell_0$.
        
        \item \textbf{Shift by approximation:} Given this improved approximation we can again translate toward the identity by its inverse, yielding an $\varepsilon_{1} = \varepsilon_{0}^{3/2}$-net for the same ball. But now we are back where we started with a better net with longer products---recursively applying this process \emph{exponentially} increases protocol length while \emph{doubly-exponentially} improving approximation error. Just as in Thm.~\ref{thm:qsp_skt} the overall complexity is derived by comparing the relation between (1) how much longer products become at each step, and (2) how much finer the net becomes.
    \end{enumerate}
\end{remark}

In the next section we translate the above tools and insights to the lifted setting, summarized by Fig.~\ref{fig:skt_lifting_version}. Note that the lifted setting really is quite different; i.e., we're analyzing nets not over SU(2) (a compact, finite dimensional space with good geometric intuition), but instead groups of SU(2)-valued functions, which are infinite dimensional and compact only when endowed with special properties (compact domains, images, and equicontinuity). Crucially it is \emph{not enough to apply the SKT in a pointwise way}, as we require the same $\Phi$ to work for every $x \in X$. We have to ensure that our nets are well-defined and that our group commutators preserve additional properties unique to QSP protocols. Rather than despairing, we hope that this extension is (to the reader) exciting in the variety of techniques it employs and spaces it studies!

\section{Constructing a Solovay–Kitaev theorem for QSP} \label{sec:qsp_skt}

\noindent We will match the form of quantum signal processing to techniques from the Solovay–Kitaev theorem by `lifting' a series of definitions and constructing a modified group commutator. A reader looking for a high-level overview of the eventual construction, mirroring the standard SKT proof (Rem.~\ref{rem:skt_proof_overview}), can skip directly to Thm.~\ref{thm:qsp_skt} and work backward. The lemmata and remarks before the main theorem concern QSP protocols constrained to be near the identity in SU(2) over all signals, and characterize the behavior of these protocols under our modified group commutator. Ultimately, we are following a guiding principle of the usual SKT (\emph{it is easier to work with Lie algebras than Lie groups}) and so we aim to simply our problem by working in the tangent space of the identity of our \emph{group of group-valued functions}.

\begin{lemma}[Nested group commutators near the SU(2) identity] \label{lem:nested_group_comm}
    Let $\varepsilon > 0$ and $\hat{m}_0, \hat{m}_1, \hat{n}_0, \hat{n}_1 \in \mathbb{R}^3$ and $\hat{\sigma} = \{\sigma_x, \sigma_y, \sigma_z\}$. Define the two \emph{group commutators} (Lem.~\ref{lem:su2_group_comm}):
        \begin{equation}
            G_k = [e^{i\varepsilon(\hat{n}_k\cdot\hat{\sigma})}, e^{i\varepsilon(\hat{m}_k\cdot\hat{\sigma})}], \quad k \in \{0, 1\}.
        \end{equation}
    Then the \emph{nested group commutator} of $G_0, G_1$ can be written in to leading order in $\varepsilon$:
        \begin{align}
            [G_0, G_1] &\equiv G_0 G_1 G_0^{-1} G_1^{-1}\\
            &= 1 - 8i\varepsilon^4 ([\hat{n}_0 \times \hat{m}_0]\times[\hat{n}_1 \times \hat{m}_1])\cdot \hat{\sigma} + \mathcal{O}(\varepsilon^5).
        \end{align}
    Proof follows immediately by application of Lem.~\ref{lem:su2_group_comm}; as expected, for small enough $\varepsilon$, nesting group commutators pushes the product closer to the identity. The group commutator we use for our main theorem will be a nested group commutator of QSP unitaries, imposing additional relations between $G_0, G_1$.
\end{lemma}

\begin{definition}[Pauli form of QSP protocols] \label{def:qsp_pauli_form}
    Let $U(\Phi, x)$ the unitary resulting from a QSP protocol with phases $\Phi \in \mathbb{R}^{n+1}$ and signal unitary $W(x) = e^{i\arccos{x}\sigma_x}$. Generically this unitary has form
        \begin{equation}
            U(\Phi, x) = 
            \begin{bmatrix}
                P(x) & iQ(x)\sqrt{1 - x^2}\\
                iQ^*(x)\sqrt{1 - x^2} & P^*(x)
            \end{bmatrix},
        \end{equation}
    where $P(x), Q(x) \in \mathbb{C}[x]$ are complex-coefficient polynomials of degree (at most) $n$ and $(n-1)$ respectively. Then the \emph{Pauli form} of this same unitary is written
        \begin{equation}
            U(\Phi, x) = \exp{\left(i\theta(x)[\hat{\phi}(x) \cdot \hat{\sigma}]\right)},
        \end{equation}
    where as before $\hat{\sigma} = \{\sigma_x, \sigma_y, \sigma_z\}$ is a tuple of the Pauli matrices, and for all $x$ we have $\theta(x) \in \mathbb{R}$ and $\hat{\phi}(x) \in \mathbb{R}^3$ is a unit vector. For convenience we sometimes drop the explicit dependence of $\theta, \hat{\phi}$ on $x$. The map between $P(x), Q(x)$ and $\theta(x), \hat{\phi}(x)$ is simple to write down:
        \begin{align}
            \theta(x) &= \arccos{(\Re[P(x)])},\label{eq:pauli_standard_map_0}\\
            \phi_x(x) &= \frac{1}{i\sin{[\theta(x)]}}\Im[iQ(x)\sqrt{1 - x^2}],\\
            \phi_y(x) &= \frac{1}{\sin{[\theta(x)]}}\Re[iQ(x)\sqrt{1 - x^2}],\\
            \phi_z(x) &= \frac{1}{i\sin{[\theta(x)]}}\Im[P(x)].\label{eq:pauli_standard_map_1}
        \end{align}
    Converting between these forms is not always convenient given the branch cuts of the $\arccos$; the benefit of the Pauli form appears when dealing with group commutator identities such as those in Lems.~\ref{lem:group_comm} and \ref{lem:nested_group_comm} (as the relation to the Pauli generators is made more obvious). Moreover, certain constraints on $P(x), Q(x)$ correspond to simple constraints on $\theta(x), \hat{\phi}(x)$, as per the following corollary.
\end{definition}

\begin{corollary}[Pauli form of symmetric QSP protocols] \label{cor:pauli_form_qsp_sym}
    Let $\Phi \in \mathbb{R}^{n + 1}$ \emph{symmetric}, i.e., it is equal to its reverse, sometimes denoted $\Phi^R = \Phi$:
        \begin{equation}
            \Phi = 
            \begin{cases}
                \{\phi_0, \phi_1, \cdots \phi_{(n - 1)/2}, \phi_{(n - 1)/2}, \cdots, \phi_1, \phi_0\} & \text{if $n$ odd},\\  
                \{\phi_0, \phi_1, \cdots \phi_{n/2 - 1}, \phi_{n/2}, \phi_{n/2 - 1}, \cdots, \phi_1, \phi_0\} & \text{if $n$ even}.\\   
            \end{cases}
        \end{equation}
    Then the Pauli form of the unitary $U(\Phi, x)$ satisfies $\phi_y = 0$ \emph{as well as} the relations given in Def.~\ref{def:qsp_pauli_form}. Equivalently, the real part of $iQ(x)$ is zero for all $x \in X$. The Bloch-sphere interpretation of SU(2) elements provides a simple geometric picture of such protocols: rotations by an angle $\theta(x)$ about a (variable) axis constrained to the $XZ$-plane with components $(\phi_x, \phi_z)$.

    Note that we will also call a \emph{symmetric QSP protocol} \emph{any} interleaved product of only (1) copies of the signal unitary, and (2) signal-independent SU(2) unitaries, which additionally satisfies that the real part of the top-left unitary element in the standard basis is zero. Equivalently, that the unitary for all signals is invariant under the action $U \mapsto \sigma_y U^\dagger\sigma_y$.
\end{corollary}

As mentioned in Sec.~\ref{sec:prior_work}, symmetric QSP protocols (which remove unnecessary degrees of freedom) have been extensively studied, and are basically as useful as their non-symmetric counterparts with the \emph{additional} benefit of uniqueness properties for $\Phi$ and good numerical properties for algorithms computing $\Phi$ \cite{sym_qsp_21}. We mainly restrict to symmetric QSP and its variants given their ubiquity, the utility of said uniqueness properties in establishing uniform approximation, and guarantees on the convergence of numerical optimization over these ansätze. In fact our modified group commutator is explicitly constructed to preserve this symmetry (or `planarity,' see below), demonstrating one construction of a \emph{compatible commutator} (Def.~\ref{def:compatible_comm}).

\begin{definition}[Planar QSP protocols] \label{def:planar_qsp_protocols}
    Symmetric QSP protocols are a single instance of a wider class of QSP protocols which we will call \emph{planar} protocols. A QSP protocol is planar if the unitary resulting from the protocol, taken in the Bloch sphere, rotates about axes constrained to a \emph{known, fixed} plane. Equivalently, $\hat{\phi}$ in the Pauli form (Def.~\ref{def:qsp_pauli_form}) of this protocol is constrained to a plane for all argument (there exists a known \emph{unit normal vector} $\hat{n}$ independent of $x$ such that $\hat{\phi}(x) \cdot \hat{n} = 0$ for all $x$). We will often refer to a planar protocol by a spanning set for the fixed plane, e.g., an XZ-planar protocol corresponds to $\hat{n} = \hat{y}$.
\end{definition}

\begin{remark}[Maps between planar QSP protocols] \label{rem:planar_qsp_map}
     QSP protocols with similar form to that discussed in Cor.~\ref{cor:pauli_form_qsp_sym}, namely that one component of $\hat{\phi}$ of their Pauli form is zero, have their rotation axis on the Bloch sphere constrained to a fixed (independent of the signal) plane. These are a subset of \emph{planar protocols} (Def.~\ref{def:planar_qsp_protocols}) and we can obliviously transform this plane. For example, if one is given an XZ-planar protocol (e.g., a symmetric QSP protocol), then this can be made into an YZ-planar protocol by the conjugation
        \begin{equation}
            \mathcal{M}_{XZ \mapsto YZ}[\ast] \equiv e^{I\sigma_z(\pi/4)}(\ast)e^{-I\sigma_z(\pi/4)}.
        \end{equation}
    Verifying correctness is simple, as the relevant quantity, $\phi_x \sigma_x + \phi_z \sigma_z$, is mapped linearly to $\phi_x \sigma_y + \phi_z \sigma_z$. Note, of course, that many conjugation $\mathcal{M}_{XZ \mapsto YZ}$ could be defined up to the O(1)-symmetry preserving the plane. These physical maps specifically have simple action among the $P(x), Q(x)$ for the same QSP protocol (e.g., for the example above, $Q(x) \mapsto iQ(x)$).
\end{remark}

\begin{definition}[Planar protocols about the SU(2) identity] \label{def:planar_protocols_iden}
    Let $\varepsilon > 0$; we say an XZ-planar protocol $U(\varepsilon) \in \text{SU(2)}$ is about the identity if its power series in $\varepsilon$ converges to its value, with form (to leading order)
        \begin{equation} \label{eq:generic_planar_perturb_fg}
            U(\varepsilon) = 
            I 
            + 
            i\varepsilon
            \begin{bmatrix}
                f & g\\
                g & -f
            \end{bmatrix}
            +
            \mathcal{O}(\varepsilon^2).
        \end{equation}
    Here $f, g \in \mathbb{R}$ are possibly functions of some underlying unknown scalar parameter but are independent of $\varepsilon$. This is simply the most generic expansion of an SU(2) operator about the identity generated by $\sigma_x, \sigma_z$ to linear order in $\varepsilon$. We show an XZ-planar protocol here for simplicity; for other planar protocols the first order expansion could have been generated by any other $\hat{\phi}$ constrained to a plane. 
\end{definition}

\begin{lemma}[Cyclic permutations of products near the SU(2) identity] \label{lem:cyclic_iden_prod}
    Let $A, B, C \in \text{SU(2)}$ such that $\lVert ABC - I\rVert < \varepsilon$ in the operator norm. Then $CAB$ is also $\varepsilon$-close to $I$ in the operator norm. Proof follows by the invariance of the operator norm under unitary conjugation, as
        \begin{equation}
            C(ABC - I)C^{-1} = C(ABC)C^{-1} - CIC^{-1} = CAB - I.
        \end{equation}
    Relatedly, if we know the product to leading order in some small parameter, e.g., $ABC = I + \varepsilon D + \mathcal{O}(\varepsilon^2)$, then the explicit relation between the leading components of the cyclic permutation and the original is $CAB = I + \varepsilon CDC^{-1} + \mathcal{O}(\varepsilon^2)$.
\end{lemma}

\begin{lemma}[Shifting planar protocols about the SU(2) identity] \label{lem:sym_perturb_iden}
    Let $A \in \text{SU(2)}$ such that the top-left matrix element for the standard-basis representation of $A$ has zero real part, viz., that the Pauli form for $A$ has $\phi_y = 0$, or that $A$ is XZ-planar (Rem.~\ref{rem:planar_qsp_map}). Let $B, B' \in \text{SU(2)}$ un-restricted save for the relations
        \begin{align}
            \langle 0| B' |0 \rangle &= \hphantom{-}\langle 0| B |0 \rangle\\
            \langle 0| B' |1 \rangle &= -\langle 0| B |1 \rangle^*
        \end{align}
    Then the products $BB'$ and $B'AB$ are both XZ-planar. Moreover, if $\lVert I - ABB'\rVert < \varepsilon$ in the operator norm, then $\lVert I - B'AB\rVert < \varepsilon$. Finally, $I - B'AB$, while not unitary, also has zero real component for its off diagonal elements in the standard basis. Proof follows from direct computation of generically parameterized SU(2) unitaries (relying on Cor.~\ref{cor:pauli_form_qsp_sym}) and Lem.~\ref{lem:cyclic_iden_prod}. A similar argument for the Pauli forms of symmetric QSP protocols is glossed in Lem.~\ref{rem:form_of_difference_planar}.
\end{lemma}

\begin{figure}
    \centering
    \includegraphics[width=\textwidth]{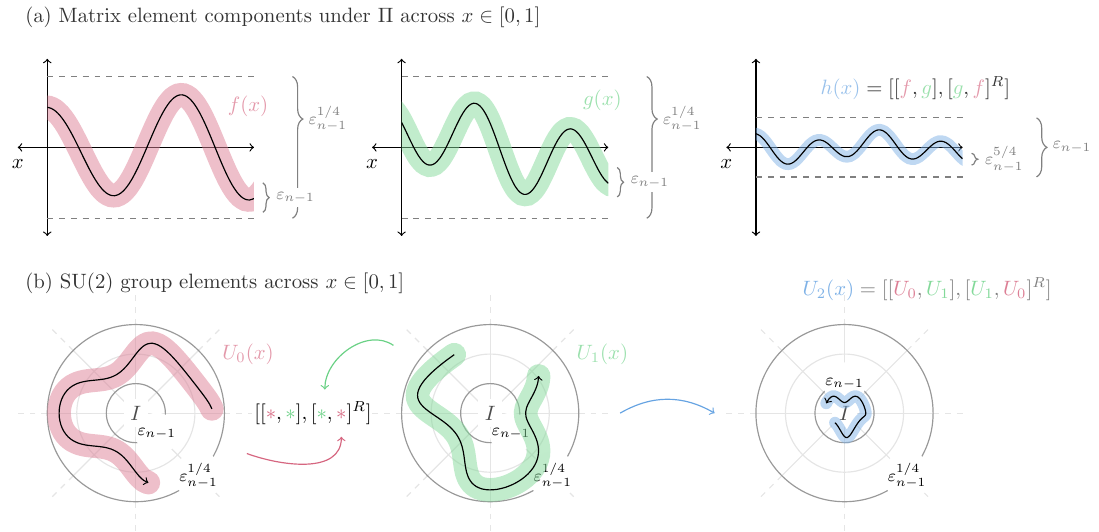}
    \caption{Two depictions of the action of the nested commutator (Lem.~\ref{lem:group_comm_planar_qsp}) about the SU(2) identity. Subfigure (a) shows a component of a unitary matrix element (e.g., $\Im[\langle 0 | U(\Phi, x)| 0\rangle]$), while subfigure (b) shows the unitary $U(\Phi, x)$ tracing out a path near the identity $I \in \text{SU(2)}$ for a range of $x$ (note when plotting across all $x \in [-1,1]$ these paths would be closed). The action of the nested commutator\footnote{We are slightly abusing notation in (a) and (b), as the nested commutator only makes sense for SU(2)-valued functions, while the notation in the third graph of (a) is under the projection $\Pi$ (see Def.~\ref{def:qsp_instruction_set}).} $[[\ast_{1}, \ast_{2}],[\ast_{2}, \ast_{1}]^R]$ (Lem.~\ref{lem:nested_group_comm}) on two protocols near the identity is both (1) to produce a new protocol closer to the identity, and (2) to allow for \emph{approximations} of such products near the identity to match the intended product better than expected (Lem.~\ref{lem:approx_nested_comm}). Specifically, we depict how two $\varepsilon_{n-1}$-approximations to protocols with $\varepsilon_{n-1}^{1/4}$-deviation from the identity produce, as detailed in Thm.~\ref{thm:qsp_skt}, $\varepsilon_{n-1}^{5/4}$-approximations (i.e., better than $\varepsilon_{n-1}$) for protocols with $\varepsilon_{n-1}$-deviation from the identity. This property underlies the `shrinking lemmas' of both the standard Solovay–Kitaev theorem and our QSP-variant (Thm.~\ref{thm:shrinking}).}
    \label{fig:qsp_net_creation}
\end{figure}

\begin{lemma}[On nested commutators of related planar QSP protocols about the SU(2) identity] \label{lem:group_comm_planar_qsp}
    Let $\varepsilon > 0$ and $U_0, U_1, U_2, U_3$ four planar QSP protocols whose Pauli forms are related in the following way to leading order in $\varepsilon$, i.e., they are planar protocols near the identity (as per Def.~\ref{def:planar_protocols_iden}):
        \begin{align}
            \theta_0 &= \varepsilon f,\quad \hat{\phi}_0 = \{\cos{t}, \sin{t}, 0\}, && \text{XY-planar protocol,}\label{eq:planar_protocols_0}\\
            \theta_1 &= \varepsilon g,\quad \hat{\phi}_0 = \{0, \cos{s}, \sin{s}\}, && \text{YZ-planar protocol,}\\
            \theta_2 &= \varepsilon f,\quad \hat{\phi}_0 = \{\cos{t}, -\sin{t}, 0\}, && \text{XY-planar protocol,}\\
            \theta_3 &= \varepsilon g,\quad \hat{\phi}_0 = \{0, \cos{s}, -\sin{s}\}, && \text{YZ-planar protocol,}\label{eq:planar_protocols_1}
        \end{align}
    where $f, g, s, t \in [-\pi, \pi)$ may in general be functions of the same underlying scalar parameter(s), here suppressed. Then the nested group commutator (Def.~\ref{lem:nested_group_comm}) has the form, to leading order in $\varepsilon$:
        \begin{equation}
            [[U_0, U_1], [U_2, U_3]] 
            = 
            I - 
            8 i\varepsilon^4 f^2 g^2
            \left[
            \sin^2{(s)}\sin{(2t)}\,\sigma_z 
            -
            \sin{(2s)}\cos^2{(t)}\,\sigma_x
            \right]
            +
            \mathcal{O}(\varepsilon^5).
        \end{equation}
    Note that the protocols $U_0, U_1, U_2, U_3$ can be built by applying conjugations to two symmetric QSP protocols. I.e., let $V_0, V_1$ symmetric QSP protocols with phases $\Phi_0,  \Phi_1$ achieving the polynomials $(P_0, Q_0)$ and $(P_1, Q_1)$, then we can relate the $V_j, j \in \{0, 1\}$ to the $U_k, k \in \{0, \cdots, 3\}$ by the following:
        \begin{align}
            U_0 &= e^{i\sigma_x  (\pi/4)}V_0 e^{-i\sigma_x  (\pi/4)}, && \text{XZ $\mapsto$ XY},\\
            U_1 &= e^{i\sigma_z  (\pi/4)}V_1 e^{-i\sigma_z  (\pi/4)}, && \text{XZ $\mapsto$ YZ},\\
            U_2 &= e^{i\sigma_z  (\pi/2)}e^{i\sigma_x  (\pi/4)}V_0 e^{-i\sigma_x  (\pi/4)}e^{i\sigma_z  (\pi/2)}, && \text{XZ $\mapsto$ -XY},\\
            U_3 &= e^{i\sigma_x  (\pi/4)} e^{i\sigma_z  (\pi/4)}V_1 e^{-i\sigma_z  (-\pi/4)}e^{-i\sigma_x  (\pi/4)}, && \text{XZ $\mapsto$ -YZ}.
        \end{align}
    In this case the relation between the $\{\theta_k, \hat{\phi}_k, k \in \{0, \cdots, 3\}\}$ and $(P_j, Q_j), j \in \{0, 1\}$ to leading order in $\varepsilon$ can be written explicitly, making use of Eqs.~(\ref{eq:pauli_standard_map_0})--(\ref{eq:pauli_standard_map_1}) and the underlying parameters $f, g, t, s$ in Eqs.~(\ref{eq:planar_protocols_0})--(\ref{eq:planar_protocols_1}):
        \begin{align}
            \Im{[P_0]} &= i\varepsilon f \cos{t},\\
            \Im{[P_1]} &= i\varepsilon g \cos{s},\\
            \Im{[Q_0]}\sqrt{1 - x^2} &= i\varepsilon f \sin{t},\\
            \Im{[Q_1]}\sqrt{1 - x^2} &= i\varepsilon g \sin{s},
        \end{align}
    This allows us to write the nested commutator in terms of the polynomials achieved by the two constituent symmetric QSP protocols. Writing the matrix form of this expression to leading order in $\varepsilon$:
        \begin{align}
            &[[U_0, U_1], [U_2, U_3]] = \nonumber\\
            &I - 
            16i\varepsilon^4
            \Biggr[
            \Im[Q_1]^2\Im[P_0]\Im[Q_0](1 - x^2)
            \,\sigma_z
            -
            \Im[P_0]^2\Im[P_1]\Im[Q_1]
            \,\sigma_x
            \Biggr]
            \sqrt{1 - x^2}
            + 
            \mathcal{O}(\varepsilon^5). \label{eq:leading_order_nested_comm}
        \end{align}
    Finally, note that this obeys the form of Def.~\ref{def:planar_protocols_iden}, i.e., that it is an XZ-planar protocol about the identity, though now it is $\mathcal{O}(\varepsilon^4)$-close; this can be verified using Lem.~\ref{lem:comm_preserves_planar}.
\end{lemma}

\begin{lemma}[The nested commutator of related planar protocols is planar] \label{lem:comm_preserves_planar}
    The observation made in Lem.~\ref{lem:group_comm_planar_qsp}, namely that a nested commutator built from two specially-oriented planar protocols is also planar is a specially engineered property of the product. Let $U_0, \dots, U_3$ SU(2) unitaries and consider their nested commutator:
        \begin{align}
            [[U_0, U_1],[U_2, U_3]] &= 
            (U_0 U_1 U_0^{\dagger} U_1^{\dagger})
            (U_2 U_3 U_2^{\dagger} U_3^{\dagger})
            (U_0 U_1 U_0^{\dagger} U_1^{\dagger})^\dagger
            (U_2 U_3 U_2^{\dagger} U_3^{\dagger})^\dagger,\\
            &= ([U_0, U_1])([U_2, U_3])([U_0, U_1])^\dagger([U_2, U_3])^\dagger,\\
            &= ([U_0, U_1])([U_2, U_3])([U_1, U_0])([U_3, U_2]).
        \end{align}
    Moreover, assume relations among the unitaries: $U_2 = U_0^\dagger$ and $U_3 = U_1^\dagger$. Then if $U_0$ and $U_1$ each satisfy $\sigma_y U_k \sigma_y = U_k^\dagger$ for $k \in \{0, 1\}$, the full expression is XZ-planar. Proof follows from simplifying the expression:
        \begin{align}
            [[U_0, U_1],[U_2, U_3]]
            &= ([U_0, U_1])([U_0^\dagger, U_1^\dagger])([U_1, U_0])([U_1^\dagger, U_0^\dagger]),\\
            &= ([U_0, U_1])([U_1, U_0])^{R}([U_1, U_0])([U_0, U_1])^R,
        \end{align}
    where we have used two identities for group commutators: $[A, B]^\dagger = [B, A]$ and $[A^\dagger, B^\dagger] = [B, A]^R$, where the superscript-$R$ indicates reversal of all four elements in the group commutator product: $ABA^{-1}B^{-1} \mapsto B^{-1}A^{-1}BA$. From this we see that the overall nested commutator is reversal-invariant: $U^R = U$.

    Finally, we examine the action of conjugation by $\sigma_y$, which will tell us if the product is XZ-planar. Each among $U_0, U_1$, the only relevant terms, transform in a special way
        \begin{align}
            U_0 \mapsto \sigma_{y}U_0\sigma_{y} = e^{-i\sigma_x  (\pi/4)}V_0^\dagger e^{i\sigma_x  (\pi/4)} &= U_0^\dagger,\\
            U_1 \mapsto \sigma_{y}U_1\sigma_{y} = e^{-i\sigma_z  (\pi/4)}V_1^\dagger e^{i\sigma_z  (\pi/4)} &= U_1^\dagger,
        \end{align}
    as each of $V_0$ and $V_1$ are XZ-planar. This means that the entire nested commutator transforms as (applying the two group commutator identities liberally)
        \begin{align}
            U \mapsto \sigma_{y}U\sigma_{y} &= \sigma_{y}([U_0, U_1][U_1, U_0]^R)(\ast)^R \sigma_{y},\\
            &= ([U_0^\dagger, U_1^\dagger][U_1^\dagger, U_0^\dagger]^R)(\ast)^R,\\
            &= ([U_0^\dagger, U_1^\dagger]^{R\dagger}[U_1^\dagger, U_0^\dagger]^{RR\dagger})^\dagger(\ast)^R,\\
            &= ([U_1, U_0]^{\dagger}[U_1^\dagger, U_0^\dagger]^{\dagger})^\dagger(\ast)^R\\
            &= ([U_0, U_1][U_0^\dagger, U_1^\dagger])^\dagger(\ast)^R,\\
            &= ([U_0, U_1][U_1, U_0]^{R})^\dagger(\ast)^R,\\
            &= U^\dagger,
        \end{align}
    where the notation $(\ast)^R$ indicates the reversed-version of the preceding terms, showing $U$ is XZ-planar, given the observation at the end of Cor.~\ref{cor:pauli_form_qsp_sym}. We could have also seen this from the fact that we can reverse the terms of $U$ for free, the conjugation of a commutator is the commutator with its arguments swapped (note this is \emph{not} the reverse of the commutator), and that $U_0, U_1 \mapsto U_0^\dagger, U_1^\dagger$ reverses each of our commutators.
\end{lemma}

\begin{lemma}[Relating operator and component norms] \label{lem:elem_op_norm}
    Let $A$ a linear operator between normed spaces, then 
    \begin{equation}
        \lVert A\rVert_{\rm op} \geq \max_{ij} \lvert A_{ij}\rvert,
    \end{equation}
    i.e., the operator norm is greater than the norm of any individual matrix element in any orthonormal basis. Consequently if the operator norm of some difference $\lVert B - C\rVert_{\rm op}$ is bounded above by $\varepsilon$, then each of the matrix elements of $B, C$ differ by at most $\varepsilon$. Proof follows from the definition of the operator norm:
        \begin{equation}
            \lVert A\rVert_{\rm op} \equiv
            \max_{|x| = 1} \lVert Ax\rVert,
        \end{equation}
    which we can see must be greater than the magnitude of any column of $A$ by choosing $x = e_{k}$, and thus greater than any element given that $\lVert v\rVert \geq |v_k|$ by the triangle inequality. This result will often be implicitly applied, allowing us to ensure element-wise uniform approximation by bounding operator norm closeness.
\end{lemma}

\begin{lemma}[Approximating nested commutators; modified from \cite{dn_skt_overview_05}] \label{lem:approx_nested_comm}
    Let $V, W, \tilde{V}, \tilde{W}$ unitaries such that $d(V, \tilde{V}), d(W, \tilde{W}) < \Delta$ and $d(I, W), d(I, V) < \delta$, where $d(A, B)$ denotes the operator norm $\lVert A - B \rVert$. Then
        \begin{equation}
            d([[V, W],[V^\dagger, W^\dagger]], [[\tilde{V}, \tilde{W}],[\tilde{V}^\dagger, \tilde{W}^\dagger]]) < 
            32 \Delta (\delta + \mathcal{O}(\delta^2) + \mathcal{O}(\Delta)).
        \end{equation}
    The proof of this fact is helped by the proof for the standard group commutator given by Dawson and Nielsen \cite{dn_skt_overview_05} (and mentioned back in Lem.~\ref{lem:approx_group_com}). Our nested commutator is a product of four commutators, each of which are identical up to renaming of $V, W$ by their conjugates or opposites, none of which change the arguments given in \cite{dn_skt_overview_05}, and whose leading-order contributions we can simply sum. We reproduce their argument for the leading-order $\Delta\delta$-term for one commutator.

    Re-writing the imperfect unitaries in terms of their perturbations, $\tilde{V} = V + \Delta_V$ and $\tilde{W} = W + \Delta_W$, we can expand the first commutator, collecting only those terms involving a single $\Delta$, of which there are four. These terms come in pairs up to exchange of $V, W$, namely
        \begin{equation} \label{eq:comm_delta_terms}
            \biggr[
            \Delta_V W V^\dagger W^\dagger 
            + VW\Delta_V^\dagger W^\dagger 
            \biggr]
            + 
            \bigg[
            V\Delta_WV^\dag W^\dag
            +
            VWV^\dag \Delta_W^\dag
            \biggr],
        \end{equation}
    the first of which we want to show has size $4\Delta \delta$ to leading order (which by symmetry the other two terms will also contribute). Here we use the special fact that $W, V$ are near the identity, expanding $W = I + \delta_W$. This induces four $\Delta\delta$ terms, e.g., of form $\Delta_V\delta_W V^\dag W^\dag$. Thus computing the operator norm of (\ref{eq:comm_delta_terms}) and applying the triangle inequality repeatedly gives the leading $\Delta\delta$-order terms, as well as $\lVert \Delta_V V^\dag + V\Delta_V^\dag\rVert$, which by the unitarity of $\tilde{V} = V + \Delta_V$ and $V$, differs from $I$ by a term of order $\Delta^2$, completing the sub-argument.

    Each of the four commutators will induce the same $8\Delta\delta$ contribution, which as they are leading order in both parameters will add linearly through the product, yielding the $32\Delta\delta$ given in the lemma. In practice, we will select $\Delta = \sqrt[4]{\varepsilon_{n - 1}}$ and $\delta = \varepsilon_{n - 1}$, showing that the nested commutator using $\varepsilon_{n - 1}$-approximations of the ideal elements can yield a $\mathcal{O}(\varepsilon^{5/4})$-approximation of the ideal nested commutator. This unexpectedly good approximation by imperfect elements is analogous to the standard commutator setting, albeit here the improvement $\varepsilon_{n - 1} \mapsto \varepsilon_{n - 1}^{5/4}$ is worse than the standard $\varepsilon_{n - 1} \mapsto \varepsilon_{n - 1}^{3/2}$.
\end{lemma}

\begin{lemma}[Planar protocols about the SU(2) identity have approximately independent components] \label{lem:independently_tunable_planar_perturb}
    Choose $R(x), S(x) \in \mathbb{R}[x]$ real-valued polynomials that can be achieved as the real part of the on-diagonal components of (possibly different) symmetric QSP protocols with phases $\Psi_0, \Psi_1$, and choose $r, s \in \mathbb{R}$ as $\mathcal{O}(1)$ constants. Then for $\varepsilon > 0$ there exist QSP protocols with the following form to leading order in $\varepsilon$:
        \begin{equation}
            U(\Phi, x) = I + 
            i\varepsilon
            \begin{bmatrix}
                R(x) + r & S(x) + s\\
                S(x) + s & -R(x) - r
            \end{bmatrix}
            + \mathcal{O}(\varepsilon^2).
        \end{equation}
    Moreover, the phases $\Phi$ relate simply to concatenations of and basic operations on the phase lists $\Psi_0, \Psi_1$.
\end{lemma}

The proof of Lem.~\ref{lem:independently_tunable_planar_perturb} is deferred in its details to Lem.~\ref{lem:explicit_qsp_planar_perturb}, but we discuss it here at a high level. First, however, note in Def.~\ref{def:planar_protocols_iden} the generic form of an XZ-planar QSP protocol about the identity. Also recall in Lem.~\ref{lem:group_comm_planar_qsp} we showed that the nested commutator of XZ-planar protocols near the identity is itself near the identity, and depends in an intricate way on the functions achieved by the inputs to the nested commutator. In the following theorem we will have to show that the image of a certain set of such protocols (a ball about the identity function) under the nested commutator obeys special properties, thus enabling the `net-refinement' step (sometimes called a shrinking lemma in standard SKT literature \cite{nc_textbook_11, dn_skt_overview_05}). A crucial observation enabling this is the ability to \emph{independently modify} the leading-order components $f, g$ for an XZ-planar protocol near the identity as they appear in (\ref{eq:generic_planar_perturb_fg}). This lemma says that this independent manipulation is indeed possible \emph{up to leading order}.

It turns out that building up the possible set of perturbations around the identity function is straightforward, as given by Lem.~\ref{lem:explicit_qsp_planar_perturb}. We proceed there by building an initial, simple set of perturbed protocols, followed by considering the closure of such protocols under finite unitary products and conjugations. Geometrically we should understand the statement of this lemma in terms of Fig.~\ref{fig:qsp_net_creation}---i.e., the action of the nested commutator takes two paths in the two-dimensional subspace of the tangent plane of the identity in SU(2) (spanned by the Pauli operators $\sigma_x, \sigma_z$) to a third path \emph{also} in this plane which has (1) smaller supremum norm, and (2) less than the expected error given known error for its inputs. This (leading order) component independence lemma allows us to show, as is discussed in the following theorem, that this map is surjective on the space of paths constituting the next step's finer net.

Having given a series of definitions and lemmata on planar protocols perturbed from the SU(2) identity, we can now analyze the effect of our nested commutator (Lem.~\ref{lem:group_comm_planar_qsp}) on nets ensured by the $\Pi$-density of our QSP instruction set (Def.~\ref{def:qsp_instruction_set}). A key sub-result in the proof of the standard SKT, as will be the case for our QSP-SKT, is a `shrinking lemma' (here raised to the level of a theorem) ensuring that, at each step in a recursive argument, good pre-images exist for any target in the relevant ball around the identity. If this can be ensured, then the approximate nested commutator result (Lem.~\ref{lem:approx_nested_comm}) immediately applies, allowing our net can be successively refined (followed by a shift closer to the identity). 

If this goes through, then writing out the full QSP-SKT theorem is mostly a matter of patching together the same three-step procedure depicted in Fig.~\ref{fig:skt_proof_overview} and analyzing the resulting gate complexity. Before stating and proving this shrinking lemma, we define (for convenience) a space of functions we refer to often.

\begin{definition}[The common QSP function space] \label{def:qsp_skt_function_space}
    For brevity in stating later results, we call $F(X, \xi) \subsetneq C(X, \mathbb{R})$ the \emph{common QSP function space}: real-valued continuous functions over $[-1,1]$ which are:
        \begin{enumerate}
            \item Bounded in magnitude by one, i.e., for all $f \in F$ and $x \in X$, we have $|f(x)| \leq 1$.
            \item Of definite parity, i.e., for all $f \in F$ and $x \in X$, there is some $k \in \{0, 1\}$ such that $f(-x) = (-1)^k f(x)$.
            \item Have bounded Lipschitz constant, i.e., in addition to being once-differentiable, we are guaranteed that for all $f \in F$ and $x \in X$, we have $|f'(x)| \leq \xi$ for a constant $\xi > 0$.
        \end{enumerate}
    It turns out that for fixed $\xi$, $F(X, \xi)$ is relatively compact (Thm.~\ref{thm:arzela_ascoli}), i.e., that its closure $\bar{F}(X, \xi)$ is a compact metric space equipped with the supremum norm. Oftentimes we will not make it explicit that we are referring to the closure of this space; additionally, we sometimes do not refer to $\xi$ explicitly when it is not used, e.g., writing $F(X)$ for expediency, though it is always assumed to be some fixed constant.
\end{definition}

\begin{theorem}[Generating nets for planar protocols; a shrinking lemma for QSP] \label{thm:shrinking}
    Let $F(X, \xi) \subsetneq C(X, \mathbb{R})$ the common function space in Def.~\ref{def:qsp_skt_function_space}. Let $\Pi: \text{SU(2)} \rightarrow \mathbb{R}$ project out the imaginary component of the top-left element of an SU(2) unitary in the computational basis. Then there exists some $\varepsilon_0 > 0$ such that for all $\varepsilon < \varepsilon_0$ the nested commutator (Lem.~\ref{lem:group_comm_planar_qsp}) can be applied to transform $\varepsilon$-nets for $S_{\varepsilon^{1/4}}(F(X, \xi))$ (the ball of functions, induced by $\Pi$, which both have supremum norm bounded above by $\varepsilon^{1/4}$ and are contained in $F(X, \xi)$; for more discussion on this set, see Thm.~\ref{thm:qsp_identity_net_form}) to $\varepsilon^{5/4}$-nets for $S_{\varepsilon^{1/4}}(F(X, c_0\xi)) \supsetneq S_{\varepsilon^{1/4}}(F(X, \xi))$ (for constant $c_0 > 1$), where elements in the second net have phase- and oracle-length at most $c_1\ell$ for $\ell$ the maximum length of products in the initial net (for constant $c_1 > 1$).

    Proof basically follows by inspection of the definitions and lemmata given above. The main contribution is to ensure that the nested group commutator devised to preserve the planar protocols \emph{always admits} pre-images at each step of the recursion. In the usual SKT this was equivalent to finding pre-images of a vector cross product. Here, our target space no longer form a compact group, and we are looking for a pre-image of a parameterized, nested cross product. We divide our argument into three steps. Note that while we prove this result for our specific case of symmetric QSP protocols and the nested commutator (Lem.~\ref{lem:group_comm_planar_qsp}), we also provide Def.~\ref{def:compatible_comm} for \emph{compatible commutators}, which allow an analogous `shrinking lemma' to go through for more general choices of $(\Pi, F, X)$.
    \begin{enumerate}[label=(\arabic*)]
        \item \emph{The nested commutator is surjective on balls about the identity function.} A key aspect of the standard SKT proof is the ability to find a preimage under the group commutator for a desired element; i.e., given $U$ to find two unitaries $V, W$ such that $VWV^\dagger W^\dagger = U$, under the assumption that $U, V, W$ are restricted to small balls around the identity in SU(2). In truth one will actually implement $\tilde{V}, \tilde{W}$ approximating $V, W$, showing through Lem.~\ref{lem:approx_group_com} that the error is mitigated by closeness to the identity.

        In the QSP setting we instead have the nested commutator with leading order behavior summarized by (\ref{eq:leading_order_nested_comm}) in Lem.~\ref{lem:group_comm_planar_qsp}; for this setting we require more than a pointwise condition at each $x \in X$, ensuring existence of preimages that are SU(2)-valued functions. This surjectivity, however, is simple to see following a couple of intermediate observations. It turns out that the real and imaginary parts of the $\sigma_x$ and $\sigma_z$-dependent perturbations for protocols close to the identity for all $x$ can be independently modified (Lem.~\ref{lem:independently_tunable_planar_perturb}) to leading order, and thus each of the three distinct terms in the coefficient of $\sigma_z$ in (\ref{eq:leading_order_nested_comm}) are uncorrelated and dense in $S_{\varepsilon^{1/4}}(F(X, \xi))$. As this class contains an identity for the monoid of pointwise SU(2) multiplication, this map must be surjective. In fact, assuming that the components $\Im[P_k], \Re[Q_k], \Im[Q_k]$ for $k \in \{0, 1\}$ are dense in $S_{\varepsilon^{1/4}}(F(X, \xi))$, the image of all valid protocols under this commutator has components dense in a $S_{\varepsilon^{1/4}}(F(X, c_0\xi)$ (still equicontinuous) with Lipschitz constant increased by a constant factor (here $c_0$). Consequently, as we continue the recursive process of refining our net, the properties of the functional class in which the relevant matrix component is dense (under $\Pi$) are modified (i.e., the functions become `spikier'), though this poses no problem as $F(X, \xi) \subsetneq F(X, c_0\xi)$.

        \item \emph{Partial constraints simplify our problem:} Remembering the caveat in Fig.~\ref{fig:skt_lifting_version} relating to the projection $\Pi$ (which specifies a component of a matrix element we want to be dense in a functional class), we now show that this weaker condition makes the problem of computing preimages under the nested commutator easier. I.e., we justify why, in the point above, we mentioned only the $\sigma_z$-proportional leading-order component of (\ref{eq:leading_order_nested_comm}) and ignored the leading order $\sigma_x$ component:\footnote{In the following equation we fold all non-$\varepsilon$-dependent constants into $(\ast)_x$ and $(\ast)_z$; in truth we know a lot about the structure of these coefficients but none of this needed for the argument we give besides their boundedness.}
            \begin{equation} \label{eq:planar_protocol_ast_terms}
                [[U_0, U_1],[U_0^\dagger, U_1^\dagger]] =
                I + \varepsilon^4
                \biggr[(\ast)_z\,\sigma_z + (\ast)_x\,\sigma_x\biggr] + \mathcal{O}(\varepsilon^5).
            \end{equation}
        The ability to ignore this additional component is addressed by what we call \emph{retroactive definition}. I.e., by finding some preimages $V(\Phi_0, x), W(\Phi_1, x)$ for a target $U(x)$, their nested commutator will induce \emph{some} $\sigma_x$ term (i.e., the coefficient $(\ast)_x$ in (\ref{eq:planar_protocol_ast_terms})) along with the desired $\sigma_z$ term (i.e., $(\ast)_z$ in (\ref{eq:planar_protocol_ast_terms})). Whatever $(\ast)_x$ term is induced by our choice of $(\ast)_z$, we can \emph{retroactively define} the target $U(x) \mapsto U'(x)$ to have an $(\ast)_x$ term that is $\mathcal{O}(\varepsilon^5)$-compatible with the one induced by our choice of $(\ast)_z$. Note, however, that to safely be able to do this we need to ensure two things.

        The first is that we are always trying to approximate our target \emph{unitary} in the \emph{operator norm}, while the statements above concern approximating target \emph{functions} using \emph{supremum norms}; this causes no issue in our setting as these will be related by a constant due to working in SU(2), and so closeness to desired $(\ast)_x, (\ast)_z$ is sufficient (see Lem.~\ref{lem:elem_op_norm}). The second issue is that we have to make sure that \emph{retroactive definition} of the off-diagonal $(\ast)_x$ components of our target at each step of our recursive argument are `backwards compatible' (i.e., they don't invalidate the existence of previously assumed nets). This is fine as, once we are near enough to the identity unitary, after the first shift, condition (2) of Def.~\ref{def:compatible_comm} applies; for symmetric QSP specifically, this follows as $\Im[\langle 0 | U(\Phi, x) |0\rangle] = \Im[P(x)]$ uniquely defines $\langle 0 | U(\Phi, x) |1\rangle = iQ(x)\sqrt{1 - x^2}$, as the latter is by definition pure imaginary, and $\sqrt{1 - |P(x)|^2}$ can only differ from $iQ(x)\sqrt{1 - x^2}$ in QSP protocols by a phase, which in this case must be trivial. In other words, symmetric QSP protocols near the identity are unique, given $P(x)$, up to a sign on $Q(x)$, and thus there is only one `branch' on which to converge.

        \item \emph{On ensuring the recursive step:} Unlike in the standard SKT proof, which considers nets over SU(2), we consider nets over SU(2)-valued functions,\footnote{As mentioned we only need to consider components of such functions, which inherit relevant topological properties.} which for these spaces to be compact must be bounded and equicontinuous. Unlike in the SU(2) setting, this compact set of functions \emph{does not form a group} under pointwise products (i.e., although the space of all QSP protocols does form a group under pointwise products, this group is not compact). Consequently the equicontinuity properties of our net will change under our nested commutator (the Lipschitz constant will grow by a constant factor at each step). However, at each step this space remains compact by application of Tychonoffs's theorem and that continuous maps take compact sets to compact sets.
    \end{enumerate}
    From these three observations above we see that for each set forming an $\varepsilon$-net for $S_{\varepsilon^{1/4}}(F(X, \xi))$ (a ball about the zero function in the supremum norm, contained within $F(X, \xi)$), application of the nested commutator to this set produces a new set which is a finer $\varepsilon^{5/4}$-net, at the cost of increasing the length of the protocol by a factor of at worst $16$, and increasing the Lipschitz constant by the same factor of $16$. As this new net satisfies the same properties as the input net, we can repeat this process (shifting back closer to the identity by applying the inverse of one step's approximation to the current target). This process is the subject of the main Thm.~\ref{thm:qsp_skt}. 
\end{theorem}

We take a small detour to state a more generic result given only a \emph{compatible group commutator} (Def.~\ref{def:compatible_comm}), in which case the net refinement process is generalized to an improvement by a factor of $\varepsilon \mapsto \varepsilon^{1 + b}$ on balls of radius $\varepsilon^{1/a}$ for constants $a > 1$ and $b > 0$. In the following definition we give sufficient conditions on `group commutators' in our lifted setting that allow them to be \emph{compatible} with proving a QSP-SKT theorem (or more precisely a shrinking lemma as in Thm.~\ref{thm:shrinking}), given $(\Pi, F, X)$. While our main theorem (Thm.~\ref{thm:qsp_skt}) concerns only the symmetric ansatz and our nested group commutator (Lem.~\ref{lem:group_comm_planar_qsp}), it is an interesting open question to understand when similar theorems could be proven for other group-valued functions. In these cases generalized group commutators may be required to extend to diverse ansäzte, providing a curious new role for QSP in understanding lifted `shrinking lemmata.'

\begin{definition}[Compatible group commutator for $(\Pi, F, X)$] \label{def:compatible_comm}
    Let $G$ a topological group,\footnote{We assume $G$ admits a representation; distances between group elements are given by the operator norm of differences of representations of these elements.} $X$ a compact metric space, $G(X)$ a subset of $C(X, G)$ the continuous $G$-valued functions on $X$, and $\Pi$ a projection from $G \rightarrow \mathbb{R}$ such that $C(X, \mathbb{R}) \supseteq \Pi\circ G(X) \supseteq F(X)$. Then $\mathfrak{G}: G(X)^{\otimes n} \rightarrow G(X)$ is a \emph{compatible group commutator} for $(\Pi, F, X)$ if it satisfies the following properties.
    \begin{enumerate}[label=(\arabic*)]
        \item \emph{Surjectivity on balls about the identity function.} There is a unique \emph{identity function} $e(x) \in G(X)$ which is the group identity for $G$ over all $X$. Consider then, for some $\varepsilon > 0$, the ball of functions
            \begin{equation}
                S_{\varepsilon} \equiv \Big\{g \in G(X) \;\Big\vert\; \sup_{x \in X} |g(x) - e(x)| < \varepsilon \Big\},
            \end{equation}
        where distance is taken again with respect to the operator norm applied to differences between representations of elements of $G(X)$.

        We require that there exists some $\varepsilon > 0$, and some $a > 1$ such that $\mathfrak{G}$ acting on elements in $S_{\varepsilon^{1/a}}$ is surjective on $S_{\varepsilon}$, i.e., that
            \begin{equation}
                \mathfrak{G}(S_{\varepsilon^{1/a}}^{\otimes n}) \supseteq S_{\varepsilon}.
            \end{equation}
        In other words, for a small enough ball about the identity function $e(x)$, we require that all functions in that ball have \emph{at least one} pre-image under $\mathfrak{G}$, all of whose elements are within a (possibly polynomially larger) ball about the identity function.

        \item \emph{Closeness under $\Pi$ near the identity function implies closeness in the ambient group.} Let $\Pi: G \rightarrow \mathbb{R}$ the projection defined above. Then we require that there exists $\varepsilon' > 0$ and a constant $C > 0$ such that
            \begin{equation}
                \sup_{x \in X}\, \lvert \Pi\,g(x) - \Pi\,h(x)\rvert < \varepsilon
                \implies 
                \sup_{x \in X}\, \lvert g(x) - h(x) \rvert < C\varepsilon,
            \end{equation}
        for all $g(x), h(x) \in S_{\varepsilon'}$. I.e., on small enough balls around the identity function, we require closeness under $\Pi$ to imply closeness (with at most a constant blow-up) between representations of group-valued functions in the operator norm.

        This property is essential to allowing property (1) to be used to show (approximate) surjectivity onto a ball of group-valued functions from surjectivity under $\Pi$, and also allows the critical `translation' step in the recursive argument used in proofs of SKT under only the assumption of closeness under $\Pi$.

        \item \emph{Shrinking property: approximate pre-images for $\mathfrak{G}$ perform better than expected near the identity function.} We require that there exists $\Delta > 0$ and $b > 0$ such that, for every $h(x) \in S_{\Delta}$, there exists at least one pre-image for $h(x)$ under $\mathfrak{G}$, i.e., $g_1(x), g_2(x), \dots, g_n(x)$ where if each $g_{k}(x)$ is replaced by some approximate $g_{k}^\prime(x)$ such that
            \begin{equation}
                \sup_{x \in X}|g_k(x) - g_k^\prime(x)| < \delta < \Delta,
            \end{equation}
        then the difference between the ideal image of $\mathfrak{G}$ and its action on perturbed inputs is bounded as
            \begin{equation}
                \sup_{x \in X}\,\lvert \mathfrak{G}(g_1(x), g_{2}(x), \dots, g_{n}(x)) - \mathfrak{G}(g_1^{\prime}(x), g_{2}^{\prime}(x), \dots, g_{n}^{\prime}(x))\rvert = \mathcal{O}(\Delta^{b} \delta).
            \end{equation}
        I.e., we require that generating approximations to images under the group commutator works better than expected. We would naïvely expect that the error of the output of $\mathfrak{G}$ would scale at best linearly with the error $\delta$ of its inputs; here we require that this error can be suppressed by at least some factor $\Delta^b$ \emph{when our image} is $\Delta$-close to the identity function (and thus our pre-images, by (1), are at worst $\Delta^{1/a}$ close to the identity function).
    \end{enumerate}
\end{definition}

Given a `commutator' with the properties of Def.~\ref{def:compatible_comm}, we are able to recover a proof of the `shrinking lemma' (Thm.~\ref{thm:shrinking}) used as a component of our QSP-SKT proof (Thm.~\ref{thm:qsp_skt}). As we will see in that proof, for the specific case that our compatible commutator is the nested commutator of Lem.~\ref{lem:group_comm_planar_qsp}, the circuit complexity of the resulting approximation can be directly related to $b$ from property (3). We note that for our purposes it was sufficient to investigate only a single, concrete commutator with two inputs. We thus provide Def.~\ref{def:compatible_comm} primarily for the purpose of future work (and to state Thm.~\ref{thm:qsp_skt} formally with weaker input conditions); higher order compatible commutators may be required in extended settings, in which case it is helpful to outline sufficient properties for an analogue of the `shrinking lemma' to go through.

Before analyzing the repeated application of the QSP shrinking lemma (Thm.~\ref{thm:shrinking}) to constitute our main Thm.~\ref{thm:qsp_skt}, we point out a few of Thm.~\ref{thm:shrinking}'s specific attributes, highlighting where problems appeared in trying to simply apply the standard SKT in a pointwise way. The first observation is that elements of our net no longer form a compact group: while the closure of the set of unitaries whose matrix elements are bounded equicontinuous functions satisfying parity constraints is \emph{compact}, this set is not closed under multiplication of its elements (and specifically under our nested commutator). We solve this problem by adding an additional parameter to our successively finer nets tracking Lipschitz constant bounds; in this way we show each net \emph{still} has preimages under our nested commutator at each step (i.e., property (1) of Def.~\ref{def:compatible_comm}). A second problem is that the double cross product induced by the nested commutator comes with some boundary conditions, as seen by the powers of $\sqrt{1 - x^2}$ multiplying $\sigma_z$ and $\sigma_x$ in (\ref{eq:leading_order_nested_comm}) to leading order. We solve this problem by noting that these functions, along with the additional ability to shift by a constant of order $\varepsilon$, are dense in the desired space of equicontinuous functions in the ball $S_{\varepsilon^{1/4}}(F(X, \xi))$. Thirdly, we had to be careful as we only wanted to control a single component of a unitary element---while this makes the problem no harder, we still had to make this treatment explicit in the form of \emph{retroactive definition} of the unspecified components at each step (i.e., property (2) of Def.~\ref{def:compatible_comm}). Fourthly, the astute reader might notice that the smoothness of our target function didn't appear anywhere in our argument. In actuality such smoothness will appear in the complexity of generating the \emph{initial, constant-precision net}; e.g., if our function has a constant-sized jump discontinuity within a $1/\delta$-sized region, or a maximum derivative of magnitude $\delta$, applications the Markov brothers' inequality and Jackson-type theorems will yield $\mathcal{O}(\delta/\varepsilon_{0})$ factors for the complexity of the initial net \cite{cs_qsvt_tang_tian}. It is only in the case that our target function is analytic on the Bernstein ellipse that the $\text{polylog}(1/\varepsilon)$ dependence from our net refinement procedure dominates (for detailed discussion, see Thm.~\ref{thm:lcu_const_space_density} and Rem.~\ref{rem:volume_lower_bounds} in Sec.~\ref{sec:skt_qsp_applications}).

In the following theorem, as mentioned, we successively apply the three major steps outlined in Fig.~\ref{fig:skt_proof_overview}, each now suitably modified to deal with SU(2)-valued functions (and their action under $\Pi$). Following construction of our `lifted' mathematical objects the main proof is now mostly a game of snapping the relevant pieces together and computing the total complexity. In the concluding Sec.~\ref{sec:discussion_open} we will discuss ways in which the constant factors that appear below could be improved.

\begin{theorem}[QSP-SKT: A Solovay–Kitaev theorem for quantum signal processing] \label{thm:qsp_skt}
    Let $X = [-1,1]$ and $\Sigma = \{W, P_0, P_1, \dots, P_n\}$ an \emph{instruction set} (Def.~\ref{def:qsp_instruction_set}) comprising an \emph{oracle} $W \in X \rightarrow \text{SU(2)}$ and \emph{phases} $P_k \in \text{SU(2)}$ whose finite products are $\Pi$-dense (Def.~\ref{def:pi_density_qsp}) in a compact function space $F(X) \subsetneq C(X, \mathbb{R})$, where $F(X)$ is some common function space as per Def.~\ref{def:qsp_skt_function_space}, and $\Pi$ is a projection from $\text{SU(2)} \rightarrow \mathbb{R}$. 
    Moreover, let $\mathfrak{G}$ a \emph{compatible commutator} for $(\Pi, F, X)$ (Def.~\ref{def:compatible_comm}).
    Then for any $\varepsilon > 0$ and any $f \in F(X)$ there exists an element in $\Sigma^*$ (finite products of elements from the instruction set) of oracle length and phase length (see Def.~\ref{def:phase_oracle_length}) $\mathcal{O}(\log^{c}{(\varepsilon^{-1})})$ (for $c$ a constant) which $\varepsilon$-uniformly approximates $f$ under $\Pi$.
    
    The proof combines lemmata given above for the manipulation of planar protocols near the SU(2) identity to generate increasingly good nets, just as in Fig.~\ref{fig:skt_proof_overview}, save for paths in the tangent space as depicted in Fig.~\ref{fig:qsp_net_creation}. These steps together, as in the standard SKT \cite{ksv_textbook_02, dn_skt_overview_05, nc_textbook_11}, technically also furnish a (bad) `constructive algorithm' for QSP phase finding up to a series of numerical caveats. As our intention is just to show the existence of a set of good phases, details on how one would compute these objects are omitted.
    \begin{enumerate}[label=(\arabic*)]
        \item \emph{Use initial density assumption:} We use Thm.~\ref{thm:informal_qsp_density}) to establish $\varepsilon_{0}$-net a about special subset of QSP protocols. This initial density assumption allows us to find a set of phases such that the resulting unitary $U_0 U_0^R$ is $\varepsilon_{0}$-close in trace distance to the target unitary $A_0$. Note that this density assumption allows us to find a protocol whose $\Im{\langle 0 | U_0 U_0^R| 0\rangle}$ component (i.e., the protocol under $\Pi$) is $\varepsilon_0$-close to the desired component, where this component is in the common QSP function space (Def.~\ref{def:qsp_skt_function_space}).
        
        \item \emph{Use initial net translate to protocols near the identity:} We have made the assumption that $\lVert I - A_0 (U_0 U_0^R)^\dagger\rVert < \varepsilon_0$. We use this along with Lem.~\ref{lem:cyclic_iden_prod} to show that $U_0^\dagger A_0 U_{0}^{R\dagger}$ must also be close to the identity \emph{as well as} an XZ-planar protocol. This establishes an $\varepsilon_0$-net over XZ-planar protocols encoding small-norm functions obeying parity and derivative constraints.

        \item \emph{Use nested group commutator to refine net about the SU(2) identity:} We look at the image of our net about the identity under the nested commutator (Lem.~\ref{lem:nested_group_comm}), which preserves XZ-planar protocols (Lem.~\ref{lem:group_comm_planar_qsp}). We use these lemmas to find pairs of QSP protocols achieving desired $\varepsilon_{n - 1}^{1/4}$-norm functions up to accuracy $\varepsilon_{n - 1}$, such that their nested group commutators achieves a $\varepsilon_{n - 1}$-norm function up to accuracy $\varepsilon_{n - 1}^{5/4}$ at the cost of increasing the total protocol by a factor of $17$. This precisely the statement of Thm.~\ref{thm:shrinking}, as our translation to the identity has taken us into $S_{\varepsilon_{n-1}^{1/4}}(F(X, \xi))$ under $\Pi$.
        
        \item \emph{Analyze our group commutator to derive protocol length:} The image of our net under the nested group commutator is described by the following modifications to its defining parameters:
            \begin{align}
                \varepsilon_n &= \varepsilon_{n - 1}^{5/4},\\
                \ell_n &= (16 + 1)\ell_n = 17\ell_n.
            \end{align}
        Following analogous arguments to those of \cite{dn_skt_overview_05}, we can determine the required length of a protocol achieving a desired precision:
            \begin{equation}
                n = \left\lceil \frac{\log{(\log{c\varepsilon^{-1}}/\log{c\varepsilon_0^{-1}})}}{\log{5/4}} \right\rceil
                \implies 
                \ell_{\varepsilon}
                =
                \mathcal{O}\left(\log^{(\log{17})/(\log{5/4})}(\varepsilon^{-1})\right),
            \end{equation}
        where $(\log{17})/(\log{5/4}) \approx 12.7$ is the approximate exponent. Note this length is proportional to $\ell_0$, which contains factors depending on the maximum derivative of the target function as well as discontinuities (see Thm.~\ref{thm:lcu_const_space_density}). The constants appearing here are worse than in the usual Solovay–Kitaev theorem, as well as those encountered in standard proofs of the complexity of QSP. This is due to our use of a commutator which self-corrects errors less optimally than its non-nested counterpart, as well as the application of less than state-of-the-art (but clearer) methods for proving the SKT. The purpose of this work is to show the possibility of applying SKT methods to QSP, as well as how these methods persist in settings where standard QSP approaches fail. Such worsening parameters also appear in other extensions to the SKT, such as the inverse-free setting \cite{bt_inverse_free_21}.
        
        \item \emph{Once the approximation is good enough, read off the protocol:} At each step we generated a new approximating unitary of the form $U_{n}U_{n}^R$, where the bifurcation into an initial protocol and its reversed version is guaranteed (past the first step) by the planarity of the nested commutator. This allows us to continually translate nearer to the identity over $X$ by a symmetry-preserving translation, i.e.,
            \begin{equation}
                (U_{n-1}^{\dagger}\cdots U_{1}^{\dagger} U_{0}^{\dagger}) A_{0}(U_{0}^{R\dagger} U_{1}^{R\dagger} \cdots U_{n-1}^{R\dagger}) = V_{n-1} \mapsto U_{n}^{\dagger}V_{n - 1} U_{n}^{R\dagger},
            \end{equation}
        which by is Lem.~\ref{lem:cyclic_iden_prod} still close to the identity and by Lem.~\ref{lem:sym_perturb_iden} still XZ-planar. At end of the recursive generation of our nets, we can unwrap this protocol to give our $\varepsilon_{n}$-approximation of $A_0$:
            \begin{equation}
                (U_0 U_1 \cdots U_{n})I(U_{n}^R\cdots U_{1}^R U_{0}^R) \approx_{\varepsilon_n} A_0.
            \end{equation}
        For the setting where the initial function has odd-parity, we will insert an additional copy of the signal oracle in the center of this protocol, which was removed at the beginning of the recursive argument to ensure even parity; this ensures no discrete oracle applications are `split' across our symmetric translation step.
    \end{enumerate}
\end{theorem}

\begin{remark}[On the form of the difference between two close planar QSP protocols] \label{rem:form_of_difference_planar}
    In many places in this work we consider `translating' QSP protocols by other protocols in a symmetric way, such that the result is about the identity (Def.~\ref{def:planar_protocols_iden}), which allows for SKT techniques to be applied in the linearized limit. Here we look at the form of these translations. 
    
    Consider a QSP protocol with Pauli form (Def.~\ref{def:qsp_pauli_form}) $U = \exp\{i\theta(\phi\cdot\sigma)\}$, where we have supressed the arguments for $\theta(x)$ and $\hat{\phi}(x)$ (as well as vector-notation) for brevity. Assume $U$ is an (even-parity) symmetric QSP protocol, in which case it can be written as the product of two protocols satisfying
        \begin{equation}
            U = VV^T = \exp\{i\theta(\phi\cdot\sigma)\}
            = \exp\{i\theta'(\phi'\cdot\sigma)\}\exp\{i\theta'(\phi'\cdot\sigma)\}^{T},
        \end{equation}
    for some related $\theta', \phi'$. More specifically, the relation between $\theta, \phi$ and their primed counterparts is simple (and can be immediately written down by identifying the transpose of a QSP unitary with (\ref{eq:pauli_standard_map_0})--(\ref{eq:pauli_standard_map_1})):
        \begin{align}
            \theta'(x) &= \hphantom{-}\theta(x),\\
            \phi'_x(x) &= \hphantom{-}\phi_x(x),\\
            \phi'_y(x) &= -\phi_y(x),\\ 
            \phi'_z(x) &= \hphantom{-}\phi_z(x).
        \end{align}
    Let $U'$ a symmetric QSP protocol perturbed from $U$, which without loss of generality can be written
        \begin{equation}
            U' = V'(V')^T =  \exp\{i[\theta' + \varepsilon\nu]([\phi' + \varepsilon\mu]\cdot\sigma)\}\exp\{i[\theta' + \varepsilon\nu]([\phi' + \varepsilon\mu]\cdot\sigma)\}^{T},
        \end{equation}
    where we have included a small parameter $\varepsilon$ multiplying $\mathcal{O}(1)$-norm functions $\nu(x), \hat{\mu}(x)$. Note also that this type of perturbation can only be valid to first order as $\hat{\phi}$ is unit norm. To this effect we can look at the difference $U(U')^\dagger$, and more specifically the \emph{cyclic permutation} of this difference, which by Lem.~\ref{lem:cyclic_iden_prod} must also be within $\mathcal{O}(\varepsilon)$-distance of the identity:
        \begin{align}
            U(U')^\dagger = VV^T [V' (V')^T]^\dagger &= VV^T ((V')^T)^\dagger (V')^\dagger,\\
            &\longmapsto \left[(V')^\dagger V\right] \left[V^T((V')^T)^\dagger\right]. \label{eq:sym_difference_perturb}
        \end{align}
    By inspection we see that this cyclically permuted product is also planar. We look at the first term in brackets in (\ref{eq:sym_difference_perturb}) first, though the same argument will go through for the second term. Expanding to first order in $\varepsilon$, this term must look like:
        \begin{align}
            \left[(V')^\dagger V\right] &= \exp\{i[\theta' + \varepsilon\nu]([\phi' + \varepsilon\mu]\cdot\sigma)\}^\dagger\exp\{i\theta'(\phi'\cdot\sigma)\},\\
            &= I + i\varepsilon\left[\theta'(\mu\cdot\sigma) + \nu(\phi'\cdot\sigma)\right] + \mathcal{O}(\varepsilon^2).
        \end{align}
    The remaining first order term from (\ref{eq:sym_difference_perturb}) must relate by the transpose, leading to a total perturbation:
        \begin{align}
            VV^T [V' (V')^T]^\dagger = I + i\varepsilon\left[\theta'[\mu + \mu^T] + \nu[\phi' + \phi'^T]\right]\cdot\sigma + \mathcal{O}(\varepsilon^2).
        \end{align}
    Here we have abused notation to say that the transpose works on the \emph{vectors} $\hat{\phi'}, \hat{\mu}$ to negate their $y$-components, meaning this perturbation is manifestly planar\footnote{An interesting observation is that if $\phi', \mu$ were \emph{already} planar (which we have not assumed here) then $\phi' + \phi'^T = 2\phi'$ and $\mu + \mu^T = 2\mu$, offering possibilities for the study of `doubly symmetric' protocols, i.e., where both $\phi = \phi^R$ (as always assumed) and $\phi' = (\phi')^R$, with $\phi = \phi' (\phi')^R$.} with only $x$ and $z$-components remaining.

    From this calculation we can make a couple of observations; the first is that it is always possible to consider planarity-preserving translations (as mentioned at the end of Thm.~\ref{thm:qsp_skt}, in the odd-parity setting, we can always reduce to the even-parity case by considering target protocols with their central oracle removed). The second is that the form of the difference depends on the linear combination of two protocols with $\theta, \phi$ simply related to the perturbation, i.e., $\theta'[\mu + \mu^T]$ and $\nu[\phi' + \phi'^T]$ as shown above. This linearized setting is what allows us to truly talk about `translations' (sums and differences) among the generating Lie algebra, simplifying analysis of the nested commutator in Thms.~\ref{thm:shrinking} and \ref{thm:qsp_skt}.
\end{remark}

In this section we have provided a lifted variant of the Solovay–Kitaev theorem for symmetric QSP protocols. The overall shape of this argument resembled that of the standard SKT proof, relying on a (specially structured, higher-order) group commutator which can refine $\varepsilon$-nets about classes of SU(2)-valued functions. By restricting to a compact subset of the space of possible functions, and choosing a group commutator respecting properties of the symmetric QSP ansatz, we could convert our problem the the approximation of a real-valued function appearing as a component of a matrix element. The primary difficulty of this argument rests in ensuring that the group commutator we provide permits preimages in the right compact function space, and `self-corrects' its errors when near the identity function---these conditions are unique to the QSP setting, and demonstrate that simply applying the standard SKT in a pointwise way is insufficient.

In the following section we focus on the problem of showing the density of particular QSP ansätze in particular classes of functions. This is equivalent to showing universality for a given QSP instruction set (Def.~\ref{def:qsp_instruction_set}, the lifted version of Def.~\ref{def:ins_set}), for the purpose of satisfying the preconditions of our QSP-SKT theorem (Thm.~\ref{thm:qsp_skt}, the lifted version of \ref{thm:skt}).

\section{Functional properties of QSP ansätze} \label{sec:skt_qsp_applications}

\noindent In the previous section we gave Thm.~\ref{thm:qsp_skt}, the formal version of Thm.~\ref{thm:informal_qsp_skt} from Sec.~\ref{sec:main_results}, establishing an SKT for QSP. In this section we examine the preconditions for Thm.~\ref{thm:qsp_skt} and show them for specific variants of QSP. This is in analogy to showing the universality of specific gate sets for the standard SKT, and demonstrates how Thm.~\ref{thm:qsp_skt} can be coupled with differing proofs of universality for different QSP instruction sets (Def.~\ref{def:qsp_instruction_set}). This culminates in Thms.~\ref{thm:abs_sum_sym_qsp_density} and \ref{thm:lcu_const_space_density} (informally collected along with Rem.~\ref{rem:known_qsp_density_properties} in Thm.~\ref{thm:informal_qsp_density}), as well as a series of related results on the functional properties of specific QSP ansätze.

Toward these ansatz-specific universality results, we also provide a series of definitions and lemmas on properties of compact function spaces. These, in turn, allow precise statements (Thm.~\ref{thm:qsp_identity_net_form}) on the permitted structure of QSP protocols about the identity used in the proof of the `shrinking lemma' (Thm.~\ref{thm:shrinking}). We also provide independently interesting arguments (Rem.~\ref{rem:volume_lower_bounds}) to analyze the expected complexity of our approximations. These arguments differ significantly from their standard SKT counterparts (used in that setting to show lower bounds).

\begin{definition}[Compactness (and variants) from \cite{ds_linear_operators_88}] \label{def:compactness}
    A \emph{covering} of a set $A$ in a topological space $X$ is a family of open sets whose union contains $A$. The space $X$ is \emph{compact} if every covering of $X$ contains a finite sub-covering. A subset $A$ of $X$ is called \emph{conditionally compact} if its closure is compact in the relative topology, i.e., if and only if every covering of $A$ by open sets contains a finite sub-cover.
\end{definition}

\begin{theorem}[Arzelà–Ascoli, §IV.6.7 in \cite{ds_linear_operators_88}] \label{thm:arzela_ascoli}
    If $S$ is compact, then a set in $C(S)$, real-valued continuous functions over $S$, is conditionally compact if and only if it is bounded and equicontinuous.

    We can also immediately derive a corollary: let $S$ a compact metric space and $K$ a bounded set in $C(S)$. Then $K$ is conditionally compact iff \emph{for every} $\varepsilon > 0$ there is $\delta > 0$ such that
        \begin{equation}
            \sup_{f \in K} |f(s) - f(t)| < \varepsilon, \quad \rho(s - t) < \delta,
        \end{equation}
    where $\rho$ is the distance measure inherent to $S$. Proof follows immediately from the definition of equicontinuity and the main theorem.
\end{theorem}

Before going further we show how the compactness of the function spaces we're interested in (e.g., the common QSP function space of Def.~\ref{def:qsp_skt_function_space}) allow us to estimate properties of pointwise products of elements of our QSP instruction set. Specifically, the existence of finite sub-covers for every open cover of our metric space permit us to define notions of `volume' for $\varepsilon$-balls around functions in our class, as well as for the whole space, from which we can recover substantially modified versions of the lower-bound methods used in standard SKT literature \cite{dn_skt_overview_05, hrc_efficient_discrete_02}. While in our modified setting these bounds are not as useful, the argument below presents instructive differences between SU(2) and compact subsets of SU(2)-valued functions. As discussed in Sec.~\ref{sec:discussion_open} other methods might allow tighter results for smoother function classes.

\begin{remark}[On lower bounds for a QSP-SKT per Thm.~\ref{thm:qsp_skt}] \label{rem:volume_lower_bounds}
    Following the methods in Sec.~5 of \cite{hrc_efficient_discrete_02} we can consider `volume-based' arguments toward lower bounds on the complexity of computing SKT-like approximations. For the standard SU(d) Solovay–Kitaev theorem one can show that $\varepsilon$-balls in SU(d) have volume (according to the Haar measure) proportional to $\varepsilon^{d^2 - 1}$, meaning that if each string corresponding to a product of group elements from our instruction set could reach one of these balls (and if they were both disjoint and covered all of SU(d)), then covering the entire group would mean the following inequality is satisfied:
        \begin{equation}
            2 |G|^n k_2 e^{d^2 - 1} > 1,
        \end{equation}
    where $|G|$ is the size of the instruction set alphabet, $n$ is the length of the considered sequences, and $k_2$ is a constant induced by the known result that for $V(r)$ the Haar measure of a ball of radius $r$ in SU(d), there exists constants $k_1$, $k_2$ such that
        \begin{equation}
            k_1 r^{d^2 - 1} < V(r) < k_2 r^{d^2 - 1},
        \end{equation}
    for all $r \in (0, r_0)$ for any $r_0$ and $d$. This holds as SU(d) is $d^2 - 1$ dimensional and $V(r)$ cannot depend on the center of the ball by the translation-invariance of the Haar measure. Together these observations place a lower bound on the required $n$:
        \begin{equation}
            n \geq \frac{d^2 - 1}{\log{(2|G|)}}\log{(\varepsilon^{-1})} - \frac{\log{k_2}}{\log{(2|G|)}}.
        \end{equation}
    In our setting we can still try to give a volume argument, but our manifold (and balls within it) look quite different. Specifically our balls are now about \emph{closed paths} in SU(2), parameterized by $x$, and satisfying certain equicontinuity constraints, as well as boundary, boundedness, and symmetry conditions. As a first approach, imagine a parameterized path $\Gamma$ in SU(2) depending on $\gamma \in [0,2\pi]$ as follows:
        \begin{equation}
            \Gamma(\gamma) = s(\gamma)I + it(\gamma)X + iu(\gamma)Y + iv(\gamma)Z,
        \end{equation}
    where we know certain constraints among the real-valued $s, t, u, v \in C([0, 2\pi], \mathbb{R})$, for instance that
        \begin{align}
            s(\gamma)^2 + t(\gamma)^2 + u(\gamma)^2 + v(\gamma)^2 &= 1,\\
            s(0) = s(2\pi), t(0) = t(2\pi), u(0) = u(2\pi), v(0) &= v(2\pi).
        \end{align}
    To make use of Arzelà-Ascoli (Thm.~\ref{thm:arzela_ascoli}) we need to consider the closure of the space of bounded \emph{and} equicontinuous functions, and so require the additional constraint
        \begin{equation}
            s'(\gamma)^2 + t'(\gamma)^2 + u'(\gamma)^2 + v'(\gamma)^2 \leq \xi^2,
        \end{equation}
    i.e., that the magnitude of $
    \gamma$-derivative of our SU(2)-valued function is bounded above by a constant $\xi > 0$. This together with the unitarity constraint implies that the closure of this set is compact. Considering this geometrically, this means that our curves have a \emph{length} whose upper bound is easily computed according to the ambient metric over SU(2)
        \begin{align}
            \ell(\Gamma) &= \oint (d\ell/d\gamma)\,d\gamma\\
            &= \int_{\gamma \in [0, 2\pi]} \sqrt{|s'(\gamma)|^2 + |t'(\gamma)|^2 + |u'(\gamma)|^2 + |v'(\gamma)|^2}\,d\gamma 
            \leq \xi \oint d\gamma \leq 2\pi \xi.
        \end{align}
    Consequently the set of paths we have to care about are a subset of those with length upper-bounded by $2\pi \xi$, balls around which, in the ambient metric, have volume proportional to
        \begin{equation}
            V(\Gamma(\varepsilon, \gamma)) \leq 2\pi \xi k_2 \varepsilon.
        \end{equation}
    The problem with immediately applying a volume argument to our problem to furnish a lower bound is that (1) our notion of balls have volume depending on length, and (2) that even beyond this length dependence two geometrically identical (but functionally distinct) paths $\Gamma(\gamma)$ and $\Gamma(\tau\circ\gamma)$ can exist, where $\tau$ is some continuous bijection
        \begin{align}
            \tau: [0, 2\pi] &\mapsto [0, 2\pi],\\
            \tau(0) &= \tau(2\pi), \quad |\tau'| \leq 1.
        \end{align}
    These are \emph{not} identical paths, though both parameterizations satisfy the boundedness and equicontinuity conditions we require. To determine the volume of our entire space we need to account for (1) valid re-parameterizations of a given path, as well as (2) the number of such paths for a given length. As such the following argument cannot rigorously establish a lower bound, which would require being able to lower-bound the volume of all paths and upper bound the volume of our space, and so we instead make an \emph{average case} argument analyzing the \emph{expected word length} for a cover. We will invoke compactness to allow all quantities involved to be finite and depending on $\varepsilon$, e.g., while the number of continuous functions is uncountably infinite, we can choose a number of $\varepsilon$-inequivalent\footnote{I.e., we consider only a smaller set of \emph{representative} functions which must differ by at least $\varepsilon$ in sup-norm, equivalently a finite subcover.} representative bounded equicontinuous functions to be finite, by Arzelà-Ascoli (Thm.~\ref{thm:arzela_ascoli}).

    The factor contributed by (1), the number of reparameterizations of a curve according to $\tau$ such that equicontinuity is not violated, and such that two curves cannot be covered by the same $\varepsilon$-ball in a finite subcover provided by Arzelà-Ascoli for a given $\varepsilon$, can be estimated\footnote{The purpose of the constraints below ensures that our functions differ by at least $\varepsilon$-from each other, are injective, identify $0$ with $2\pi$, and have Lipschitz constant at most $\xi$.} by counting the number of paths on a grid of size $\lceil(2\pi/\varepsilon)\rceil\times\lceil(2\pi/\varepsilon)\rceil$ such that we (a) start in the bottom left square, (b) end in the top right, (c) only move either right or up, and (d) we never move \emph{up} more than $\lceil\xi\rceil$ steps consecutively.

    This is a variant of a classic problem in combinatorics, equivalent to asking how many \emph{balanced} binary strings of length $2\lceil(2\pi/\varepsilon)\rceil = 2r$ do not contain a run of more than $\lceil\xi\rceil = s$ consecutive zeros. The total number of balanced strings of a given length easy to compute:
        \begin{equation} \label{eq:balanced_str_num}
            |\{s \mid \text{balanced, $\ell = 2r$}\}| = \binom{2r}{r} \approx \frac{2^r}{\sqrt{\pi r}}.
        \end{equation}
    We ask for an additional condition, removing strings with more that $\xi$ consecutive zeros; while there may be lighter combinatorial tools, the generating-function based method of Goulden and Jackson \cite{gj_cluster_79, gj_cluster_book_83,nz_gj_cluster_method_99} applies, and gives a curious method for analyzing the number of discretized, $\xi$-Lipschitz non-decreasing functions on our grid. Specifically, one can follow their method to build the multivariable function
        \begin{equation}
            G(t, u, v, \eta) = 
            \left[\frac{tu - 1}{t^\eta u^\eta - 1} - tv\right]^{-1}.
        \end{equation}
    Then the series generated in $t$ by $G(t, u, v, \eta)$ has the property that the (integral) coefficient of the term $u^a v^b t^c$ (which must satisfy $a + b = c$, as is manifest) is the number of binary strings of length $c$ with $a$ zeros and $b$ ones \emph{such that} the string does not contain $\eta$ or more consecutive zeros. Setting $c = 2r$ and $a = b = r$ gives us our path number for $\eta = \lceil \xi \rceil$. It turns out this only asymptotically reduces the exponent of the expression in (\ref{eq:balanced_str_num}) by a constant proportional to $\xi$.

    For the contribution from (2), the number of $\varepsilon$-inequivalent paths of length $2\pi\xi$, again an exact answer is difficult to compute, but we know it can be no worse than simply choosing $\lceil 2\pi\xi(1/\varepsilon)\rceil$-balls from our finite cover of SU(2) of size $\lceil(2\pi/\varepsilon)^2\rceil$, leading to a factor of order $(c/\varepsilon)^{c/\varepsilon}$, applying Stirlings' approximation.

    Multiplying the above factors together means that in order to cover our space of functions with disjoint balls about functions corresponding to our collection of words we have to satisfy the rough relation\footnote{We note again that this remark is not a proof of a lower bound, as our $\varepsilon$-balls no longer have uniform size, and our volume argument is only correct to leading order, but instead an average-case analysis for the expected sufficient length to cover our (compact) function space.}
        \begin{equation}
            c_0 \xi \varepsilon |A|^{n} > e^{c_1\xi/\varepsilon}\frac{1}{\text{poly}(\varepsilon)},
        \end{equation}
    which, rearranging, implies that the expected length of words to cover our space goes as
        \begin{equation}
            n = \mathcal{O}\left[\frac{\xi}{\varepsilon}\text{polylog}{(\xi^{-1})}\text{polylog}{(\varepsilon^{-1})}\right].
        \end{equation}
    While at first this might seem much worse than the usual pure $\text{polylog}{(\varepsilon^{-1})}$ scaling expected from QSP, said scaling is only for functions analytic on the domain of the signal.\footnote{In truth for functions analytic on the slightly larger Bernstein ellipse; for further discussion on the efficiency of polynomial approximation see \cite{cs_qsvt_tang_tian, trefethen_approx_19}.} For our setting, where we only assume singly-differentiable functions with $\xi$-bounded Lipschitz constant, the worst-case lower bound for the polynomial degree required to uniformly approximate said functions is given by Jackson-type theorems \cite{trefethen_approx_19} which require degree $n \approx \varepsilon^{-1/r}$ for $r$-times differentiable targets, matching our bound; similarly, by the Markov-brothers' inequality \cite{trefethen_approx_19}, a linear $\xi$-dependence is also generally optimal (we find a result slightly below linear $\xi$-dependence because this is not a lower bound but an average-case argument, as we cannot lower bound the volume of an average word).
\end{remark}

We now consider a more structured class of QSP ansatz (specifically symmetric QSP \cite{sym_qsp_21}), and provide a partial characterization of the structure of protocols which are $\varepsilon$-near the identity over all signals $x$. It turns out that this partial characterization can be bootstrapped to a result strong enough to satisfy the requirements of our QSP `shrinking lemma' (Thm.~\ref{thm:shrinking}), specifically the approximate element independence used in Lem.~\ref{lem:independently_tunable_planar_perturb}. We break this argument into three parts, with the simpler case in Lem.~\ref{lem:explicit_qsp_planar_perturb}, discussion of the closure of this case under finite products in Rem.~\ref{rem:possible_qsp_perturb}, followed by a concise overview in Thm.~\ref{thm:qsp_identity_net_form}.

\begin{lemma}[On a special class of planar QSP protocols about the identity] \label{lem:explicit_qsp_planar_perturb}
    A sufficient condition for a symmetric QSP protocol\footnote{Again we should note that we're being concise here and ignoring odd-order symmetric protocols; by $U(\Phi_1 \cup \Phi_2, x)$ we mean the product $U(\Phi_1, x)U(\Phi_2, x)$. In the odd order case the last phase of $\Phi_1$ and the first phase of $\Phi_2$ are split by the application of an oracle unitary.} to be the identity over all signals $x$ is that it has the form
        \begin{align}
            U(\Phi\cup\Phi^{R}, x)U([-\Phi]\cup[-\Phi^{R}], -x) 
            &= U(\Phi, x)U(\Phi^{R}, x)U(-\Phi, -x)U(-\Phi^{R}, -x)\\
            &= V(x)W(x)W(x)^\dagger V(x)^\dagger,\\
            &= I,
        \end{align}
    where $U(\Phi, x) \mapsto U(\Phi, -x)$ can be induced by $W(x) \mapsto ZW(x)Z$ for each oracle call, and reversing the phases $(\Phi \mapsto \Phi^{R})$ or negating $(\Phi \mapsto -\Phi)$ is straightforward. Cyclic permutations of unitaries which multiply to the identity must still multiply to the identity, i.e.,
        \begin{equation} \label{eq:sym_qsp_iden_prod}
            U(-\Phi^R, -x) U(\Phi, x) U(\Phi^{R}, x)U(-\Phi, -x) = I.
        \end{equation}
    It's not difficult to see that the above protocol must be planar as its phases are invariant under reversal; we now investigate a simple perturbation of the above product preserving this planarity. Let $\Phi = \{\phi_0, \phi_1, \dots, \phi_n\}$, and consider the perturbation $\Phi \mapsto \Phi'$ that only changes the last phase by a small amount, $\phi_n \mapsto \phi_n + \varepsilon$, for the two outer terms of the expression on the left side of (\ref{eq:sym_qsp_iden_prod}). This induces an XZ-planar QSP protocol within $\mathcal{O}(\varepsilon)$ of the identity for all arguments. Explicitly, the leading order (XZ-planar) perturbation breaks into two terms, each contributed by the expansion $\exp{(i(\phi_n + \varepsilon)Z)} \approx \exp{(i(\phi_n)Z)}(I + i\varepsilon Z)$, with an overall form:
        \begin{align}
            U(-\Phi^{'R}, -x) &U(\Phi, x)U(\Phi^{R}, x)U(-\Phi^{'}, -x)\nonumber\\
            &= I - 
            iZ\varepsilon 
            \left[
               U(\Phi'', x) +
               U(-\Phi''^{R}, -x)
            \right]
            + \mathcal{O}(\varepsilon^2), \label{eq:perturbed_protocol_leading}
        \end{align}
    where $\Phi''$ is an \emph{symmetric} list of $2n$ phases with form
        \begin{equation}
            \Phi'' = \{\phi_0, \phi_1, \dots, \phi_{n-1},2\phi_{n}, \phi_{n-1}, \dots, \phi_{1}, \phi_{0}\}.
        \end{equation}
    We can actually check the form of this sum, as we know that symmetric QSP protocols have $Q(x)$ is purely imaginary while $P(x)$ can have non-trivial real and imaginary components. Relatedly, one can also show (Sec.~1.5.2 in \cite{rossi_fqa_thesis_24}) that $\Phi \mapsto \Phi^R$ takes $(P, Q)$ to $(P, -Q^*)$, while $\Phi \mapsto -\Phi$ takes $(P, Q)$ to $(P^*, -Q^*)$, and thus their composition takes $(P, Q) \mapsto (P^*, Q)$. Taken together this means that the perturbation above has a \emph{purely imaginary} on-diagonal component and a \emph{purely real} off-diagonal component, which is sufficient to establish that the perturbation is planar. Note that as the perturbed protocol itself has phases invariant under reversal this must have been the case, but it is pleasant to see that the leading order perturbation is exactly a sum of two explicitly characterized QSP protocols.

    I.e., let $\Psi$ some symmetric list of phases, i.e., $\Psi^{R} = \Psi$, and let $S(x) \in \mathbb{C}[x]$, $R(x) \in \mathbb{R}[x]$ be the on and off-diagonal polynomials constituting the unitary achieved by these phases. Then there exists an XZ-planar QSP protocol with phases relating to $\Psi$ (i.e., the exact protocol in (\ref{eq:perturbed_protocol_leading}) with $\Phi = \Psi$) that has the leading order form
        \begin{equation}
            V(\Psi, x) = 
            I - i\varepsilon
            \begin{bmatrix}
                S(x) + S^*(x) & 0\\
                0 & -S(x) - S^*(x)
            \end{bmatrix}
            + \mathcal{O}(\varepsilon^2),
        \end{equation}
    where the contributions due to $R(x)$ (which absorbed the factor of $\sqrt{1 - x^2}$ here for brevity) have cancelled themselves out as $x \mapsto -x$ imparts a sign on the diagonal, while $\Phi \mapsto -\Phi^R$ leaves the off diagonal unchanged, causing the two terms to cancel. This perturbation has manifestly the correct form, with a imaginary on-diagonal given $S(x) \in \mathbb{C}[X]$ as noted. We also have a simple description of the possible on-diagonal perturbations we can achieve, as the real part of $S(x)$ for a symmetric protocol can be driven to any bounded, definite parity\footnote{In this case we are dealing only with even-parity protocols, and thus $\Phi$ of length $2n + 1$, but a similar analysis could be done in the odd-parity case by inserting an oracle at the central position to maintain planarity.} polynomial with the boundary condition that this real component of $S(x)$ must take the value $1$ at $x = \pm1$. Note that this matches with the requirement that our perturbation must approach $\exp{(-2i\varepsilon Z)}$ when $x = \pm1$ as the other phases by definition add to zero; in the following Rem.~\ref{rem:possible_qsp_perturb} we show that this boundary condition can be easily circumvented.
\end{lemma}

\begin{remark}[On the set of possible planar pertubations for QSP protocols] \label{rem:possible_qsp_perturb}
    As shown in Ex.~\ref{lem:explicit_qsp_planar_perturb}, it is simple to achieve QSP protocols which perturb around the identity for all arguments, and one form of this perturbation can be written exactly in terms of the sum of two related planar QSP protocols (weighted by the magnitude of the perturbation). More excitingly, given two or more such protocols we know that their \emph{products} must also be perturbations around the identity, and moreover that to leading order these perturbations simply sum. With this in mind, we can bootstrap the example in Ex.~\ref{lem:explicit_qsp_planar_perturb} to a wider class of planar protocols about the identity. 

    We know that, given a planar QSP protocol we can induce certain simple transformations in its elements by conjugations by exponentials of Paulis. If we want to induce a sign on the diagonal or off diagonal components we can conjugate by $X$ or $Z$ respectively, allowing us to induce (assuming $\beta$ is pure imaginary as required)
        \begin{align}
            \begin{bmatrix}
                \alpha & \beta\\
                -\beta^* & \alpha^*
            \end{bmatrix}
            &\longmapsto
            Z
            \begin{bmatrix}
                \alpha & \beta\\
                -\beta^* & \alpha^*
            \end{bmatrix}
            Z
            =
            \begin{bmatrix}
                \alpha & -\beta\\
                \beta^* & \alpha^*
            \end{bmatrix},
            &&\text{($Z$-conjugation)},\\
            &\longmapsto
            \makebox[0pt][l]{X}\phantom{Z}
            \begin{bmatrix}
                \alpha & \beta\\
                -\beta^* & \alpha^*
            \end{bmatrix}
            \makebox[0pt][l]{X}\phantom{Z}
            =
            \begin{bmatrix}
                -\alpha & \beta\\
                -\beta^* & -\alpha^*
            \end{bmatrix}, &&\text{($X$-conjugation)}.
        \end{align}
    This, along with the linear action of these perturbations under multiplication, mean we can achieve the following perturbations where $\Psi$ is a set of symmetric phases that achieve $S(x), R(x)$:
        \begin{align}
            XV(\Psi, x)X &= I + i\varepsilon 
            \begin{bmatrix}
                S(x) + S^*(x) & 0\\
                0 & -S(x) - S^*(x)
            \end{bmatrix} + \mathcal{O}(\varepsilon^2),\\
            HV(\Psi, x)H &= I - i\varepsilon 
            \begin{bmatrix}
                0 & S(x) + S^*(x)\\
                S(x) + S^*(x) & 0
            \end{bmatrix} + \mathcal{O}(\varepsilon^2),
        \end{align}
    where in the last example we used the XZ-preserving Hadamard gate to swap the action of $X$ and $Z$ to generate an off-diagonal perturbation. Note that there are also products similar to the above which are not manifestly planar, and so would need to be made planar by employing the cyclic identities given in Lem.~\ref{lem:cyclic_iden_prod}. The takeaway  is that we can control the on and off-diagonal components of protocols near the identity approximately independently \emph{up to conditions imposed on the real part of the on-diagonal elements of symmetric QSP protocols}. In fact, we can do even a little bit more than this, as multiplying the above protocols on both sides by $e^{i\varepsilon Z/2}$ or  $e^{i\varepsilon X/2}$ can shift the on or off-diagonal components by a small constant, alleviating pointwise boundary conditions, e.g.,
        \begin{align}
            e^{i\varepsilon Z/2}V(\Psi, x)e^{i\varepsilon Z/2} &= 
            I - i\varepsilon
            \begin{bmatrix}
                S(x) + S^*(x) + 1 & 0\\
                0 & -S(x) - S^*(x) - 1
            \end{bmatrix}
            + \mathcal{O}(\varepsilon^2),\\
            e^{i\varepsilon X/2}V(\Psi, x)e^{i\varepsilon X/2} &= 
            I - i\varepsilon
            \begin{bmatrix}
                S(x) + S^*(x) & 1\\
                1 & -S(x) - S^*(x)
            \end{bmatrix}
            + \mathcal{O}(\varepsilon^2).
        \end{align}
    Numerous generalizations to these perturbation methods exist\footnote{E.g., while we perturb a single phase, one could choose to perturb multiple phases, in which case the perturbation would have the form of a sum of many terms that look like QSP protocols.} which may allow for more efficient nets. For our purposes, however, the above arguments are sufficient to generate any desired perturbation. This argument also exemplifies one of the great benefits of using Solovay–Kitaev methods to analyze the functional properties of QSP protocols: we get to work in the \emph{tangent space of the identity for all $x$}, meaning that the closure under finite multiplications is simple.
\end{remark}

The lemma and remark above allow us to relate the density properties of the symmetric QSP ansatz to symmetric QSP protocols close to the identity in SU(2) for all signals. This is a slightly different statement than what is implied by just the density of an ansatz in some compact space of SU(2)-valued functions. While we use that density statement to translate (via shifting, as in (c) of Fig.~\ref{fig:skt_proof_overview}) to a protocol near the identity for all $x$, in following steps we use the existence of the protocols we've just characterized about the identity (and summarized in Thm.~\ref{thm:qsp_identity_net_form}) to make sure that the nested group commutator we use in Thm.~\ref{thm:shrinking} and the main SKT-QSP Thm.~\ref{thm:qsp_skt} is surjective. Following the initial translation, subsequent analysis is done in a small neighborhood of the identity in SU(2), which greatly simplifies our analysis.

\begin{theorem}[Relation between general QSP density and density about identity] \label{thm:qsp_identity_net_form}
    Let $\Sigma$ a QSP instruction set (Def.~\ref{def:qsp_instruction_set}) for $(\Pi, F, X)$ where $F$ is the common QSP function space (Def.~\ref{def:qsp_skt_function_space}) with Lipschitz constant $\xi$, $X = [-1, 1]$, and $\Pi$ projects onto the imaginary part of the top-left matrix element in the computational basis. Then $\Sigma^\ast$ (under $\Pi$) contains a $\varepsilon$-net for $S_{\varepsilon^{1/4}}(F(X, \xi))$ under $\Pi$, the subset of $F(X, \xi)$ whose elements have distance at most $\varepsilon^{1/4}$ from the identically-zero function on $X$ in the supremum norm. Moreover, each element of this net has a \emph{pre-image} (w.r.t.~$\Pi$) such that this pre-image's distance from the identity is bounded above by $\mathcal{O}(\varepsilon^{1/4})$ in the operator norm.
    \begin{proof}
        The purpose of this statement is to usefully handle the fact that, while $\Sigma$ being an instruction set means that it can achieve arbitrary approximations to members of $F(X, \xi)$ under $\Pi$, this \emph{does not necessarily mean} that some $\sigma \in \Sigma^\ast$ achieving some $f \in S_{\varepsilon}(F(X, \xi))$ (functions with supremum norm bounded above by $\varepsilon$) is necessarily $\mathcal{O}(\varepsilon)$-close to $I \in SU(2)$ over all $x \in X$ in the operator norm. Equivalently
            \begin{equation}
                \lVert\Pi (\sigma - \sigma')\rVert_{\rm sup} = \lVert f - f'\rVert_{\rm sup} < \varepsilon
                \;\not\!\!\!\implies
                \lVert\sigma - \sigma'\rVert_{\rm op} = \mathcal{O}(\varepsilon).
            \end{equation}
        In other words, our protocols might permit enough freedoms that two element- or component-wise close QSP protocols need not be close in the operator norm. However, as we only care about approximating a single component, we can always translate close to the identity using our net under $\Pi$ by \emph{retroactive definition} (see Thm.~\ref{thm:shrinking}, and property (2) in Def.~\ref{def:compatible_comm}). Consequently all we need to show is that density for $\Sigma$ under $\Pi$ in $F(X, \xi)$ implies density in $S_{\varepsilon^{1/4}}(F(X, \xi))$ \emph{with the additional property} that all elements have pre-images (w.r.t.~$\Pi$) that are $\mathcal{O}(\varepsilon^{1/4})$-close to $I$ in the \emph{operator norm} over all $x \in X$.

        But this is what we showed in Lem.~\ref{lem:explicit_qsp_planar_perturb} and the following Rem.~\ref{rem:possible_qsp_perturb}! Namely there we built perturbations from QSP protocols which would otherwise be identically $I$ over all $x \in X$ to induce QSP protocols which were $\varepsilon^{1/4}$-near to $I$ in the operator norm over all $x \in X$, \emph{and} whose image under $\Pi$ were exactly the ($\varepsilon^{1/4}$-scaled) imaginary parts of the top left elements of arbitrary symmetric QSP protocols (i.e., dense in $F(X, \xi)$ for any finite $\xi$). Consequently density (under $\Pi$) in $F(X, \xi)$ can be bootstrapped to density (under $\Pi$) in $S_{\varepsilon^{1/4}}(F(X, \xi))$ with the additional promise that there exists pre-images (for $\Pi$) for all elements in this ball whose difference from $I$ over all $x \in X$ is bounded above by $\mathcal{O}(\varepsilon)$ in the operator norm.
    \end{proof}
\end{theorem}

For the latter half of this section we investigate the properties of specific QSP circuit ansätze, providing methods for establishing the density of components of their matrix elements in different classes of functions. Beyond providing the preconditions for QSP-SKT theorems (e.g., those of the main Thm.~\ref{thm:qsp_skt}), these results show that establishing density is (depending on the ansatz and chosen functional class) oftentimes significantly easier than recovering the entire theory of QSP. This allows for alternative techniques (which might be sub-optimal otherwise) to be applied (e.g., the composition of several simpler protocols) to generate constant-precision nets. This allows us to realize novel resource tradeoffs by combining different methods for manipulating block encodings (e.g., LCU methods in Thm.~\ref{thm:lcu_const_space_density} to generate constant [rather than logarithmic] space protocols dense in spaces of supremum-norm bounded [rather than one-norm bounded] functions). Many of these tradeoffs are non-trivial, and suggest new interpolations between block encoding manipulation methods, as well as detours around previously significant roadblocks in analyzing extensions to QSP/QSVT.

\begin{theorem}[Density of symmetric QSP for absolutely summable functions] \label{thm:abs_sum_sym_qsp_density}
    Let $g(x)$ a function in the common QSP space $F(X)$ (Def.~\ref{def:qsp_skt_function_space}) with the additional restriction that the one-norm of its Chebyshev coefficients are upper bounded by a constant $C$,
        \begin{equation}
            g(x) = \sum_{k = 0}^{\infty} c_k T_k(x), \quad \text{s.t.}\quad \sum_{k = 0}^{\infty} |c_k| < C,
        \end{equation}
    and the function is subnormalized by $(1 - \delta)$. Let $G(X)$ the closure of the set of such $g(x)$ w.r.t.~the supremum norm. Then the symmetric QSP ansatz is $\Pi$-dense in the compact subset $G(X) \subsetneq F(X) \subsetneq C(X, \mathbb{R})$, where $\Pi$ is projection onto the imaginary component of the top-right matrix element in the computational basis.

    \begin{proof}
        It is known from \cite{szego_nlfa_qsp_24, amt_23} that in the small $\lVert \Phi \rVert_1$ limit,\footnote{This is mentioned in multiple locations, e.g., p.~5 of the standard lecture reference \cite{tt_nlfa_notes_12}.} the achieved function, i.e., the imaginary component of the \emph{top-right} element of the unitary achieved by the \emph{symmetric} QSP protocol \cite{sym_qsp_21} with phases $\Phi$, is simply the \emph{discrete inverse linear Fourier transform} according to these phases:
            \begin{equation} \label{eq:linear_ft_sym_qsp_limit}
                U(\cos{\theta}, \Phi) =
                \begin{bmatrix}
                    \ast & \ast + i \left[\sum_{k} \phi_k e^{2i\pi k \theta}\right]\\
                    \ast & \ast 
                \end{bmatrix}
                +
                \mathcal{O}(\lVert \Phi \rVert_1^2).
            \end{equation}
        Here we have made the implicit substitution $x = \cos{\theta}$ or equivalently $x = (e^{i\theta} + e^{-i\theta})/2$, which identifies the Chebyshev coefficients of our target function in $x$ with its Fourier coefficients in $\theta$. In this way we see that a constant bound on the one norm of the Fourier coefficients is related (by a constant) to a one norm bound on the QSP phases.

        To show density we pick some $\varepsilon > 0$ and some $g(x) \in G(x)$. We rescale $g(x)$ to some $g'(x) = g(x)/\alpha$ such that the sub-leading terms in (\ref{eq:linear_ft_sym_qsp_limit}) have operator norm bounded above by $\varepsilon/\alpha$ (i.e., such that $(C/\alpha)^2 < \varepsilon/\alpha$ or else $\alpha > C^2/\varepsilon$. We can then use LCU, with controlled access to our protocol (and our protocol with its phases reversed, thus embedding $-Q^*(x)$ in its top-right element), using a single additional qubit, to prepare the modified block encoding which isolates only this imaginary component as the top-left element:
            \begin{equation} \label{eq:lcu_linear_qsp_limit}
                V(\cos{\theta}, \Phi) =
                \begin{bmatrix}
                    \sum_{k} (\phi_k/\alpha) e^{2i\pi k \theta} & \ast\\
                    \ast & \ast 
                \end{bmatrix}
                +
                \mathcal{O}(\varepsilon/\alpha).
            \end{equation}
        Note that we have not yet ensured the existence of such $\phi_k$ such that their inverse Fourier transform is the function of interest $g'(\theta)$, which will require an additional observation. Namely, we can apply to this block encoding the simple QSVT protocol with \emph{identically zero phases} of length $a = \mathcal{O}(\alpha)$, which transforms the top-left component as
            \begin{equation} \label{eq:qsvt_composition_limit}
                W(\cos{\theta}, \Phi) =
                \begin{bmatrix}
                    T_{a}\left[\sum_{k} (\phi_k/\alpha) e^{2i\pi k \theta}\right] & \ast\\
                    \ast & \ast 
                \end{bmatrix}
                +
                \mathcal{O}(\varepsilon).
            \end{equation}
        More precisely we choose an (odd) integer $a$ such that the image under $T_a(x)$ (the $a$-th Chebyshev polynomial of the first kind) of the image of $g'(x)$ over $x \in [-1,1]$ is the same as the image of $g(x)$ over $x \in [-1,1]$, or else that the range of $g'(x) = g(x)/\alpha$ is (non-uniformly) re-expanded to its original range by $T_a(x)$ over $x \in \text{range}(g') \subseteq [-1/\alpha, 1/\alpha]$.

        Having constructed this series of composed protocols: (1) QSP in the small phase limit, followed by (2) constant-space LCU, followed by (3) application of the trivial-phase QSVT protocol, we can now relate the required phases $\phi_k$ to the Fourier coefficients of $g(\theta)$ simply. Specifically, the phases for protocol (1) are just $\phi_k' = \phi_k/\alpha$ where the $\phi_k'$ are the Fourier coefficients of $T_{a}^{-1}(g(\theta))$, where this inverse is well-defined on the range $\cos{\theta} \in [-1/\alpha, 1/\alpha]$; that this is possible is ensured by sub-normalizing our target function by $(1 - \delta)$ for $\delta > 0$, ensuring that the domain-restricted) inverse of the Chebyshev polynomial is not singular. The composition of these three steps thus has both (a) simply computable phases guaranteed by the invertability of the Fourier transform on absolutely summable functions, and (b) the property that its achieved function is $\varepsilon$-close to the desired $g(x)$ in the operator norm.
    \end{proof}
\end{theorem}

\begin{theorem}[Density for constant space QSVT through LCU] \label{thm:lcu_const_space_density}
    Let $F(X)$ the common QSP function space (Def.~\ref{def:qsp_skt_function_space}). Then for any $\varepsilon > 0$ and for any fixed $f(x) \in F(X)$ analytically continuable to the interior of the Bernstein ellipse $E_\rho = \{(1/2)(z + z^{-1}) : |z| = \rho \}$ for $\rho > 1$, there exists a QSVT protocol of length
        \begin{equation} \label{eq:cont_space_lcu_qsvt_length}
            \ell(f,\varepsilon) = 
            \mathcal{O}\left(\frac{1}{\delta}\frac{1}{\log{\rho}}\log^{c}\left[\frac{1}{(\rho - 1)\varepsilon}\right]\right)
        \end{equation}
    using constant additional space whose top-left matrix element in the computational basis $\varepsilon$-uniformly approximates $f(x)$ over $x \in X$.

    \begin{proof}
        Proof follows by a direct application of LCU to generate constant-precision block encodings for the initially assumed net in Thm.~\ref{thm:qsp_skt}. More concretely, for a given $f(x)$ and finite $\varepsilon_{0}$, the degree of the polynomial that $\varepsilon_0$-uniformly approximates $f(x)$ for $x \in [-1,1]$ is known if $f(x)$ is analytically continuable to the interior of the Bernstein ellipse defined in the theorem statement. Namely, if $|f(x)| \leq M$ for $x \in X$ is also known, then $|a_0| \leq M$ and $|a_k| \leq 2M\rho^{-k}$ (Thms.~8.1 and 8.2 from \cite{trefethen_approx_19}, and exposited in the context of QSP in Thm.~20 of \cite{cs_qsvt_tang_tian}). The coefficients' exponential decay means for a desired precision an approximation $f_n$ keeping the first $n$ terms converges uniformly to $f$ at a rate bounded as
            \begin{equation}
                \lVert f - f_n \rVert_{[-1,1]} \leq \frac{2M \rho^{-n}}{\rho - 1},
            \end{equation}
        meaning that keeping the first $n$ terms of our Chebyshev expansion allows $\varepsilon_0$-uniform approximation precisely when
            \begin{equation}
                n = \left\lceil\frac{1}{\log\rho}\log\frac{2M}{(\rho - 1)\varepsilon_0}\right\rceil.
            \end{equation}
        It is known that LCU requires $\mathcal{O}(\log{n})$ additional qubits to prepare block encodings applying degree-$n$ polynomials, and that we are constrained to the $M = 1$ case. Once this $\varepsilon_0$-net is generated, we satisfy the conditions of Thm.~\ref{sec:qsp_skt}, meaning that $\varepsilon$-uniform approximation can be achieved by only polylogarithmic blowup in length dependent only on $\varepsilon$ (as we've assume analyticity inside the Bernstein ellipse). This gives circuits of the length quoted in (\ref{eq:cont_space_lcu_qsvt_length}), using \emph{constant} $a(f)$ additional qubits scaling as
            \begin{equation}
                a(f) = \mathcal{O}
                \left(
                \log\left[\frac{1}{\log\rho}\log\frac{1}{(\rho - 1)}\right]
                \right).
            \end{equation}
    \end{proof}
\end{theorem}

It's worth examining the statement of Thm.~\ref{thm:lcu_const_space_density}; compared to simply using LCU for the entire protocol (which can achieve arbitrary polynomial functions in the input block encoding) here we are guaranteed, for a fixed function $f$, the use of only \emph{constant} additional space instead of \emph{logarithmic} additional space in the \emph{number of terms} in the resulting $\varepsilon$-uniform polynomial approximation. In terms of approximation error, this means that LCU would have required $\log{\log{(\varepsilon^{-1})}}$-space for analytic functions on the Bernstein ellipse, or $\log{(\varepsilon^{-1})}$-space for norm-bounded continuous functions generally.

To be clear, the above result is weaker than what we already know about the functional properties of the standard QSP ansatz, but it has been shown in an entirely disjoint way, and provides a curious interpolation between the known (time and space) complexities of QSP/QSVT and LCU. Moreover, the density statements of Thm.~\ref{thm:abs_sum_sym_qsp_density} (which used constant space irrespective of the applied function under the restriction to absolute summability) and Thm.~\ref{thm:lcu_const_space_density} (which used constant space in terms of the desired \emph{precision} for all supremum-norm bounded functions), were far easier to prove than the full functional properties of standard QSP/QSVT, as in both cases the phases required relate \emph{directly} to the Fourier/Chebyshev coefficients of the target function. This simplicity follows from our desire only to show \emph{density}, not \emph{efficient density} (which is left to the QSP-SKT). For ansatz modifications as mentioned in Sec.~\ref{sec:discussion_open}, the techniques in the two theorems above survive much better than the techniques used to prove the remark below. To remind the reader of the ideal density properties we would like to recover for the standard symmetric QSP ansatz, we roll them into the following remark.

\begin{remark}[Known density properties for symmetric QSP] \label{rem:known_qsp_density_properties}
    Here we restate that we already know comparatively strong density properties of the standard symmetric QSP ansatz. Consider the form of QSP protocols from Def.~\ref{def:qsp},
        \begin{equation}
            U(\Phi, x) 
            \equiv
            e^{i\phi_0\sigma_z}
            \prod_{j = 1}^{k} 
            \left[
            \underbrace{
            \begin{bmatrix}
                x & i\sqrt{1 - x^2}\\
                i\sqrt{1 - x^2} & x
            \end{bmatrix}
            }_{\textstyle e^{i\arccos{(x)}\,\sigma_x}}
            \underbrace{
            \begin{bmatrix}
                e^{i\phi_j} & 0\\
                0 & e^{-i\phi_j}
            \end{bmatrix}
            }_{\textstyle e^{i\phi_j\sigma_z}}
            \right]
            =
            \begin{bmatrixcolor}[black!30!white]
                P(x) & \textcolor{black!30!white}{\ast\;\;}\\
                \textcolor{black!30!white}{\ast} & \textcolor{black!30!white}{\ast\;\;}
            \end{bmatrixcolor}.
        \end{equation}
    Under the additional constraint that $\Phi = \Phi^R$, that is, that the QSP phases are invariant under reversal, then it can be shown that off-diagonal matrix element becomes purely real for all arguments (as implied by Cor.~\ref{cor:pauli_form_qsp_sym}), and moreover that the achievable $\Im[P(x)]$ are simply described. Specifically, for all real-valued polynomials $\tilde{P}(x)$ which are of (1) definite parity, (2) bounded in norm by one for $x \in [-1,1]$, there exists a set of symmetric QSP phase factors of length equal to the degree of $\tilde{P}(x)$ such that it appears as the imaginary part of $P(x)$ \cite{sym_qsp_21}. By simple application of the Stone-Weierstrass theorem we thus see that the symmetric QSP ansatz is $\Pi$-dense in $F(X) \subsetneq C^0[-1,1]$ (following Def.~\ref{def:pi_density_qsp}) where $F(X)$ is the space of real-bounded, norm-bounded, definite-parity, continuous functions on $X = [-1,1]$.
\end{remark}



\begin{theorem}[QSP without phases, QSP without polynomials, and QSP from garbage] \label{thm:qsp_garbage}
    Consider the (infinite) QSP instruction set $\Sigma = \{W, P_\phi\}$, where by $P_\phi$ represents all $\sigma_z$-rotations by \emph{any} real angle $\phi$, and $W = \exp\{i\arccos{x}\,\sigma_x\}$ is the standard oracle. Then the class of functions in which $\Sigma$ is dense under $\Pi$ is unchanged by any of the following modifications:
        \begin{enumerate}[label=(\arabic*)]
            \item Replacing $P_\phi$ with any instruction set (Def.~\ref{def:ins_set}) for the standard SKT, i.e., one which is dense in SU(2), with $\exp\{i(\pi/2)\sigma_z\}$ additionally appended. We call this \emph{QSP without phases}.
            
            \item Replacing $W$ by another oracle $W' = \exp\{if(x)\,\sigma_x\}$ for $f(x) : [-1,1] \rightarrow [-\pi, \pi)$ smooth, invertible, and with finite non-zero derivative everywhere on $[-1,1]$. We call this \emph{QSP without polynomials}.
        \end{enumerate}
    \begin{proof}
        The proof of the first item is straightforward, as replacing our original set (all permissible $\sigma$ rotations) by a constant-precision discretized version (e.g., by using the methods of \cite{rs_ancilla_free_approx_16}), requires only density in SU(2); the additional $\pi/2$ rotation has been included to invert the oracle. This immediately implies the informal Thm.~\ref{thm:informal_qsp_density}. While this statement probably holds even in the absence of inverses, showing this is involved even in the non-QSP setting \cite{bt_inverse_free_21}. 

        The proof of the second item is a little more complex, and starts by re-writing the oracle in terms of a modified $f'(x)$, namely
            \begin{equation}
                e^{if(x)\,\sigma_x} =
                \begin{bmatrix}
                    g(x) & i\sqrt{1 - g(x)^2}\\
                    i\sqrt{1 - g(x)^2} & g(x)
                \end{bmatrix},
            \end{equation}
        where this form is guaranteed as $f(x)$ is real valued and contained in  $[-\pi, \pi)$; specifically, we note that $g(x) = \cos{[f(x)]}$ takes values in $[-1,1]$, and is invertible on $[-\pi,\pi)$ \emph{iff} $f(x)$ is invertible on $[-\pi, \pi)$. Furthermore, we can take $g(x)$ and $h(x) = \sqrt{1 - g(x)^2}$ and re-write them in terms of complex exponentials of $f(x)$:
            \begin{align}
                \begin{bmatrix}
                    g(x) & ih(x)\\
                    ih(x) & g(x)
                \end{bmatrix}
                &= 
                \frac{1}{2}
                \begin{bmatrix}
                    (g + ih(x)) + (g - ih(x)) & (g + ih(x)) - (g - ih(x))\\
                    \ast & \ast
                \end{bmatrix}\\
                &= 
                \frac{1}{2}
                \begin{bmatrix}
                    e^{if(x)} + e^{-if(x)} & e^{if(x)} - e^{-if(x)} \\
                    \ast & \ast
                \end{bmatrix},
            \end{align}
        as is assured by our use of $\sigma_x$ rotations (reëxpressing in terms of coefficients from the $|\pm\rangle$ basis). Consequently the linear Fourier transform observed in Thm.~\ref{thm:abs_sum_sym_qsp_density} will be directly replaced by a transform with respect to the basis $e^{in\,f(x)}$ for $n \in \mathbb{Z}$. By a change of variables $y = f(x)$, and our assumption of the injectivity and smoothness of $f$ on its domain, this modification will change the Fourier coefficients (computed by integrals against plane-waves) only by the addition of a (positive, non-singular, non-zero, proportional to $1/f'(x)^{-1}$) Jacobian to the integrand; consequently we will still have complete basis for absolutely summable functions, leaving our density result unchanged. Proof of this fact is by contradiction, considering some (non-zero) test function which integrates to zero against all $e^{in\,f(x)}$ for $n \in \mathbb{Z}$, which by the properties of our Jacobian must also integrate to zero against $e^{in\,x}$ for $n \in \mathbb{Z}$, contradicting their completeness.
    \end{proof}
\end{theorem}

We have given the two properties in Thm.~\ref{thm:qsp_garbage} the names `QSP without phases' and `QSP without polynomials' because they have each allowed us to make \emph{useful, concrete} statements about the property of alternating circuit ansätze which don't exhibit the typical $z$-rotation structure and polynomial matrix element character of standard QSP. Further relaxations of these ansätze which do not modify density statements (and thus cannot hinder the application of a QSP-SKT) are an interesting starting point for further study. We affectionately call these new avenues, where a wide class of perturbed \emph{QSP-like} circuits can be analyzed, `QSP from garbage.' I.e., almost \emph{any structured} product of unitaries seems to permit \emph{some} useful functional characterization, as these density properties are quite `squishy' (i.e., amenable to invariance under small perturbations intuitively).

Before some final discussion we clarify one last point on the difference between our `lifting' of the SKT to QSP, and the usual `lifting' argument used to go from QSP to QSVT. Further extensions and relaxations of these lifting arguments offer numerous open questions for understanding quantum algorithms (and quantum information processing subroutines) in terms of common category-theoretic constructions. In our case the benefits of these lifting arguments were practical: we can use various SKT-methods to solve difficult problems in QSP, just as one is able to use various simpler QSP methods to solve problems in transforming block encoded linear operators by QSVT.

\begin{remark}[On two notions of `lifting' for QSP] \label{rem:two_notions_lifting_qsp}
    Much of this work has been summarized by saying that QSP `lifts' the SKT in a qualified, mathematical sense (Def.~\ref{def:lift}), summarized in Fig.~\ref{fig:skt_lifting_version}. For those familiar with the quantum singular value transformation, they might also remember that QSVT is said to be a `lifted' version of QSP \cite{gslw_19}. In fact, these two lifting arguments, while distinct, are instances of the same mathematical notion given in Def.~\ref{def:lift}. We can thus depict both diagramatically in the same scheme:
    \begin{equation} \label{eq:two_lifting_args}
        \vcenter{\hbox{\includegraphics[width=0.85\textwidth]{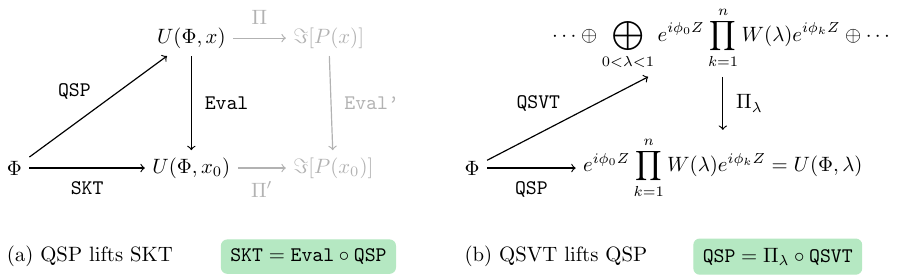}}}
    \end{equation}
    That is, while QSP can be seen to simultaneously implement a continuum of gate approximation problems accessed by $\texttt{Eval}$ at a particular $x = x_0$, QSVT can be seen to simultaneously implement a (finite) number of instances of QSP whose `signal' is a singular value $\lambda$ of a block encoded linear operator. For QSVT these simultaneous instances of QSP happen in direct sum within two-dimensional invariant subspaces labeled by each $\lambda$ (the so called `qubitized' subspaces guaranteed by Jordan's lemma \cite{jordan_75} or the cosine-sine decomposition \cite{pw_cs_decomp_94, cs_qsvt_tang_tian}), here accessed by the projectors $\Pi_\lambda$ whose image is said subspace. 
    
    To review this latter lifting argument in concrete terms, QSVT considers as its oracle a \emph{block encoding} of a linear operator $A$, i.e., some unitary $U$ and two orthogonal projectors $\Pi, \tilde{\Pi}$ that locate (a possibly rescaled, possibly approximate) $A = \tilde{\Pi}U\Pi$ inside $U$. It turns out that the action of $U$ can be summarized in terms of its action on special subspaces with direct-sum structure:
        \begin{equation}
            U = \cdots \oplus \bigoplus_{0 < \lambda < 1}
            \begin{bmatrix}
                \lambda & \sqrt{1 - \lambda^2}\\
                -\sqrt{1 - \lambda^2} & \lambda 
            \end{bmatrix}
            \oplus \cdots,
        \end{equation}
    that is, $U$ is the signal operator implied in (\ref{eq:two_lifting_args}) with a few differences, namely that it is a direct sum of reflections about axes relating to each singular value $\lambda$ of $A$ between the two-dimensional subspaces spanned by $\{|\ell_\lambda\rangle, |\ell_\lambda^\perp\rangle\}$ and $\{|r_\lambda\rangle, |r_\lambda^\perp\rangle\}$ (here the perpendicular subspaces are generated by applying $(I - \tilde{\Pi})U$ and $(I - \Pi)U^\dagger$ respectively to the \emph{opposite-sided} singular vector for $\lambda$ and normalizing), where
        \begin{equation}
            A = \sum_{\lambda} \lambda |\ell_\lambda\rangle\langle r_\lambda| + \cdots,
        \end{equation}
    is the SVD of $A$, and where we have left off terms with SVs equal to zero or one (which are dealt with slightly differently but simply). Note here that $\tilde{\Pi}$ and $\Pi$ are projectors onto the spans of the $|\ell\rangle$ and $|r\rangle$ respectively. The core insight of QSVT is that by using $U, U^\dagger$, $\tilde{\Pi}$ and $\Pi$-controlled NOT gates, and single qubit rotations, exactly QSP can be performed in each of these invariant subspaces, thus transforming 
        \begin{equation}
            A = \sum_{\lambda} \lambda |\ell_\lambda\rangle\langle r_\lambda| + \cdots \mapsto 
            P^{\rm SV}(A) \equiv \sum_{\lambda} P(\lambda) |\ell_\lambda\rangle\langle r_\lambda| + \cdots
        \end{equation}
    as a block encoding for any $P(\ast)$ achievable by QSP. Crucially, the cost of preparing the block encoding and performing this transformation can be logarithmic in the dimension of $A$, allowing for certain extremely efficient transformations outperforming their classical counterparts, and capturing BQP-complete problems like local Hamiltonian simulation \cite{lc_19_qubitization, gslw_19}.

    The main takeaway is that these two notions of lifting do not `talk' to each-other; in other words, the morphisms depicted can be further composed (i.e., \texttt{SKT} can be lifted to \texttt{QSVT} through \texttt{QSP}\footnote{We use monospace names for these morphisms as they have formal definitions and type signatures, as discussed in Fig.~\ref{fig:proof_scheme_summary}.}) with basically no additional bookkeeping. It is for this reason that we largely consider only QSP in our work, as all results we show here can be lifted to QSVT immediately.
\end{remark}

\section{Discussion and outlook} \label{sec:discussion_open}

\noindent In this work we constructed one version of a Solovay–Kitaev theorem (SKT) for quantum signal processing (QSP), showing that the universality of a \emph{QSP instruction set} (Def.~\ref{def:qsp_instruction_set}) implies the existence of good, short protocols---i.e., a unitary with a component of a matrix element that uniformly approximates a desired real-valued function of the signal. We show this universality for a variety of ansätze and a variety of classes of functions in a simple way (Thms.~\ref{thm:abs_sum_sym_qsp_density} and Thms.~\ref{thm:lcu_const_space_density}), satisfying conditions for the application of our main Thm.~\ref{thm:qsp_skt} as well as a variety of corollaries.

The standard SKT and our QSP-SKT are quite distinct. The usual SKT considers the approximation of elements of a finite-dimensional compact group (e.g., SU(2)) by finite products from a finite instruction set known to be dense in SU(2). For the QSP-SKT, in contrast, our instruction set is endowed with both \emph{fixed} and \emph{parameterized} elements, where the intent is now to uniformly approximate members of a \emph{class of functions} in these parameters appearing as components of matrix elements of finite pointwise products from our instruction set. The ambient spaces in each case, e.g., SU(2) versus SU(2)-valued functions, have entirely different character (the most obvious being their dimension and compactness properties).

Nevertheless, by carefully defining certain compact subsets of function spaces to uniformly approximate, and by analyzing generalized group commutators compatible (Def.~\ref{def:compatible_comm}) with SKT techniques, something resembling standard SKT proof methods can be recovered in the QSP setting to rapidly improve the properties of $\varepsilon$-nets for spaces of functions. Ultimately, just as for the usual SKT, this allows us to bootstrap good, short protocols from constant-precision nets.

From a mathematical point of view we `lift' (Fig.~\ref{fig:skt_lifting_version}, and different from `lifting' to QSVT, see Rem.~\ref{rem:two_notions_lifting_qsp}) the Solovay–Kitaev theorem, showing that a continuum of gate approximation problems can be \emph{simultaneously satisfied} by the same `program' of QSP phases (i.e., that a morphism built on the SKT can be made to `factor through' a morphism built on QSP). While the theory of standard QSP implicitly shows this is possible for a fixed ansatz, its proof methods are brittle and rely on fragile algebraic properties that generalize poorly to modified ansätze. In contrast, standard SKT proof methods are rich with geometric intuition (working mainly in certain tangent spaces of the manifold of interest), make no reference to a privileged functional basis, and decouple the proof of the universality of an instruction set from the proof of the efficiency of this approximation. In this vein our contribution shows (1) the insufficiency of merely applying the standard SKT in a pointwise way (as we require a \emph{single} program and certain continuity constraints), and (2) that QSP-like algorithms can be usefully analyzed in `linearized limits' (tangent spaces) where questions on Lie groups reduce to simpler questions on generating Lie algebras.

From an algorithmic point of view, new proof techniques for QSP-like ansätze are welcome. The constructive algorithms embedded in proofs of QSP properties are numerically unstable and unused in practice, with current leading methods for QSP phase-finding based on simple Newton's method-like optimization (performing extremely well, even in the absence of general proofs of convergence). If such optimization methods are used in practice anyway, then the algorithmist is free to employ more flexible methods showing only the \emph{existence} of good, short \emph{QSP-like} protocols. For settings in which standard QSP proof methods fail entirely (e.g., the multivariable setting \cite{rc_m_qsp_22, glw_m_qsp_24, laneve_m_qsp_25}, the continuous variable setting \cite{liu_cv_qsp_24}, or the probabilistic setting \cite{martyn_rall_halving_24}, etc.) such SKT-based techniques may be the only viable approach to formally analyze the analytic properties of these ansätze. It remains an interesting open question whether other powerful techniques, for instance the recently developed harmonic analytic connection between QSP and nonlinear Fourier analysis (NLFA) \cite{amt_23, szego_nlfa_qsp_24, laneve_qsp_nlfa_25}, can offer further flexible tools to understand the functional properties of wide classes of parameterized ansätze.

Beyond presenting independently interesting mathematical questions on uniform approximation in group-valued function spaces and encouraging algorithmic interpretation of wider classes of alternating ansätze, this work also leads to immediate questions on widening the intersection between quantum signal processing and techniques from gate approximation (two highly productive sub-fields which have largely remained separate). In the following final subsection we motivate a selection of open questions which may appeal more immediately to people well-versed in either of these sub-fields.

\subsection{Open problems}

\begin{enumerate}[label=(\arabic*)]
    \item Many of the constants that appear in our results are inherited from basic SKT techniques similar to the `net-refinement' appearing in \cite{dn_skt_overview_05}. It is observed in that work as well as others \cite{nc_textbook_11, hrc_efficient_discrete_02} that careful error analysis, tuned instruction sets \cite{sarnak_letter_15, lps_ramanujan_graphs_88}, as well as the use of non-uniform step sizes and higher-order commutators \cite{kuperberg_skt_23, et_shortest_comm_13} can improve the constant power $c$ to which the $\log{(\varepsilon^{-1})}$ term is raised in the overall gate complexity. Can we make use of similar methods?
    
    \item We show our result for the symmetric QSP ansatz \cite{sym_qsp_21} because its phases are especially easy to numerically compute, and the components of the resulting unitary are simply constrained. However, our methods should extend to a family of paired statements showing (a) that an ansatz is $\Pi$-dense in a class of functions, followed by (b) a net-refinement procedure with a compatible group commutator (Def.~\ref{def:compatible_comm}) showing that the universality in (a) implies the existence of short protocols. Such results would be novel in the multivariable setting \cite{rc_m_qsp_22, rcc_modular_qsp_23}, the G-QSP setting \cite{mw_gqsp_24}, the parallel and randomized setting \cite{mrclc_parallel_qsp_24, martyn_rall_halving_24}, and the continuous variable setting \cite{liu_cv_qsp_24}.

    \item As mentioned in Rem.~\ref{rem:volume_lower_bounds}, it is difficult to use an argument like that of \cite{hrc_efficient_discrete_02} to lower bounds on the word-length required to approximate our class of functions. Moreover, the space we consider (Def.~\ref{def:qsp_skt_function_space}), while compact, is often too `large.' In general QSP/QSVT work especially well when the target function is analytic on the Bernstein ellipse, in which case uniform approximation requires protocol of only \emph{polylogarithmic} length in $\varepsilon^{-1}$ \cite{trefethen_approx_19, cs_qsvt_tang_tian, shao_montanaro_query_24, cwsba_matrix_func_complex_24}. Can we make better use of smoothness assumptions to bound the complexity our initial constant-precision net?
    
    \item In the spirit of (2), which looks to QSP extensions that might permit `SKTs,' we can also ask for SKT techniques with novel quantum algorithmic interpretations when lifted. E.g., very little is known about SU(d) variants of QSP \cite{laneve_qsp_su_n_24} while the SU(d) SKT is standard \cite{dn_skt_overview_05}. More recently, work on inverse-free SKTs \cite{bt_inverse_free_21} have appeared. Can these results be reinterpreted in the QSP setting?
    
    \item Great effort has gone into proving the stability of classical algorithms for finding QSP phases (e.g., the linear-systems based methods of \cite{amt_23, szego_nlfa_qsp_24, ny_fast_phases_24}), with the benefit that these algorithms, even without formal guarantees, perform extremely well in practice \cite{dong_efficient_phases_21}. This work expands those structured ansätze for which we can show the \emph{existence} of short protocols. Empirically, how well do numerical phase finding algorithms work for these extended ansäzte?
    
    \item The SKT appears implicitly in the error-corrected setting, where the instruction set is not only discretized but has special structure (e.g., Clifford+T). QSP in discretized settings is largely unstudied.\footnote{Limited work has been done in the coherently noisy setting, where QSP \emph{can} self-correct certain errors \cite{tltc_alec_23}.} One work along these lines, however, appears in the approximation of $Z$-rotations  \cite{rs_ancilla_free_approx_16}, which might be seen as a variant of QSP where polynomial coefficients and arguments are restricted to certain extensions\footnote{They even have a variant of what in standard QSP is solved by the Fejér-Riesz lemma \cite{polya_szego_analysis_98}, which Ross and Selinger recognize as a Diophantine equation.} of $\mathbb{Z}$. What is the `right' way to discretize QSP? Are there benefits for error correction?
\end{enumerate}

\noindent Having laid out open questions, we will wrap up with the message that, if nothing else is taken from this work, we hope the reader is convinced that
    \begin{equation*}
        \text{QSP}\cap\text{SKT} \neq \varnothing.
    \end{equation*}
Moreover, this abstract intersection between QSP and SKT methods is primed for the application of novel mathematical techniques, with the \emph{practical} benefit of providing insight into the complexity theoretic properties of structured quantum circuits, beyond the reach of standard QSP proof techniques.

\section{Acknowledgments}

\noindent The author is grateful to Rahul Sarkar, Lin Lin, John Martyn, and Isaac Chuang for discussions over an extended period on the topic of this work. ZMR acknowledges funding from the Japan Society for the Promotion of Science (JSPS) Postdoctoral Fellowship for Research in Japan.

\bibliography{main}

\newcommand{\etalchar}[1]{$^{#1}$}
\begin{thebibliography}{DLNW24b}

\bibitem[ALL23]{dll_lchs_23}
Dong An, Jin-Peng Liu, and Lin Lin.
\newblock Linear combination of {H}amiltonian simulation for nonunitary dynamics with optimal state preparation cost.
\newblock {\em Phys. Rev. Lett.}, 131:150603, 2023.

\bibitem[ALM{\etalchar{+}}24]{szego_nlfa_qsp_24}
Michel Alexis, Lin Lin, Gevorg Mnatsakanyan, Christoph Thiele, and Jiasu Wang.
\newblock Infinite quantum signal processing for arbitrary {Szegö} functions.
\newblock {\em arXiv preprint, arXiv:2407.05634}, 2024.

\bibitem[AMT23]{amt_23}
Michel Alexis, Gevorg Mnatsakanyan, and Christoph Thiele.
\newblock Quantum signal processing and nonlinear {Fourier} analysis.
\newblock {\em arXiv preprint, arXiv:2310.12683}, 2023.

\bibitem[BCK15]{bck_ham_lcu_15}
Dominic~W. Berry, Andrew~M. Childs, and Robin Kothari.
\newblock Hamiltonian simulation with nearly optimal dependence on all parameters.
\newblock In {\em {2015 IEEE 56th Annual Symposium on Foundations of Computer Science (FOCS)}}, pages 792--809. IEEE, 2015.

\bibitem[BGT21]{bt_inverse_free_21}
Adam Bouland and Tudor Giurgica-Tiron.
\newblock Efficient universal quantum compilation: An inverse-free {Solovay-Kitaev} algorithm.
\newblock {\em arXiv preprint, arXiv:2112.02040}, 2021.

\bibitem[CDG{\etalchar{+}}20]{chao_machine_prec_20}
Rui Chao, Dawei Ding, Andras Gilyen, Cupjin Huang, and Mario Szegedy.
\newblock Finding angles for quantum signal processing with machine precision.
\newblock {\em arXiv preprint arXiv:2003.02831}, 2020.

\bibitem[CWS{\etalchar{+}}24]{cwsba_matrix_func_complex_24}
Santiago Cifuentes, Samson Wang, Thais~L. Silva, Mario Berta, and Leandro Aolita.
\newblock Quantum computational complexity of matrix functions.
\newblock {\em arXiv preprint, arXiv:2410.13937}, 2024.

\bibitem[DLNW24a]{dlnw_infinite_22}
Yulong Dong, Lin Lin, Hongkang Ni, and Jiasu Wang.
\newblock Infinite quantum signal processing.
\newblock {\em {Quantum}}, 8:1558, 2024.

\bibitem[DLNW24b]{dlnw_robust_iter_23}
Yulong Dong, Lin Lin, Hongkang Ni, and Jiasu Wang.
\newblock Robust iterative method for symmetric quantum signal processing in all parameter regimes.
\newblock {\em SIAM J. Sci. Comput.}, 46(5):A2951--A2971, 2024.

\bibitem[DMWL21]{dong_efficient_phases_21}
Yulong Dong, Xiang Meng, K.~Birgitta Whaley, and Lin Lin.
\newblock Efficient phase-factor evaluation in quantum signal processing.
\newblock {\em Phys. Rev. A}, 103(4), 2021.

\bibitem[DN06]{dn_skt_overview_05}
Christopher~M. Dawson and Michael~A. Nielsen.
\newblock The {Solovay-Kitaev} algorithm.
\newblock {\em Quantum Info. Comput.}, 6(1):81–95, 2006.

\bibitem[DPS90]{top_groups_90}
D.~N. Dikranjan, I.~R. Prodanov, and Luchezar Stoyanov.
\newblock {\em Topological Groups: Characters, Dualities, and Minimal Group Topologies}.
\newblock Monographs and textbooks in pure and applied mathematics. Marcel Dekker, 1990.

\bibitem[DS88]{ds_linear_operators_88}
N.~Dunford and J.~T. Schwartz.
\newblock {\em Linear Operators, Part 1: General Theory}.
\newblock Wiley Classics Library. Wiley, 1988.

\bibitem[ET13]{et_shortest_comm_13}
Abdelrhman Elkasapy and Andreas Thom.
\newblock On the length of the shortest non-trivial element in the derived and the lower central series.
\newblock {\em arXiv preprint, arXiv:1311.0138}, 2013.

\bibitem[GJ79]{gj_cluster_79}
I.~Goulden and D.~M. Jackson.
\newblock An inversion theorem for cluster decompositions of sequences with distinguished subsequences.
\newblock {\em J. London Math. Soc.}, 2(20):567--576, 1979.

\bibitem[GJ83]{gj_cluster_book_83}
I.~Goulden and D.~M. Jackson.
\newblock {\em Combinatorial Enumeration}.
\newblock John Wiley, New York, 1983.

\bibitem[GLW24]{glw_m_qsp_24}
Niladri Gomes, Hokiat Lim, and Nathan Wiebe.
\newblock Multivariable {QSP} and bosonic quantum simulation using iterated quantum signal processing.
\newblock {\em arXiv preprint, arXiv:2408.03254}, 2024.

\bibitem[GSLW19]{gslw_19}
András Gilyén, Yuan Su, Guang~Hao Low, and Nathan Wiebe.
\newblock Quantum singular value transformation and beyond: exponential improvements for quantum matrix arithmetics.
\newblock {\em Proceedings of the 51st Annual ACM SIGACT Symposium on Theory of Computing}, 2019.

\bibitem[Haa19]{haah_2019}
Jeongwan Haah.
\newblock Product decomposition of periodic functions in quantum signal processing.
\newblock {\em Quantum}, 3:190, 2019.

\bibitem[HRC02]{hrc_efficient_discrete_02}
Aram~W. Harrow, Benjamin Recht, and Isaac~L. Chuang.
\newblock Efficient discrete approximations of quantum gates.
\newblock {\em J. Math. Phys}, 43(9):4445–4451, 2002.

\bibitem[Jor75]{jordan_75}
Camille Jordan.
\newblock Essai sur la g{\'e}om{\'e}trie {\`a} $ n $ dimensions.
\newblock {\em Bulletin de la Soci{\'e}t{\'e} math{\'e}matique de France}, 3:103--174, 1875.

\bibitem[{Kit}97]{kitaev_q_algs_ecc_97}
A.~Yu {Kitaev}.
\newblock {Quantum computations: algorithms and error correction}.
\newblock {\em Russ. Math. Surv.}, 52(6):1191--1249, 1997.

\bibitem[KSV02]{ksv_textbook_02}
A.~Y. Kitaev, A.~Shen, and M.~N. Vyalyi.
\newblock {\em Classical and Quantum Computation}.
\newblock Graduate studies in mathematics. AMS, 2002.

\bibitem[Kup23]{kuperberg_skt_23}
Greg Kuperberg.
\newblock Breaking the cubic barrier in the {Solovay-Kitaev} algorithm.
\newblock {\em arXiv preprint, arXiv:2306.13158}, 2023.

\bibitem[Lan24]{laneve_qsp_su_n_24}
Lorenzo Laneve.
\newblock Quantum signal processing over {SU(N)}.
\newblock {\em arXiv preprint, arXiv:2311.03949}, 2024.

\bibitem[Lan25]{laneve_qsp_nlfa_25}
Lorenzo Laneve.
\newblock {Generalized Quantum Signal Processing and Non-Linear Fourier Transform are equivalent}.
\newblock {\em arXiv preprint, arXiv:2503.03026}, 2025.

\bibitem[LC17]{lc_17_simulation}
G.~H. Low and I.~L. Chuang.
\newblock Optimal {H}amiltonian simulation by quantum signal processing.
\newblock {\em Phys. Rev. Lett.}, 118:010501, 2017.

\bibitem[LC19]{lc_19_qubitization}
G.~H. Low and I.~L. Chuang.
\newblock Hamiltonian simulation by qubitization.
\newblock {\em Quantum}, 3:163, 2019.

\bibitem[Llo95]{lloyd_gate_set_universal_95}
Seth Lloyd.
\newblock Almost any quantum logic gate is universal.
\newblock {\em Phys. Rev. Lett.}, 75:346--349, 1995.

\bibitem[LMSS{\etalchar{+}}24]{liu_cv_qsp_24}
Yuan Liu, John~M. Martyn, Jasmine Sinanan-Singh, Kevin~C. Smith, Steven~M. Girvin, and Isaac~L. Chuang.
\newblock Toward mixed analog-digital quantum signal processing: Quantum {AD/DA} conversion and the {Fourier} transform.
\newblock {\em arXiv preprint, arXiv:2408.14729}, 2024.

\bibitem[LPS88]{lps_ramanujan_graphs_88}
Alexander Lubotzky, Ralph Phillips, and Peter Sarnak.
\newblock Ramanujan graphs.
\newblock {\em Combinatorica}, 8(3):261--277, 1988.

\bibitem[LS24]{low_su_qevt_24}
Guang~Hao Low and Yuan Su.
\newblock Quantum eigenvalue processing.
\newblock In {\em {2024 IEEE 65th Annual Symposium on Foundations of Computer Science (FOCS)}}, page 1051–1062. IEEE, 2024.

\bibitem[LW25]{laneve_m_qsp_25}
Lorenzo Laneve and Stefan Wolf.
\newblock On multivariate polynomials achievable with quantum signal processing.
\newblock {\em {Quantum}}, 9:1641, 2025.

\bibitem[LYC16]{lyc_16_equiangular_gates}
G.~H. Low, T.~J. Yoder, and I.~L. Chuang.
\newblock Methodology of resonant equiangular composite quantum gates.
\newblock {\em Phys. Rev. X}, 6:041067, 2016.

\bibitem[MF24]{mf_recursive_24}
Kaoru Mizuta and Keisuke Fujii.
\newblock Recursive quantum eigenvalue and singular-value transformation: analytic construction of matrix sign function by {Newton} iteration.
\newblock {\em Phys. Rev. Res.}, 6:L012007, 2024.

\bibitem[MFM23]{mori_m_qsp_comment_23}
Hitomi Mori, Keisuke Fujii, and Kaoru Mizuta.
\newblock Comment on ``{M}ultivariable quantum signal processing ({M-QSP}): prophecies of the two-headed oracle''.
\newblock {\em arXiv preprint arXiv:2310.00918}, 2023.

\bibitem[MR25]{martyn_rall_halving_24}
John~M. Martyn and Patrick Rall.
\newblock Halving the cost of quantum algorithms with randomization.
\newblock {\em npj Quantum Inf.}, 11(47), 2025.

\bibitem[MRC{\etalchar{+}}24]{mrclc_parallel_qsp_24}
John~M. Martyn, Zane~M. Rossi, Kevin~Z. Cheng, Yuan Liu, and Isaac~L. Chuang.
\newblock Parallel quantum signal processing via polynomial factorization.
\newblock {\em arXiv preprint, arXiv:2409.19043}, 2024.

\bibitem[MRTC21]{mrtc_21}
John~M. Martyn, Zane~M. Rossi, Andrew~K. Tan, and Isaac~L. Chuang.
\newblock Grand unification of quantum algorithms.
\newblock {\em {PRX} Quantum}, 2(4), 2021.

\bibitem[MS24]{shao_montanaro_query_24}
Ashley Montanaro and Changpeng Shao.
\newblock Quantum and classical query complexities of functions of matrices.
\newblock In {\em Proceedings of the 56th Annual ACM Symposium on Theory of Computing (STOC)}, page 573–584, New York, NY, USA, 2024. Association for Computing Machinery.

\bibitem[MW24]{mw_gqsp_24}
Danial Motlagh and Nathan Wiebe.
\newblock Generalized quantum signal processing.
\newblock {\em PRX Quantum}, 5:020368, 2024.

\bibitem[NC11]{nc_textbook_11}
Michael~A. Nielsen and Isaac~L. Chuang.
\newblock {\em Quantum {C}omputation and {Q}uantum {I}nformation}.
\newblock Cambridge University Press, USA, 10th edition, 2011.

\bibitem[NKK{\etalchar{+}}23]{nemeth_m_qsp_23}
Bal{\'a}zs N{\'e}meth, Blanka K{\"o}v{\'e}r, Bogl{\'a}rka Kulcs{\'a}r, Roland~Botond Mikl{\'o}si, and Andr{\'a}s Gily{\'e}n.
\newblock On variants of multivariate quantum signal processing and their characterizations.
\newblock {\em arXiv preprint arXiv:2312.09072}, 2023.

\bibitem[NY24]{ny_fast_phases_24}
Hongkang Ni and Lexing Ying.
\newblock Fast phase factor finding for quantum signal processing.
\newblock {\em arXiv preprint, arXiv:2410.06409}, 2024.

\bibitem[NZ99]{nz_gj_cluster_method_99}
John Noonan and Doron Zeilberger.
\newblock The {G}oulden-{J}ackson cluster method: extensions, applications and implementations.
\newblock {\em J. Differ. Equ. Appl.}, 5(4-5):355--377, 1999.

\bibitem[PS98]{polya_szego_analysis_98}
George Polya and Gabor Szegö.
\newblock {\em Problems and Theorems in Analysis II}.
\newblock Springer, Berlin, 1998.

\bibitem[PW94]{pw_cs_decomp_94}
Christopher~C. Paige and Musheng Wei.
\newblock History and generality of the {CS} decomposition.
\newblock {\em Linear Algebra Appl.}, 208:303--326, 1994.

\bibitem[RC22]{rc_m_qsp_22}
Zane~M. Rossi and Isaac~L. Chuang.
\newblock Multivariable quantum signal processing ({M}-{QSP}): prophecies of the two-headed oracle.
\newblock {\em {Quantum}}, 6:811, 2022.

\bibitem[RCC23]{rcc_modular_qsp_23}
Zane~M. Rossi, Jack~L. Ceroni, and Isaac~L. Chuang.
\newblock Modular quantum signal processing in many variables.
\newblock {\em arXiv preprint, arXiv:2309.16665}, 2023.

\bibitem[Ros24]{rossi_fqa_thesis_24}
Z.~M. Rossi.
\newblock {\em Functional quantum algorithms: a mélange of methods for matrix functions}.
\newblock PhD thesis, Massachusetts Institute of Technology, 2024.

\bibitem[RS16]{rs_ancilla_free_approx_16}
Neil~J. Ross and Peter Selinger.
\newblock Optimal ancilla-free {Clifford+T} approximation of z-rotations.
\newblock {\em Quantum Info. Comput.}, 16(11–12):901–953, 2016.

\bibitem[Sar15]{sarnak_letter_15}
Peter Sarnak.
\newblock Letter to {S}cott {A}aronson and {A}ndy {P}ollington on the {S}olovay–{K}itaev {T}heorem and golden gates, 2015.
\newblock \url{https://publications.ias.edu/sites/default/files/Letter%20-%20golden%20gates%20march_0.pdf}.

\bibitem[TLTC23]{tltc_alec_23}
Andrew~K. Tan, Yuan Liu, Minh~C. Tran, and Isaac~L. Chuang.
\newblock Error correction of quantum algorithms: Arbitrarily accurate recovery of noisy quantum signal processing.
\newblock {\em arXiv preprint arXiv:2301.08542}, 2023.

\bibitem[Tre19]{trefethen_approx_19}
Lloyd~N. Trefethen.
\newblock {\em Approximation theory and approximation practice, extended edition}.
\newblock SIAM, 2019.

\bibitem[TT12]{tt_nlfa_notes_12}
Terence Tao and Christoph Thiele.
\newblock Nonlinear {Fourier} analysis.
\newblock {\em arXiv preprint, arXiv:1201.5129}, 2012.

\bibitem[TT23]{cs_qsvt_tang_tian}
Ewin Tang and Kevin Tian.
\newblock A {CS} guide to the quantum singular value transformation.
\newblock {\em arXiv preprint arXiv:2302.14324}, 2023.

\bibitem[WDL22]{sym_qsp_21}
Jiasu Wang, Yulong Dong, and Lin Lin.
\newblock On the energy landscape of symmetric quantum signal processing.
\newblock {\em {Quantum}}, 6:850, 2022.

\bibitem[YLC14]{ylc_14}
Theodore~J. Yoder, Guang~Hao Low, and Isaac~L. Chuang.
\newblock Fixed-point quantum search with an optimal number of queries.
\newblock {\em Phys. Rev. Lett.}, 113(21):210501, 2014.

\end{thebibliography}

\end{document}